\documentclass[11pt]{article}

\usepackage{pstricks}
\usepackage{pst-plot}
\usepackage{pst-node}
\usepackage{amsmath}
\usepackage{pst-grad}
\usepackage{amssymb}
\usepackage{xspace}
\usepackage{pst-math}
\usepackage{pst-xkey}
\usepackage{pst-coil}
\usepackage{pst-3dplot}
\usepackage[T1]{fontenc}
\usepackage{ae}
\usepackage[ansinew]{inputenc}

\newrgbcolor{darkred}{0.8 0 0}
\newrgbcolor{darkerred}{0.7 0 0}
\newrgbcolor{darkestred}{0.5 0 0}
\newrgbcolor{darkgreen}{0 0.5 0.5}
\newrgbcolor{darkergreen}{0 0.6 0}
\newrgbcolor{darkestgreen}{0 0.5 0}
\newrgbcolor{darkblue}{0 0 0.6}
\newrgbcolor{darkestblue}{0 0 0.5}
\newrgbcolor{redgreen}{0.8 0.8 0}
\newrgbcolor{bluegreen}{0 0.4 0.7}
\newrgbcolor{bluewhite}{0 0.9 1}
\newrgbcolor{lightgreen}{0.6 1 0.6}
\newrgbcolor{lightred}{1 0.8 0.8}
\newrgbcolor{lightyred}{1 0.4 0.4}
\newrgbcolor{lightyblue}{0.5 0.5 1}
\newrgbcolor{lightblue}{0.8 1 1}
\newrgbcolor{lilas}{0.5 0.3 0.7}
\newrgbcolor{lilaswhite}{1 0.8 1}
\newrgbcolor{redy}{1 0.6 0.6}
\newrgbcolor{bluish}{0.6 0.6 1}
\newrgbcolor{sand}{0.9 0.9 0.7}
\newrgbcolor{brown}{0.4 0.4 0}
\newrgbcolor{blue1}{0 0.35 0.65}
\newrgbcolor{blue2}{0 0.4 0.6}
\newrgbcolor{blue3}{0 0.45 0.55}
\newrgbcolor{blue4}{0 0.5 0.5}
\newrgbcolor{blue5}{0 0.55 0.45}
\newrgbcolor{blue6}{0 0.6 0.4}
\newrgbcolor{blue7}{0 0.65 0.35}

\newcommand{\R }{\mathbb R}
\newcommand{\N }{\mathbb N}
\newcommand{\Z }{\mathbb Z}
\newcommand{\Q }{\mathbb Q}
\newcommand{\C }{\mathbb C}
\newcommand{\sen }{\ \! {\rm sen}\ \! }
\newcommand{\tg }{\ \! {\rm tg}\ \! }
\newcommand{\cosec }{\ \! {\rm cosec}\ \! }
\newcommand{\cotg }{\ \! {\rm cotg}\ \! }
\newcommand{\dx }{\ \! dx}
\newcommand{\e }{\ \! e}
\newcommand{\senh }{\ \! {\rm senh}\ \! }
\newcommand{\tgh }{\ \! {\rm tgh}\ \! }
\newcommand{\cotgh }{\ \! {\rm cotgh}\ \! }
\newcommand{\sech }{\ \! {\rm sech}\ \! }
\newcommand{\cosech }{\ \! {\rm cosech}\ \! }
\newcommand{\arcsen }{\ \! {\rm arcsen}\ \! }
\newcommand{\arctg }{\ \! {\rm arctg}\ \! }
\newcommand{\arccotg }{\ \! {\rm arccotg}\ \! }
\newcommand{\arcsenh }{\ \! {\rm arcsenh}\ \! }
\newcommand{\arccosh }{\ \! {\rm arccosh}\ \! }
\newcommand{\arctgh }{\ \! {\rm arctgh}\ \! }
\newcommand{\arccosech }{\ \! {\rm arccosech}\ \! }
\newcommand{\arcsech }{\ \! {\rm arcsech}\ \! }
\newcommand{\arcsec }{\ \! {\rm arcsec}\ \! }
\newcommand{\arccosec }{\ \! {\rm arccosec}\ \! }
\renewcommand{\arctgh }{\ \! {\rm arctgh}\ \! }
\newcommand{\arccotgh }{\ \! {\rm arccotgh}\ \! }
\newcommand{\nega }{\neg \ }
\newcommand{\eq }{\Leftrightarrow }

\begin{document}

\title{Pruning a Minimum Spanning Tree}

\author{Leonidas Sandoval Junior \\ \\ Insper, Instituto de Ensino e Pesquisa}

\maketitle

\begin{abstract}
This work employs some techniques in order to filter random noise from the information provided by minimum spanning trees obtained from the correlation matrices of international stock market indices prior to and during times of crisis. The first technique establishes a threshold above which connections are considered affected by noise, based on the study of random networks with the same probability density distribution of the original data. The second technique is to judge the strengh of a connection by its survival rate, which is the amount of time a connection between two stock market indices endure. The idea is that true connections will survive for longer periods of time, and that random connections will not. That information is then combined with the information obtained from the first technique in order to create a smaller network, where most of the connections are either strong or enduring in time.
\end{abstract}

\section{Introduction}

Minimum spanning trees are networks of nodes that are all connected by at least one edge so that the sum of the edges is minimum, and which present no loops. This kind of tree is particularly useful for representing complex networks, filtering the information about the correlations between all nodes and presenting it in a planar graph. Because of this simplicity, minimum spanning trees have been widely used to represent some important financial structures, namely the structures of stock exchanges \cite{h1}-\cite{h42}, of currency exchange rates \cite{h54}-\cite{h39}, of world trade \cite{h33}-\cite{h36}, commodities \cite{commodity}, of GDPs (Gross Domestic Products) \cite{h24}, corporations \cite{corp}-\cite{corpnet} and, most important for this article, of the world financial markets \cite{h13}, \cite{h29}, \cite{h19}-\cite{wtm5}.

Like any other representation of real world interactions, it is subject to a great amount of random noise, which is sometimes difficult to isolate. This work employs some techniques in order to filter random noise from the information provided by a minimum spanning tree. The networks that are represented are obtained from the correlations between international stock exchange indices prior to and during years of well known international financial crises, namely the 1987 Black Monday, the 1997 Asian Financial Crisis, the 1998 Russian Crisis, the crisis after September, 11, 2001, and the Subprime Mortgage crisis of 2008. The reason for choosing international stock exchange indices is because one has some idea of which indices should be more correlated based on information that comes from other sources that are not the correlation matrix, such as geographical proximity, cultural affinity, and strength of commercial relations \cite{h33}, and also because one can then follow the interactions of stock markets at different periods of time and in different volatility regimes.

One of the techniques is to establish a threshold above which connections are considered affected by noise based on the study of random networks with the same probability density distribution of the original data, what is obtained by shuffling the original data so that correlations between the various nodes are basically random. This eliminates both random connections and true ones, what can be shown by using arguments of geographical, cultural, and financial links between the stock exchanges. This effect is particularly strong for emerging or small financial markets.

The second technique is to judge the strengh of a connection by its survival rate, which is the amount of time a connection between two stock market indices endure. The idea is that true connections will survive for longer periods of time, and that random connections will not. That information is then combined with the information coming from the first technique in order to create a smaller network, where most of the connections are either strong or enduring in time. The resulting network is a k-minimum spanning tree \cite{kmst1}-\cite{kmst3}, with some additional nodes and edges attached to it, and the result is then studied using the tools of network analysis.

This study brings out important information, like the existence of an American and of an European cluster, both strongly connected in themselves but weakly connected with one another. The study also brings out the importance of the Netherlands in the 80's and 90's, and the emergence of France as the main hub for the European cluster. The study also shows how a Pacific Asian cluster was formed during the last two decades, and of how smaller, weakly interacting clusters, like the one formed by some members of the former Yugoslavia, or the one of Arab indices, have formed in the last decade.

\section{Correlations and distance}

The time series of financial market indices encode an enormous amount of information about the way they relate to each other. Part of this information may be captured by the correlation matrix of their log-returns, which are defined as
\begin{equation}
S_t=\ln (P_t)-\ln (P_{t-1})\approx \frac{P_t-P_{t-1}}{P_t}\ ,
\end{equation}
where $P_t$ is the value of a certain index at time $t$ and $P_{t-1}$ is the value of the same index at time $t-1$.

There are many measures of correlation between elements of time series, the most popular being the Pearson correlation coefficient. The drawback of this correlation measure is that it only detects linear relationships between two variables. Two different measures that can detect nonlinear relations are the Spearman and the Kendall tau rank correlations, which measure the extent to which the variation of one variable affects other variable, withouth that relation necessarily being linear. In this work, we choose Spearman's rank correlation, for it is fairly fast to calculate.

The correlation matrix between log-returns of financial market indices may be used to classify the markets from which it is made into clusters, or into ``neighbors''. For that, we use the process first developed in \cite{h1} and used in most articles, which considers a suitable distance function between the markets based on the correlation function, and then uses the Minimum Spanning Tree technique in order to establish links between them.

An Euclidean metric must fulfill the axioms $d_{ij}=0\Leftrightarrow i=j$, $d_{ij}=d_{ji}$, and $d_{ij}\leq d_{ik}+d_{kj}$. In this work, I shall use a different metric from reference \cite{h1}, which is a nonlinear mapping of the Pearson correlation coefficients between stock returns. The metric to be considered here differs from the aforementioned metric because it is a linear realization of the Spearman rank correlation coefficient between the indices that are being studied:
\begin{eqnarray}
\label{metric2}
 & & d_{ij}=1-c_{ij}\ ,
\end{eqnarray}
where $c_{ij}$ are elements of the correlation matrix calculated using Sperman's rank correlation. This distance goes from the minimum value 0 (correlation 1) to the maximum value 2 (correlation -1). The linear mapping is used in order to best identify the relation between correlation and distance.

Using the metric given by (\ref{metric2}), one obtains a distance matrix, over which can be applied the Minimum Spanning Tree technique, which consists of choosing one of the indices (nodes) and finding the next node which is closest to that one, and then to link them. One then considers the cluster formed by those two nodes and finds the next node that is closest to any of the members of that cluster, and links it to the node it is closest to inside the cluster. The process then goes on until there are no free nodes left.

As an example, using the correlation matrix for 1987, starting from the S\&P 500 index, the Nasdaq is the closest index to it. So, one establishes a link between them and looks for the next node (index) that is closest to one of the two indices in the cluster. It turns out to be the S\&P TSX (Canada), which is closest to the node Nasdaq. So, one establishes a link between Nasdaq and S\&P TSX. Following this procedure, one obtains a planar graph with no loops that shows relationships between the markets that are being considered.

Although the minimum spanning tree is a valuable tool for the study of connections between different assets or indices, it has some limitations. One of them is that it sometimes overstates connections that are dim, leading sometimes to the illusion that two indices are strongly connected when they are not. Some countries that have long distances from any other country and are actually wanderers in the landscape of the correlation ties are connected almost at random in a minimum spanning tree.

One way to decide wether a connection made by the Minimum Spanning Tree technique is random or not is to compare the results of real data with those obtained for a minimium spanning tree that is built from random interactions. This is discussed in the next section.

\section{Simulations with random matrices}

Figure 1 shows the minimum spanning tree obtained from the Spearman correlation of 40 randomly generated returns, based on a Gaussian distribution with zero mean and standard deviation 2. One caracteristic of this graph is that distances vary little, averaging $0.8$. Figure 2 displays the frequency distribution of distances for this particular minimum spanning tree.

The other remarkable fact is the distribution of node degrees, where a node degree is the number of connections a particular node has with other nodes (as an example, node 14 has three connections, so it has node degree 3). Figure 3 shows the frequency distribution for this particular minimum spanning tree realization of the random data.

\begin{pspicture}(-9,-4.5)(1,7.2)
\psset{xunit=2,yunit=2} \scriptsize
\psline(0,0)(-0.851,0) 
\psline(-0.851,0)(-1.684,0) 
\psline(0,0)(0.790,0) 
\psline(0.790,0)(1.629,0) 
\psline(1.629,0)(2.467,0) 
\psline(0,0)(0,0.790) 
\psline(0,0.790)(0.859,0.790) 
\psline(0,0.790)(-0.814,0.790) 
\psline(-0.814,0.790)(-1.707,0.790) 
\psline(2.467,0)(2.467,0.754) 
\psline(1.629,0)(1.629,0.862) 
\psline(1.629,0)(1.629,-0.808) 
\psline(0.790,0)(1.344,-0.554) 
\psline(0.790,0)(0.790,-0.808) 
\psline(0.790,-0.808)(0.790,-1.627) 
\psline(0,0)(0,-0.821) 
\psline(0,-0.821)(0,-1.622) 
\psline(0,-0.821)(-0.767,-0.821) 
\psline(0,-1.622)(0.599,-2.221) 
\psline(0,-1.622)(-0.554,-2.176) 
\psline(0.599,-2.221)(1.417,-2.221) 
\psline(1.417,-2.221)(2.271,-2.221) 
\psline(-0.554,-2.176)(-1.336,-2.176) 
\psline(-1.336,-2.176)(-2.160,-2.176) 
\psline(-0.814,0.790)(-0.814,1.591) 
\psline(-0.814,1.591)(-1.615,1.591) 
\psline(-1.615,1.591)(-2.429,1.591) 
\psline(-0.814,1.591)(-0.024,1.591) 
\psline(-0.024,1.591)(0.742,1.591) 
\psline(0.742,1.591)(0.742,2.483) 
\psline(-0.024,1.591)(-0.024,2.384) 
\psline(-1.615,1.591)(-1.615,2.361) 
\psline(-1.615,2.361)(-0.866,2.361) 
\psline(-1.615,2.361)(-2.424,2.361) 
\psline(-2.424,2.361)(-3.238,2.361) 
\psline(-2.424,2.361)(-2.424,3.202) 
\psline(-2.424,2.361)(-3.017,1.768) 
\psline(-3.238,2.361)(-3.238,3.147) 
\psline(-3.238,2.361)(-3.238,1.486) 
\psdot[linecolor=blue,linewidth=1.2pt](0,0) 
\psdot[linecolor=blue,linewidth=1.2pt](-0.851,0) 
\psdot[linecolor=blue,linewidth=1.2pt](-1.684,0) 
\psdot[linecolor=blue,linewidth=1.2pt](0.790,0) 
\psdot[linecolor=blue,linewidth=1.2pt](1.629,0) 
\psdot[linecolor=blue,linewidth=1.2pt](2.467,0) 
\psdot[linecolor=blue,linewidth=1.2pt](0,0.790) 
\psdot[linecolor=blue,linewidth=1.2pt](0.859,0.790) 
\psdot[linecolor=blue,linewidth=1.2pt](-0.814,0.790) 
\psdot[linecolor=blue,linewidth=1.2pt](-1.707,0.790) 
\psdot[linecolor=blue,linewidth=1.2pt](2.467,0.754) 
\psdot[linecolor=blue,linewidth=1.2pt](1.629,0.862) 
\psdot[linecolor=blue,linewidth=1.2pt](1.629,-0.808) 
\psdot[linecolor=blue,linewidth=1.2pt](1.344,-0.554) 
\psdot[linecolor=blue,linewidth=1.2pt](0.790,-0.808) 
\psdot[linecolor=blue,linewidth=1.2pt](0.790,-1.627) 
\psdot[linecolor=blue,linewidth=1.2pt](0,-0.821) 
\psdot[linecolor=blue,linewidth=1.2pt](0,-1.622) 
\psdot[linecolor=blue,linewidth=1.2pt](-0.767,-0.821) 
\psdot[linecolor=blue,linewidth=1.2pt](-0.554,-2.176) 
\psdot[linecolor=blue,linewidth=1.2pt](0.599,-2.221) 
\psdot[linecolor=blue,linewidth=1.2pt](-1.336,-2.176) 
\psdot[linecolor=blue,linewidth=1.2pt](-2.160,-2.176) 
\psdot[linecolor=blue,linewidth=1.2pt](1.417,-2.221) 
\psdot[linecolor=blue,linewidth=1.2pt](2.271,-2.221) 
\psdot[linecolor=blue,linewidth=1.2pt](-0.814,1.591) 
\psdot[linecolor=blue,linewidth=1.2pt](-1.615,1.591) 
\psdot[linecolor=blue,linewidth=1.2pt](-2.429,1.591) 
\psdot[linecolor=blue,linewidth=1.2pt](-0.024,1.591) 
\psdot[linecolor=blue,linewidth=1.2pt](0.742,1.591) 
\psdot[linecolor=blue,linewidth=1.2pt](0.742,2.483) 
\psdot[linecolor=blue,linewidth=1.2pt](-0.024,2.384) 
\psdot[linecolor=blue,linewidth=1.2pt](-1.615,2.361) 
\psdot[linecolor=blue,linewidth=1.2pt](-0.866,2.361) 
\psdot[linecolor=blue,linewidth=1.2pt](-2.424,2.361) 
\psdot[linecolor=blue,linewidth=1.2pt](-3.238,2.361) 
\psdot[linecolor=blue,linewidth=1.2pt](-2.424,3.202) 
\psdot[linecolor=blue,linewidth=1.2pt](-3.017,1.768) 
\psdot[linecolor=blue,linewidth=1.2pt](-3.238,3.147) 
\psdot[linecolor=blue,linewidth=1.2pt](-3.238,1.486) 
\rput(0.15,0.15){30}
\rput(-0.851,0.15){25}
\rput(-1.684,0.15){1}
\rput(0.790,0.15){22}
\rput(1.779,0.15){24}
\rput(2.617,0.15){35}
\rput(0,0.940){3}
\rput(0.859,0.940){17}
\rput(-0.664,0.940){15}
\rput(-1.707,0.940){20}
\rput(2.467,0.904){39}
\rput(1.629,1.012){18}
\rput(1.629,-0.958){9}
\rput(1.344,-0.704){29}
\rput(0.640,-0.808){13}
\rput(0.790,-1.777){7}
\rput(0.15,-0.821){6}
\rput(0.15,-1.572){23}
\rput(-0.767,-0.671){11}
\rput(-0.704,-2.026){21}
\rput(0.749,-2.026){36}
\rput(-1.336,-2.026){32}
\rput(-2.160,-2.026){38}
\rput(1.417,-2.026){40}
\rput(2.271,-2.026){37}
\rput(-0.814,1.741){14}
\rput(-1.765,1.741){26}
\rput(-2.429,1.741){5}
\rput(-0.174,1.741){2}
\rput(0.892,1.741){16}
\rput(0.742,2.633){4}
\rput(-0.024,2.534){10}
\rput(-1.615,2.511){28}
\rput(-0.866,2.511){31}
\rput(-2.274,2.511){33}
\rput(-3.088,2.511){34}
\rput(-2.424,3.352){27}
\rput(-3.017,1.618){19}
\rput(-3.238,3.297){8}
\rput(-3.238,1.336){12}
\end{pspicture}

\begin{center}
Figure 1: minimum spanning tree for randomly generated data with forty vertices.
\end{center}

\vskip 0.3 cm

In general, minimum spanning trees generated from real data will differ significantly from this behavior: the distances vary over a larger scope than the distances calculated for random data, and the node degree frequency distribution is also remarkably different, with nodes connecting to many more nodes than expected for random connections. Nevertheless, some of the distances obtained from real data will be the result of random noise, and comparison with the results obtained for random data shall be useful in order to separate what are real connections from random ones.

Actually, we shall not use minimum spanning trees based on random Gaussian returns, for the distributions of returns of financial assets are rarely Gaussian. Instead, we shall follow the approach already used by many authors [refs] and work with flushed data, which is obtained by reordering the time series of each index at random so as to break down any correlation between indices but preserve the same probability distribution.

\begin{pspicture}(-1,0)(3.5,6.1)
\psset{xunit=1,yunit=0.2}
\pspolygon*[linecolor=lightblue](0.7,0)(0.7,2)(1.4,2)(1.4,0)
\pspolygon*[linecolor=lightblue](1.4,0)(1.4,10)(2.1,10)(2.1,0)
\pspolygon*[linecolor=lightblue](2.1,0)(2.1,14)(2.8,14)(2.8,0)
\pspolygon*[linecolor=lightblue](2.8,0)(2.8,8)(3.5,8)(3.5,0)
\pspolygon*[linecolor=lightblue](3.5,0)(3.5,6)(4.2,6)(4.2,0)
\pspolygon*[linecolor=lightblue](4.2,0)(4.2,2)(4.9,2)(4.9,0)
\pspolygon[linecolor=blue](0.7,0)(0.7,2)(1.4,2)(1.4,0)
\pspolygon[linecolor=blue](1.4,0)(1.4,10)(2.1,10)(2.1,0)
\pspolygon[linecolor=blue](2.1,0)(2.1,14)(2.8,14)(2.8,0)
\pspolygon[linecolor=blue](2.8,0)(2.8,8)(3.5,8)(3.5,0)
\pspolygon[linecolor=blue](3.5,0)(3.5,6)(4.2,6)(4.2,0)
\pspolygon[linecolor=blue](4.2,0)(4.2,2)(4.9,2)(4.9,0)
\psline{->}(0,0)(5.5,0) \psline{->}(0,0)(0,25) \rput(6.5,0){distance} \rput(1.1,25){frequency} \scriptsize \psline(0.7,-0.5)(0.7,0.5) \rput(0.7,-1.5){0.73} \psline(1.4,-0.5)(1.4,0.5) \rput(1.4,-1.5){0.76} \psline(2.1,-0.5)(2.1,0.5) \rput(2.1,-1.5){0.79} \psline(2.8,-0.5)(2.8,0.5) \rput(2.8,-1.5){0.82} \psline(3.5,-0.5)(3.5,0.5) \rput(3.5,-1.5){0.85} \psline(4.2,-0.5)(4.2,0.5) \rput(4.2,-1.5){0.88} \psline(4.9,-0.5)(4.9,0.5) \rput(4.9,-1.5){0.91} \psline(-0.1,5)(0.1,5) \rput(-0.4,5){$5$} \psline(-0.1,10)(0.1,10) \rput(-0.4,10){$10$} \psline(-0.1,15)(0.1,15) \rput(-0.4,15){$15$} \psline(-0.1,20)(0.1,20) \rput(-0.4,20){$20$}
\small \rput(3.5,-4.4){Figure 2: frequency distribution of the distances} \rput(3.33,-6.7){for a minimum spanning tree obtained from 40} \rput(1,-9){random returns.}
\end{pspicture}
\begin{pspicture}(-5,0)(3.5,6.1)
\psset{xunit=1,yunit=0.2}
\pspolygon*[linecolor=lightred](0.5,0)(0.5,19)(1.5,19)(1.5,0)
\pspolygon*[linecolor=lightred](1.5,0)(1.5,8)(2.5,8)(2.5,0)
\pspolygon*[linecolor=lightred](2.5,0)(2.5,9)(3.5,9)(3.5,0)
\pspolygon*[linecolor=lightred](3.5,0)(3.5,4)(4.5,4)(4.5,0)
\pspolygon[linecolor=red](0.5,0)(0.5,19)(1.5,19)(1.5,0)
\pspolygon[linecolor=red](1.5,0)(1.5,8)(2.5,8)(2.5,0)
\pspolygon[linecolor=red](2.5,0)(2.5,9)(3.5,9)(3.5,0)
\pspolygon[linecolor=red](3.5,0)(3.5,4)(4.5,4)(4.5,0)
\psline{->}(0,0)(5,0) \psline{->}(0,0)(0,25) \rput(6.3,0){node degree} \rput(1.1,25){frequency} \scriptsize \psline(1,-0.5)(1,0.5) \rput(1,-1.5){1} \psline(2,-0.5)(2,0.5) \rput(2,-1.5){2} \psline(3,-0.5)(3,0.5) \rput(3,-1.5){3}  \psline(4,-0.5)(4,0.5) \rput(4,-1.5){4} \psline(-0.1,5)(0.1,5) \rput(-0.4,5){$5$} \psline(-0.1,10)(0.1,10) \rput(-0.4,10){$10$} \psline(-0.1,15)(0.1,15) \rput(-0.4,15){$15$} \psline(-0.1,20)(0.1,20) \rput(-0.4,20){$20$}
\small \rput(3.5,-4){Figure 3: frequency distribution of the node degrees} \rput(3.1,-6.3){for a minimum spanning tree obtained from 40} \rput(0.77,-8.6){random returns.}
\end{pspicture}

\vskip 2.2 cm

\section{Minimum spanning trees}

In what follows, we shall calculate the minimum spanning trees for a diversity of international stock exchange indices around key periods in the past decades. The data were chosen so as to present a variety of economies and also offer a diversity of volatility regimes. More specifically, we chose the years of the last major world financial crises, namely the 1987 Black Monday, the 1997 Asian Financial Crisis, the 1998 Russian Crisis, the crisis after September, 11, 2001, and the Subprime Mortgage crisis of 2008, and the years that preceded them (with the exception of the Asian Financial Crisis). As those crises usually occurred in the second and sometimes in the first semester of a year, we separated data into two semesters for each year that was studied, so that we may then study the evolution of the interactions among markets previous to and during financial crises when seen through the eyes of minimum spanning trees. The data used in this article is essentially the same as in \cite{leocorr}, and details of the indices used, and their codes, can be found in that article (or in Appendix A of the present article), which also offers a comprehensive list of bibliography on the use of Random Matrix Theory in Finance.

\subsection{1987 - Black Monday}

We first analyze the first and second semesters of 1986 and 1987. The crisis that occurred in October, 1987, caused the loss of trillions of dollars worldwide, as some markets lost some 30\% of their values in a few days. We use, for this study, the indices of 16 markets, more specifically, those from the USA (S\&P 500 and Nasdaq), Canada, Brazil, UK, Germany, Austria, Netherlands, India, Sri Lanka, Japan, Hong Kong, Taiwan, South Korea, Malaysia, and Indonesia. The names of the indices and stock exchanges to which they belong are listed in Appendix A. The data comprise small as well as very large markets, and offer a variety of nations and cultures across three continents.

Each minimum spanning tree is drawn with the indices represented by 2 to 4 letters (also listed in Appendix A) that make it easier to recognize the country each index belongs to. The vertices are also colored according to geographical region: orange for North America, green for South America, blue for Europe, brown for central Asia, and red for Pacific Asia. For each minimum spanning tree, we generated flushed data based on the same data used for calculating the minimum spanning tree. Distances that are within the region predicted for randomized data are represented in dashed lines, the other distances being represented by solid lines. All distances shown in the graphics are in scale, and the same scale is preserved throughout the article, so indices with strong connections appear much closer than ones which have low correlation, and the many minimum spanning trees can be easily compared.

Figure 4 shows the minimum spanning tree obtained for the first semester of 1986. For this interval of time, the average randomized distance is $0.81\pm 0.01$ (mean plus or minus the standard deviation), the result of 1000 simulations, with average minimum $0.77\pm 0.04$ and average maximum $0.92\pm 0.02$. So, we represent distances below 0.77 as solid lines and, above this threshold, as dashed lines. Only the connections between S\&P 500 and Nasdaq, Nasdaq and Canada, Nasdaq and the UK, Germany and Netherlands, can be trusted. Such connections like Canada and Sri Lanka, or Netherlands and Hong Kong, should be ignored, but other connections, which seem true ones, like between Japan, South Korea, and Malaysia, are also considered as probably random if the same standards are used. It is possible that distances between indices are inside the interval considered as random noise but are, nevertheless, true connections. Our method of establishing a threshold and ignoring any distance above it may eliminate some of those true connections.

\begin{pspicture}(-7.9,-4.2)(1,3.2)
\psset{xunit=3,yunit=3} \scriptsize
\psline(-0.240,0)(0,0) 
\psline[linestyle=dashed](-0.240,0)(-1.089,0) 
\psline[linestyle=dashed](-0.240,0)(-0.859,-0.619) 
\psline(0,0)(0.413,0) 
\psline(0,0)(0,0.698) 
\psline[linestyle=dashed](0,0)(0,-0.765) 
\psline[linestyle=dashed](0,0)(0.552,-0.552) 
\psline[linestyle=dashed](0.413,0)(1.290,0) 
\psline[linestyle=dashed](1.336,-0.552)(0.552,-0.552) 
\psline(0,-1.167)(0,-0.765) 
\psline[linestyle=dashed](0.552,-1.348)(0.552,-0.552) 
\psline[linestyle=dashed](0,-0.765)(-0.559,-1.324) 
\psline[linestyle=dashed](0.552,-0.552)(1.353,-1.014) 
\psline[linestyle=dashed](0.552,-0.552)(0.973,-1.281) 
\psline[linestyle=dashed](0.973,-1.281)(1.826,-1.281) 
\psdot[linecolor=orange,linewidth=1.2pt](-0.240,0) 
\psdot[linecolor=orange,linewidth=1.2pt](0,0) 
\psdot[linecolor=orange,linewidth=1.2pt](0.413,0) 
\psdot[linecolor=green,linewidth=1.2pt](1.336,-0.552) 
\psdot[linecolor=blue,linewidth=1.2pt](0,0.698) 
\psdot[linecolor=blue,linewidth=1.2pt](0,-1.167) 
\psdot[linecolor=blue,linewidth=1.2pt](0.552,-1.348) 
\psdot[linecolor=blue,linewidth=1.2pt](0,-0.765) 
\psdot[linecolor=brown,linewidth=1.2pt](-1.089,0) 
\psdot[linecolor=brown,linewidth=1.2pt](1.290,0) 
\psdot[linecolor=red,linewidth=1.2pt](0.552,-0.552) 
\psdot[linecolor=red,linewidth=1.2pt](-0.559,-1.324) 
\psdot[linecolor=red,linewidth=1.2pt](1.353,-1.014) 
\psdot[linecolor=red,linewidth=1.2pt](0.973,-1.281) 
\psdot[linecolor=red,linewidth=1.2pt](1.826,-1.281) 
\psdot[linecolor=red,linewidth=1.2pt](-0.859,-0.619) 
\rput(-0.240,0.1){S\&P}
\rput(0.13,0.1){Nasd}
\rput(0.413,0.1){Cana}
\rput(1.486,-0.552){Braz}
\rput(0,0.798){UK}
\rput(0,-1.267){Germ}
\rput(0.552,-1.448){Autr}
\rput(-0.15,-0.765){Neth}
\rput(-1.239,0){Indi}
\rput(1.440,0){SrLa}
\rput(0.702,-0.452){Japa}
\rput(-0.559,-1.424){HoKo}
\rput(1.503,-1.014){Taiw}
\rput(0.973,-1.381){SoKo}
\rput(1.976,-1.281){Mala}
\rput(-0.859,-0.719){Indo}
\end{pspicture}

\begin{center}
Figure 4: minimum spanning tree for the first semester of 1986.
\end{center}

Figure 5 does the same for data regarding the second semester of 1986. Solid lines represent distances below $0.77\pm 0.04$ (the result of yet more 1000 simulations) and dashed lines represent distances above that. Again, the lines connectiong S\&P, Nasdaq, Canada, and the UK are solid, so it is a cluster that kept its stability over the two timespans that have been studied until now. Other solid connection is again between Germany and the Netherlands, and one new solid connection appears between Canada and the Netherlands. From the dashed lines of figure 4, only the connection between S\&P and Indonesia remains, what speaks against the stability of the weaker connections of the minimum spanning trees.

\begin{pspicture}(-6.9,-3)(1,3.2)
\psset{xunit=3,yunit=3} \scriptsize
\psline(0,0)(-0.266,0) 
\psline(0,0)(0.491,0) 
\psline[linestyle=dashed](0,0)(0,-0.822) 
\psline(-0.266,0)(-0.266,0.543) 
\psline[linestyle=dashed](-0.266,0)(-0.266,-0.825) 
\psline[linestyle=dashed](-0.266,0)(-1.165,0) 
\psline(0.491,0)(1.055,0) 
\psline[linestyle=dashed](0.491,0)(0.491,0.847) 
\psline[linestyle=dashed](-1.165,-0.900)(-1.165,0) 
\psline(1.597,0)(1.055,0) 
\psline[linestyle=dashed](1.597,0)(2.372,0) 
\psline[linestyle=dashed](1.055,0)(1.055,0.859) 
\psline[linestyle=dashed](1.055,0.859)(1.940,0.859) 
\psline[linestyle=dashed](0,-0.822)(0.851,-0.822) 
\psline[linestyle=dashed](0.851,-0.822)(1.682,-0.822) 
\psdot[linecolor=orange,linewidth=1.2pt](0,0) 
\psdot[linecolor=orange,linewidth=1.2pt](-0.266,0) 
\psdot[linecolor=orange,linewidth=1.2pt](0.491,0) 
\psdot[linecolor=green,linewidth=1.2pt](-1.165,-0.900) 
\psdot[linecolor=blue,linewidth=1.2pt](-0.266,0.543) 
\psdot[linecolor=blue,linewidth=1.2pt](1.597,0) 
\psdot[linecolor=blue,linewidth=1.2pt](0.491,0.847) 
\psdot[linecolor=blue,linewidth=1.2pt](1.055,0) 
\psdot[linecolor=brown,linewidth=1.2pt](-0.266,-0.825) 
\psdot[linecolor=brown,linewidth=1.2pt](1.055,0.859) 
\psdot[linecolor=red,linewidth=1.2pt](2.372,0) 
\psdot[linecolor=red,linewidth=1.2pt](1.682,-0.822) 
\psdot[linecolor=red,linewidth=1.2pt](-1.165,0) 
\psdot[linecolor=red,linewidth=1.2pt](1.940,0.859) 
\psdot[linecolor=red,linewidth=1.2pt](0.851,-0.822) 
\psdot[linecolor=red,linewidth=1.2pt](0,-0.822) 
\rput(0,0.1){S\&P}
\rput(-0.416,0.1){Nasd}
\rput(0.641,0.1){Cana}
\rput(-1.165,-1.000){Braz}
\rput(-0.266,0.643){UK}
\rput(1.597,0.1){Germ}
\rput(0.491,0.947){Autr}
\rput(1.205,0.1){Neth}
\rput(-0.266,-0.925){Indi}
\rput(1.055,0.959){SrLa}
\rput(2.372,0.1){Japa}
\rput(1.682,-0.722){HoKo}
\rput(-1.165,0.1){Taiw}
\rput(1.940,0.959){SoKo}
\rput(0.851,-0.722){Mala}
\rput(0,-0.922){Indo}
\end{pspicture}

\begin{center}
Figure 5: minimum spanning tree for the second semester of 1986.
\end{center}

Figures 6 and 7 show the minimum spanning trees obtained from data concerning 23 indices, from the USA (S\&P and Nasdaq), Canada, Brazil, the UK, Ireland, Germany, Austria, Netherlands, Sweden, Finland, Spain, Greece, India, Sri Lanka, Bangladesh, Japan, Hong Kong, Taiwan, South Korea, Malaysia, Indonesia, and Philipines. All distances below a certain threshold ($0.75\pm 0.03$ for both the first and the second semesters) are shown in solid lines, and the ones below it are shown in dashed lines.

One thing to notice is that, for the first semester of 1987, the connections of what we may call a North American cluster, comprised of S\&P, Nasdaq, Canada, and the UK (although it is an European country), are maintained as solid lines. The only other connections which are considered solid are again the one between Germany and Netherlands, and now the one between the UK and Hong Kong. The other connections (dashed lines) seem, again, unstable in time.

The second semester of 1987 witnessed the crisis dubbed the Black Monday, and the minimum spanning tree reflects some effects of that crisis. First, one can notice the distances have shrank, a consequence of tighter correlations between the world indices. The second effect is that more of these distances are now below the threshold which is calculated with simulations of flushed data from the same second semester of 1987. Together with the solid lines of the North American indices, and between Germany and Netherlands, we also have more connections between European indices (Ireland-UK-Netherlands, Germany-Sweden, and Austria-Spain-Greece), between Pacific Asian indices (Japan-Malaysia), and between them with European indices (Japan-Sweden and Malaysia-Spain). No stability is seen in distances represented as dashed lines bewteeen the first and the second semesters.

\begin{pspicture}(-5,-5)(1,7.6)
\psset{xunit=3,yunit=3} \scriptsize
\psline(-0.232,0)(0,0) 
\psline(0,0)(0.388,0) 
\psline(0,0)(0,0.658) 
\psline[linestyle=dashed](0,0)(0,-0.827) 
\psline[linestyle=dashed](1.354,-0.827)(0.465,-0.827) 
\psline[linestyle=dashed](0,0.658)(0,1.436) 
\psline(0,0.658)(0.740,0.658) 
\psline[linestyle=dashed](0,1.436)(0.859,1.436) 
\psline[linestyle=dashed](0,1.436)(0,2.269) 
\psline[linestyle=dashed](0,1.436)(-0.780,1.436) 
\psline(0.465,-0.827)(0,-0.827) 
\psline[linestyle=dashed](0,-0.827)(0,-1.592) 
\psline[linestyle=dashed](0,-0.827)(-0.753,-0.827) 
\psline[linestyle=dashed](1.575,0.658)(2.361,0.658) 
\psline[linestyle=dashed](1.575,0.658)(0.740,0.658) 
\psline[linestyle=dashed](2.361,0.658)(2.361,1.465) 
\psline[linestyle=dashed](2.361,0.658)(2.361,-0.211) 
\psline[linestyle=dashed](2.361,1.465)(2.361,2.300) 
\psline[linestyle=dashed](3.146,-0.211)(3.146,0.643) 
\psline[linestyle=dashed](3.146,-0.211)(2.361,-0.211) 
\psline[linestyle=dashed](3.146,-0.211)(3.146,-1.066) 
\psline[linestyle=dashed](2.361,2.300)(1.493,2.300) 
\psdot[linecolor=orange,linewidth=1.2pt](-0.232,0) 
\psdot[linecolor=orange,linewidth=1.2pt](0,0) 
\psdot[linecolor=orange,linewidth=1.2pt](0.388,0) 
\psdot[linecolor=green,linewidth=1.2pt](1.354,-0.827) 
\psdot[linecolor=blue,linewidth=1.2pt](0,0.658) 
\psdot[linecolor=blue,linewidth=1.2pt](0,1.436) 
\psdot[linecolor=blue,linewidth=1.2pt](0.465,-0.827) 
\psdot[linecolor=blue,linewidth=1.2pt](0.859,1.436) 
\psdot[linecolor=blue,linewidth=1.2pt](0,-0.827) 
\psdot[linecolor=blue,linewidth=1.2pt](0,-1.592) 
\psdot[linecolor=blue,linewidth=1.2pt](1.575,0.658) 
\psdot[linecolor=blue,linewidth=1.2pt](2.361,0.658) 
\psdot[linecolor=blue,linewidth=1.2pt](2.361,1.465) 
\psdot[linecolor=brown,linewidth=1.2pt](0,2.269) 
\psdot[linecolor=brown,linewidth=1.2pt](3.146,-0.211) 
\psdot[linecolor=brown,linewidth=1.2pt](2.361,2.300) 
\psdot[linecolor=red,linewidth=1.2pt](-0.753,-0.827) 
\psdot[linecolor=red,linewidth=1.2pt](0.740,0.658) 
\psdot[linecolor=red,linewidth=1.2pt](3.146,0.643) 
\psdot[linecolor=red,linewidth=1.2pt](-0.780,1.436) 
\psdot[linecolor=red,linewidth=1.2pt](2.361,-0.211) 
\psdot[linecolor=red,linewidth=1.2pt](3.146,-1.066) 
\psdot[linecolor=blue,linewidth=1.2pt](1.493,2.300) 
\rput(-0.382,0){S\&P}
\rput(0.15,0.1){Nasd}
\rput(0.568,0){Cana}
\rput(1.354,-0.727){Braz}
\rput(0.15,0.758){UK}
\rput(0.15,1.536){Irel}
\rput(0.465,-0.727){Germ}
\rput(1.009,1.436){Autr}
\rput(0.15,-0.727){Neth}
\rput(0,-1.692){Swed}
\rput(1.575,0.758){Finl}
\rput(2.511,0.658){Spai}
\rput(2.511,1.465){Gree}
\rput(0,2.369){Indi}
\rput(3.296,-0.211){SrLa}
\rput(2.511,2.300){Bang}
\rput(-0.753,-0.727){Japa}
\rput(0.740,0.758){HoKo}
\rput(3.146,0.743){Taiw}
\rput(-0.980,1.436){SoKo}
\rput(2.361,-0.311){Mala}
\rput(3.146,-1.166){Indo}
\rput(1.343,2.300){Phil}
\end{pspicture}

\begin{center}
Figure 6: minimum spanning tree for the first semester of 1987.
\end{center}

\begin{pspicture}(-8.7,-6)(1,4.5)
\psset{xunit=3,yunit=3} \scriptsize
\psline(-0.575,0.413)(-0.365,0.413) 
\psline[linestyle=dashed](-0.575,0.413)(-0.575,1.288) 
\psline(-0.365,0.413)(-0.071,0.413) 
\psline(-0.365,0.413)(-0.365,0) 
\psline(-0.365,0.413)(-0.365,1.089) 
\psline[linestyle=dashed](-0.609,-1.990)(0.237,-1.990) 
\psline(-0.775,0)(-1.217,0) 
\psline(-0.775,0)(-0.365,0) 
\psline(0,0)(-0.365,0) 
\psline(0,0)(0.474,0) 
\psline(0,0)(-0.450,-0.450) 
\psline[linestyle=dashed](0,0)(0.548,-0.548) 
\psline(0.471,-1.255)(1.028,-1.255) 
\psline[linestyle=dashed](0.471,-1.255)(-0.314,-1.255) 
\psline(0.474,0)(1.028,0) 
\psline[linestyle=dashed](0.237,-1.990)(1.028,-1.990) 
\psline(1.028,-1.255)(1.028,-1.990) 
\psline(1.028,-1.255)(1.028,-0.648) 
\psline[linestyle=dashed](1.028,-1.990)(1.861,-1.990) 
\psline[linestyle=dashed](-2.014,-1.255)(-1.166,-1.255) 
\psline(1.028,0)(1.028,-0.648) 
\psline[linestyle=dashed](-0.314,-1.255)(-1.166,-1.255) 
\psdot[linecolor=orange,linewidth=1.2pt](-0.575,0.413) 
\psdot[linecolor=orange,linewidth=1.2pt](-0.365,0.413) 
\psdot[linecolor=orange,linewidth=1.2pt](-0.071,0.413) 
\psdot[linecolor=green,linewidth=1.2pt](-0.609,-1.990) 
\psdot[linecolor=blue,linewidth=1.2pt](-0.775,0) 
\psdot[linecolor=blue,linewidth=1.2pt](-1.217,0) 
\psdot[linecolor=blue,linewidth=1.2pt](0,0) 
\psdot[linecolor=blue,linewidth=1.2pt](0.471,-1.255) 
\psdot[linecolor=blue,linewidth=1.2pt](-0.365,0) 
\psdot[linecolor=blue,linewidth=1.2pt](0.474,0) 
\psdot[linecolor=blue,linewidth=1.2pt](0.237,-1.990) 
\psdot[linecolor=blue,linewidth=1.2pt](1.028,-1.255) 
\psdot[linecolor=blue,linewidth=1.2pt](1.028,-1.990) 
\psdot[linecolor=brown,linewidth=1.2pt](-0.365,1.089) 
\psdot[linecolor=brown,linewidth=1.2pt](-2.014,-1.255) 
\psdot[linecolor=brown,linewidth=1.2pt](1.861,-1.990) 
\psdot[linecolor=red,linewidth=1.2pt](1.028,0) 
\psdot[linecolor=red,linewidth=1.2pt](-0.450,-0.450) 
\psdot[linecolor=red,linewidth=1.2pt](-0.314,-1.255) 
\psdot[linecolor=red,linewidth=1.2pt](-1.166,-1.255) 
\psdot[linecolor=red,linewidth=1.2pt](1.028,-0.648) 
\psdot[linecolor=red,linewidth=1.2pt](-0.575,1.288) 
\psdot[linecolor=red,linewidth=1.2pt](0.548,-0.548) 
\rput(-0.725,0.413){S\&P}
\rput(-0.215,0.513){Nasd}
\rput(0.109,0.413){Cana}
\rput(-0.609,-1.890){Braz}
\rput(-0.775,0.1){UK}
\rput(-1.217,0.1){Irel}
\rput(0,0.1){Germ}
\rput(0.471,-1.155){Autr}
\rput(-0.515,0.1){Neth}
\rput(0.474,0.1){Swed}
\rput(0.237,-1.890){Finl}
\rput(1.178,-1.255){Spai}
\rput(1.178,-1.890){Gree}
\rput(-0.365,1.189){Indi}
\rput(-2.014,-1.155){SrLa}
\rput(1.861,-1.890){Bang}
\rput(1.178,0){Japa}
\rput(-0.450,-0.550){HoKo}
\rput(-0.314,-1.155){Taiw}
\rput(-1.166,-1.155){SoKo}
\rput(1.208,-0.648){Mala}
\rput(-0.575,1.388){Indo}
\rput(0.548,-0.648){Phil}
\end{pspicture}

\begin{center}
Figure 7: minimum spanning tree for the second semester of 1987.
\end{center}

\subsection{1997 and 1998 - Asian Financial Crisis and Russian Crisis}

Now, we jump in time to the next crisis: the Asian Financial Crisis, which started with the devaluation of the Thai baht, the currency of Thailand, and spread to most of the Pacific Asian countries. This lead to the Russian crisis, which began with the economic crisis in Russia due to the fall of its exports, and also because of the war in Chechnya, and spread to many more countries. Now we have data from 57 indices for 1997 - S\&P, Nasdaq, Canada, and Mexico (North America, represented in orange), Costa Rica, Bermuda, and Jamaica (Central America and the Carribbean, represented in darkgreen), Brazil, Argentina, Chile, Venezuela, and Peru (South America, in green), UK, Ireland, France, Germany, Switzerland, Austria, Belgium, Netherlands, Sweden, Denmark, Finland, Norway, Iceland, Spain, Portugal, Greece, Czech Republic, Slovakia, Hungary, Poland, and Estonia (Europe, in blue), Turkey (Eurasia, in lightblue), Israel, Lebanon, Saudi Arabia, Ohman, Pakistan, India, Sri Lanka, and Bangladesh (West and Central Asia, in brown), Japan, Hong Kong, China, Taiwan, South Korea, Thailand, Malaysia, Indonesia, and Philipines (Pacific Asia, in red), Australia (Oceania, in black), Morocco, Ghana, Kenya, South Africa, and Mauritius (Africa, in magenta). For 1997, we have no data about Russia, which is added to the data concerning 1998 (which then has 58 indices). All indices and symbols used for them are best explained in Appendix A.

Figures 8 and 9 show clear clusters, one formed by European indices, and the other formed by American ones. For the first semester of 1997, Netherlands is the main hub for the European cluster, and there is a subdivision of the American cluster between North and South American indices. Australia and South Africa are also clearly connected with the European cluster, as is the UK, which tended to belong with the American cluster before. Canada connects both clusters in both periods of time. Malaysia and Indonesia are connected by a solid connection just under the threshold. We also have a solid connection between Hong Kong and Austria and another between Iceland and Mauritius, both apparently random in nature. The threshold for the first semester of 1997 is $0.70\pm 0.03$.

\begin{pspicture}(-8.7,-6.4)(1,7)
\psset{xunit=3,yunit=3} \scriptsize
\psline(-0.301,0)(-0.555,0) 
\psline(-0.301,0)(0,0) 
\psline(-0.555,0)(-0.555,0.592) 
\psline[linestyle=dashed](-0.555,0)(-1.334,0) 
\psline(0,0)(0,0.502) 
\psline(0,0)(0.619,0) 
\psline(0.671,0.502)(0,0.502) 
\psline[linestyle=dashed](0.671,0.502)(1.439,0.502) 
\psline[linestyle=dashed](-1.966,1.202)(-1.193,1.202) 
\psline[linestyle=dashed](-1.966,1.202)(-2.747,1.202) 
\psline[linestyle=dashed](-1.966,1.202)(-1.966,0.334) 
\psline[linestyle=dashed](-1.812,-0.683)(-2.672,-0.683) 
\psline[linestyle=dashed](-1.812,-0.683)(-0.961,-0.683) 
\psline(-0.555,0.592)(-0.555,1.202) 
\psline(-0.555,1.202)(-1.193,1.202) 
\psline[linestyle=dashed](2.127,0)(1.415,0) 
\psline(0.265,-0.478)(0.619,-0.683) 
\psline(0.227,-1.116)(0.227,-0.683) 
\psline[linestyle=dashed](0.227,-1.116)(-0.360,-1.703) 
\psline[linestyle=dashed](0.227,-1.116)(0.227,-1.863) 
\psline(1.012,0)(0.619,0) 
\psline(1.012,0)(1.415,0) 
\psline(0.227,-0.683)(-0.213,-0.683) 
\psline(0.227,-0.683)(0.619,-0.683) 
\psline(0.227,-0.683)(-0.189,-1.099) 
\psline(0.619,-0.365)(0.619,-0.683) 
\psline(0.619,-0.365)(0.619,0) 
\psline[linestyle=dashed](0.619,-0.365)(1.411,-0.365) 
\psline(-0.213,-0.683)(-0.603,-0.293) 
\psline[linestyle=dashed](-0.213,-0.683)(-0.783,-1.253) 
\psline[linestyle=dashed](-0.213,-0.683)(-0.961,-0.683) 
\psline(0.317,-0.858)(0.619,-0.683) 
\psline(0.619,-0.683)(0.394,-1.074) 
\psline(0.619,-0.683)(0.985,-0.683) 
\psline(0.619,-0.683)(0.958,-0.878) 
\psline(0.619,-0.683)(0.384,-0.276) 
\psline[linestyle=dashed](0.619,-0.683)(0.619,-1.436) 
\psline(0.619,-0.683)(0.922,-1.209) 
\psline[linestyle=dashed](0.985,-0.683)(1.758,-0.683) 
\psline(0.985,-0.683)(1.464,-1.162) 
\psline[linestyle=dashed](2.119,1.247)(2.119,2.051) 
\psline(2.119,1.247)(1.439,1.247) 
\psline[linestyle=dashed](1.758,-0.683)(2.531,-0.683) 
\psline[linestyle=dashed](-1.268,-2.178)(-0.738,-1.648) 
\psline[linestyle=dashed](2.531,-0.683)(2.531,-1.487) 
\psline[linestyle=dashed](1.464,-1.162)(1.464,-1.939) 
\psline[linestyle=dashed](2.187,0.502)(1.439,0.502) 
\psline[linestyle=dashed](-1.700,-0.414)(-0.961,-0.683) 
\psline[linestyle=dashed](-2.337,-1.240)(-1.518,-1.240) 
\psline[linestyle=dashed](-1.546,-1.648)(-0.738,-1.648) 
\psline[linestyle=dashed](0.653,1.247)(0.653,2.047) 
\psline[linestyle=dashed](0.653,1.247)(1.439,1.247) 
\psline(-1.455,-0.189)(-0.961,-0.683) 
\psline[linestyle=dashed](-0.961,-0.683)(-1.518,-1.240) 
\psline[linestyle=dashed](-0.738,-1.648)(-0.189,-1.099) 
\psline[linestyle=dashed](1.439,0.502)(1.439,1.247) 
\psdot[linecolor=orange,linewidth=1.2pt](-0.301,0) 
\psdot[linecolor=orange,linewidth=1.2pt](-0.555,0) 
\psdot[linecolor=orange,linewidth=1.2pt](0,0) 
\psdot[linecolor=orange,linewidth=1.2pt](0.671,0.502) 
\psdot[linecolor=darkgreen,linewidth=1.2pt](-1.966,1.202) 
\psdot[linecolor=darkgreen,linewidth=1.2pt](-1.812,-0.683) 
\psdot[linecolor=darkgreen,linewidth=1.2pt](-2.672,-0.683) 
\psdot[linecolor=green,linewidth=1.2pt](-0.555,0.592) 
\psdot[linecolor=green,linewidth=1.2pt](0,0.502) 
\psdot[linecolor=green,linewidth=1.2pt](-0.555,1.202) 
\psdot[linecolor=green,linewidth=1.2pt](2.127,0) 
\psdot[linecolor=green,linewidth=1.2pt](-1.193,1.202) 
\psdot[linecolor=blue,linewidth=1.2pt](0.265,-0.478) 
\psdot[linecolor=blue,linewidth=1.2pt](0.227,-1.116) 
\psdot[linecolor=blue,linewidth=1.2pt](1.012,0) 
\psdot[linecolor=blue,linewidth=1.2pt](0.227,-0.683) 
\psdot[linecolor=blue,linewidth=1.2pt](0.619,-0.365) 
\psdot[linecolor=blue,linewidth=1.2pt](-0.213,-0.683) 
\psdot[linecolor=blue,linewidth=1.2pt](0.317,-0.858) 
\psdot[linecolor=blue,linewidth=1.2pt](0.619,-0.683) 
\psdot[linecolor=blue,linewidth=1.2pt](0.619,0) 
\psdot[linecolor=blue,linewidth=1.2pt](0.394,-1.074) 
\psdot[linecolor=blue,linewidth=1.2pt](0.985,-0.683) 
\psdot[linecolor=blue,linewidth=1.2pt](0.958,-0.878) 
\psdot[linecolor=blue,linewidth=1.2pt](2.119,1.247) 
\psdot[linecolor=blue,linewidth=1.2pt](1.415,0) 
\psdot[linecolor=blue,linewidth=1.2pt](0.384,-0.276) 
\psdot[linecolor=blue,linewidth=1.2pt](1.758,-0.683) 
\psdot[linecolor=blue,linewidth=1.2pt](-1.268,-2.178) 
\psdot[linecolor=blue,linewidth=1.2pt](2.531,-0.683) 
\psdot[linecolor=blue,linewidth=1.2pt](0.619,-1.436) 
\psdot[linecolor=blue,linewidth=1.2pt](1.464,-1.162) 
\psdot[linecolor=blue,linewidth=1.2pt](2.187,0.502) 
\psdot[linecolor=bluish,linewidth=1.2pt](-1.700,-0.414) 
\psdot[linecolor=brown,linewidth=1.2pt](-2.337,-1.240) 
\psdot[linecolor=brown,linewidth=1.2pt](-1.334,0) 
\psdot[linecolor=brown,linewidth=1.2pt](2.119,2.051) 
\psdot[linecolor=brown,linewidth=1.2pt](-1.546,-1.648) 
\psdot[linecolor=brown,linewidth=1.2pt](1.411,-0.365) 
\psdot[linecolor=brown,linewidth=1.2pt](0.653,1.247) 
\psdot[linecolor=brown,linewidth=1.2pt](1.464,-1.939) 
\psdot[linecolor=brown,linewidth=1.2pt](-0.360,-1.703) 
\psdot[linecolor=red,linewidth=1.2pt](0.227,-1.863) 
\psdot[linecolor=red,linewidth=1.2pt](-0.603,-0.293) 
\psdot[linecolor=red,linewidth=1.2pt](-0.783,-1.253) 
\psdot[linecolor=red,linewidth=1.2pt](-2.747,1.202) 
\psdot[linecolor=red,linewidth=1.2pt](0.653,2.047) 
\psdot[linecolor=red,linewidth=1.2pt](2.531,-1.487) 
\psdot[linecolor=red,linewidth=1.2pt](-1.455,-0.189) 
\psdot[linecolor=red,linewidth=1.2pt](-0.961,-0.683) 
\psdot[linecolor=red,linewidth=1.2pt](-0.738,-1.648) 
\psdot[linecolor=black,linewidth=1.2pt](-0.189,-1.099) 
\psdot[linecolor=magenta,linewidth=1.2pt](1.439,0.502) 
\psdot[linecolor=magenta,linewidth=1.2pt](-1.966,0.334) 
\psdot[linecolor=magenta,linewidth=1.2pt](-1.518,-1.240) 
\psdot[linecolor=magenta,linewidth=1.2pt](0.922,-1.209) 
\psdot[linecolor=magenta,linewidth=1.2pt](1.439,1.247) 
\rput(-0.301,0.1){S\&P}
\rput(-0.705,0.1){Nasd}
\rput(0.15,0.1){Cana}
\rput(0.671,0.602){Mexi}
\rput(-1.966,1.302){CoRi}
\rput(-1.812,-0.583){Berm}
\rput(-2.672,-0.583){Jama}
\rput(-0.705,0.592){Braz}
\rput(0,0.602){Arge}
\rput(-0.555,1.302){Chil}
\rput(2.127,0.1){Vene}
\rput(-1.193,1.302){Peru}
\rput(0.265,-0.378){UK}
\rput(0.077,-1.116){Irel}
\rput(1.012,0.1){Fran}
\rput(0.227,-0.583){Germ}
\rput(0.769,-0.265){Swit}
\rput(-0.143,-0.583){Autr}
\rput(0.337,-0.958){Belg}
\rput(0.759,-0.583){Neth}
\rput(0.619,0.1){Swed}
\rput(0.394,-1.174){Denm}
\rput(1.005,-0.583){Finl}
\rput(0.958,-0.978){Norw}
\rput(2.269,1.247){Icel}
\rput(1.415,0.1){Spai}
\rput(0.384,-0.176){Port}
\rput(1.758,-0.583){Gree}
\rput(-1.448,-2.178){CzRe}
\rput(2.531,-0.583){Slok}
\rput(0.619,-1.536){Hung}
\rput(1.614,-1.162){Pola}
\rput(2.187,0.602){Esto}
\rput(-1.850,-0.414){Turk}
\rput(-2.337,-1.340){Isra}
\rput(-1.334,0.1){Leba}
\rput(2.119,2.151){SaAr}
\rput(-1.736,-1.648){Ohma}
\rput(1.561,-0.365){Paki}
\rput(0.503,1.247){Indi}
\rput(1.464,-2.039){SrLa}
\rput(-0.530,-1.703){Bang}
\rput(0.227,-1.963){Japa}
\rput(-0.783,-0.293){HoKo}
\rput(-0.933,-1.253){Chin}
\rput(-2.747,1.302){Taiw}
\rput(0.653,2.147){SoKo}
\rput(2.531,-1.587){Thai}
\rput(-1.605,-0.189){Mala}
\rput(-0.811,-0.583){Indo}
\rput(-0.888,-1.548){Phil}
\rput(-0.339,-1.099){Aust}
\rput(1.589,0.602){Moro}
\rput(-1.966,0.234){Ghan}
\rput(-1.518,-1.340){Keny}
\rput(0.922,-1.309){SoAf}
\rput(1.589,1.347){Maur}
\end{pspicture}

\begin{center}
Figure 8: minimum spanning tree for the first semester of 1997.
\end{center}

For the second semester of 1997, structures are very similar, but with Germany taking over itself some of the centrality of Netherlands. Some other differences are Hungary and Israel connecting with Europe, Hong Kong connecting now with Germany, Japan connecting with Australia, and the formation of a Pacific Asian cluster comprised of Hong Kong, Malaysia, Indonesia, and Philipines. The threshold now is $0.71\pm 0.03$. Some connections, like Peru and Austria, seem to be random.

\begin{pspicture}(-4.2,-5.8)(1,10)
\psset{xunit=3,yunit=3} \scriptsize
\psline(-1.045,0)(-1.205,0) 
\psline(-1.045,0)(-0.737,0) 
\psline(-1.045,0)(-1.045,0.400) 
\psline(-1.045,0)(-1.045,-0.342) 
\psline(-0.737,0)(-0.280,0) 
\psline[linestyle=dashed](-1.045,0.400)(-1.045,1.142) 
\psline[linestyle=dashed](1.070,-0.855)(1.070,0) 
\psline[linestyle=dashed](0.359,2.200)(0.671,1.660) 
\psline[linestyle=dashed](4.081,0)(3.235,0) 
\psline(-1.045,-0.685)(-1.045,-0.342) 
\psline[linestyle=dashed](-1.045,-0.342)(-0.688,-0.959) 
\psline(-0.280,-0.481)(-0.280,0) 
\psline(0.192,1.339)(0.192,0.686) 
\psline(1.070,0)(0.516,0) 
\psline(1.070,0)(1.658,0) 
\psline(0.124,-0.509)(0.124,-0.215) 
\psline(0.694,-0.307)(0.516,0) 
\psline(0.192,0.439)(0.192,0.192) 
\psline(0.192,0.439)(0.192,0.686) 
\psline(0.253,0)(0.516,0) 
\psline(0.253,0)(0,0) 
\psline(0.253,0)(0.671,0.418) 
\psline(0.253,0)(0.552,-0.518) 
\psline(0,-0.199)(0,0) 
\psline(0,-0.199)(0,-0.629) 
\psline(0.124,-0.215)(0,0) 
\psline(0,0)(0.192,0.192) 
\psline(0,0)(-0.280,0) 
\psline(-0.590,-0.179)(-0.280,0) 
\psline(-0.280,0)(-0.481,-0.348) 
\psline(-0.280,0)(-0.058,0.384) 
\psline(-0.280,0)(-0.532,0.437) 
\psline[linestyle=dashed](2.784,2.153)(1.926,2.153) 
\psline[linestyle=dashed](-0.688,-0.959)(-0.688,-1.808) 
\psline[linestyle=dashed](2.162,0.504)(1.658,0) 
\psline[linestyle=dashed](-0.768,1.614)(-0.340,0.872) 
\psline(-0.058,0.384)(-0.340,0.872) 
\psline[linestyle=dashed](-0.058,0.384)(-0.058,1.161) 
\psline[linestyle=dashed](1.658,0)(1.658,-0.723) 
\psline[linestyle=dashed](1.658,0)(1.128,0.530) 
\psline[linestyle=dashed](1.658,0)(2.462,0) 
\psline[linestyle=dashed](-0.488,1.907)(-0.058,1.161) 
\psline[linestyle=dashed](1.658,-0.723)(1.658,-1.611) 
\psline[linestyle=dashed](1.459,1.660)(0.671,1.660) 
\psline[linestyle=dashed](0.552,-1.861)(0.552,-1.066) 
\psline[linestyle=dashed](3.235,0)(2.462,0) 
\psline[linestyle=dashed](1.926,2.153)(1.164,2.153) 
\psline[linestyle=dashed](-0.058,1.961)(-0.058,1.161) 
\psline[linestyle=dashed](1.164,2.961)(1.164,2.153) 
\psline(0.552,-1.066)(0.552,-0.518) 
\psline(0.671,0.418)(0.671,1.008) 
\psline[linestyle=dashed](1.907,1.219)(1.112,1.219) 
\psline(1.164,2.153)(0.671,1.660) 
\psline(0.671,1.008)(0.671,1.660) 
\psline(0.671,1.660)(1.112,1.219) 
\psline[linestyle=dashed](0.671,1.660)(0.671,2.410) 
\psdot[linecolor=orange,linewidth=1.2pt](-1.045,0) 
\psdot[linecolor=orange,linewidth=1.2pt](-1.205,0) 
\psdot[linecolor=orange,linewidth=1.2pt](-0.737,0) 
\psdot[linecolor=orange,linewidth=1.2pt](-1.045,0.400) 
\psdot[linecolor=darkgreen,linewidth=1.2pt](1.070,-0.855) 
\psdot[linecolor=darkgreen,linewidth=1.2pt](0.359,2.200) 
\psdot[linecolor=darkgreen,linewidth=1.2pt](4.081,0) 
\psdot[linecolor=green,linewidth=1.2pt](-1.045,-0.685) 
\psdot[linecolor=green,linewidth=1.2pt](-1.045,-0.342) 
\psdot[linecolor=green,linewidth=1.2pt](-0.280,-0.481) 
\psdot[linecolor=green,linewidth=1.2pt](0.192,1.339) 
\psdot[linecolor=green,linewidth=1.2pt](1.070,0) 
\psdot[linecolor=blue,linewidth=1.2pt](0.124,-0.509) 
\psdot[linecolor=blue,linewidth=1.2pt](0.694,-0.307) 
\psdot[linecolor=blue,linewidth=1.2pt](0.192,0.439) 
\psdot[linecolor=blue,linewidth=1.2pt](0.253,0) 
\psdot[linecolor=blue,linewidth=1.2pt](0,-0.199) 
\psdot[linecolor=blue,linewidth=1.2pt](0.516,0) 
\psdot[linecolor=blue,linewidth=1.2pt](0.124,-0.215) 
\psdot[linecolor=blue,linewidth=1.2pt](0,0) 
\psdot[linecolor=blue,linewidth=1.2pt](0.192,0.192) 
\psdot[linecolor=blue,linewidth=1.2pt](-0.590,-0.179) 
\psdot[linecolor=blue,linewidth=1.2pt](-0.280,0) 
\psdot[linecolor=blue,linewidth=1.2pt](-0.481,-0.348) 
\psdot[linecolor=blue,linewidth=1.2pt](2.784,2.153) 
\psdot[linecolor=blue,linewidth=1.2pt](0.192,0.686) 
\psdot[linecolor=blue,linewidth=1.2pt](0,-0.629) 
\psdot[linecolor=blue,linewidth=1.2pt](-0.688,-0.959) 
\psdot[linecolor=blue,linewidth=1.2pt](2.162,0.504) 
\psdot[linecolor=blue,linewidth=1.2pt](-0.768,1.614) 
\psdot[linecolor=blue,linewidth=1.2pt](-0.058,0.384) 
\psdot[linecolor=blue,linewidth=1.2pt](1.658,0) 
\psdot[linecolor=blue,linewidth=1.2pt](-0.488,1.907) 
\psdot[linecolor=bluish,linewidth=1.2pt](1.658,-0.723) 
\psdot[linecolor=brown,linewidth=1.2pt](-0.340,0.872) 
\psdot[linecolor=brown,linewidth=1.2pt](1.459,1.660) 
\psdot[linecolor=brown,linewidth=1.2pt](0.552,-1.861) 
\psdot[linecolor=brown,linewidth=1.2pt](3.235,0) 
\psdot[linecolor=brown,linewidth=1.2pt](1.926,2.153) 
\psdot[linecolor=brown,linewidth=1.2pt](-1.045,1.142) 
\psdot[linecolor=brown,linewidth=1.2pt](-0.058,1.961) 
\psdot[linecolor=brown,linewidth=1.2pt](1.164,2.961) 
\psdot[linecolor=red,linewidth=1.2pt](0.552,-1.066) 
\psdot[linecolor=red,linewidth=1.2pt](0.671,0.418) 
\psdot[linecolor=red,linewidth=1.2pt](1.907,1.219) 
\psdot[linecolor=red,linewidth=1.2pt](-0.058,1.161) 
\psdot[linecolor=red,linewidth=1.2pt](1.128,0.530) 
\psdot[linecolor=red,linewidth=1.2pt](1.164,2.153) 
\psdot[linecolor=red,linewidth=1.2pt](0.671,1.008) 
\psdot[linecolor=red,linewidth=1.2pt](0.671,1.660) 
\psdot[linecolor=red,linewidth=1.2pt](1.112,1.219) 
\psdot[linecolor=black,linewidth=1.2pt](0.552,-0.518) 
\psdot[linecolor=magenta,linewidth=1.2pt](2.462,0) 
\psdot[linecolor=magenta,linewidth=1.2pt](1.658,-1.611) 
\psdot[linecolor=magenta,linewidth=1.2pt](0.671,2.410) 
\psdot[linecolor=magenta,linewidth=1.2pt](-0.532,0.437) 
\psdot[linecolor=magenta,linewidth=1.2pt](-0.688,-1.808) 
\rput(-0.895,-0.1){S\&P}
\rput(-1.205,0.1){Nasd}
\rput(-0.737,0.1){Cana}
\rput(-1.195,0.400){Mexi}
\rput(1.070,-0.955){CoRi}
\rput(0.359,2.300){Berm}
\rput(4.081,0.1){Jama}
\rput(-1.045,-0.785){Braz}
\rput(-0.895,-0.342){Arge}
\rput(-0.280,-0.581){Chil}
\rput(0.192,1.439){Vene}
\rput(1.070,0.1){Peru}
\rput(0.124,-0.609){UK}
\rput(0.694,-0.407){Irel}
\rput(0.052,0.439){Fran}
\rput(0.133,-0.1){Germ}
\rput(-0.15,-0.199){Swit}
\rput(0.516,0.1){Autr}
\rput(0.264,-0.215){Belg}
\rput(-0.15,0.1){Neth}
\rput(0.042,0.192){Swed}
\rput(-0.600,-0.279){Denm}
\rput(-0.460,0.1){Finl}
\rput(-0.481,-0.448){Norw}
\rput(2.784,2.253){Icel}
\rput(0.052,0.686){Spai}
\rput(0,-0.729){Port}
\rput(-0.538,-0.959){Gree}
\rput(2.162,0.604){CzRe}
\rput(-0.768,1.714){Slok}
\rput(-0.228,0.384){Hung}
\rput(1.508,-0.1){Pola}
\rput(-0.488,2.007){Esto}
\rput(1.508,-0.723){Turk}
\rput(-0.490,0.872){Isra}
\rput(1.609,1.660){Leba}
\rput(0.552,-1.961){SaAr}
\rput(3.235,0.1){Ohma}
\rput(1.926,2.253){Paki}
\rput(-1.045,1.242){Indi}
\rput(-0.058,2.061){SrLa}
\rput(1.164,3.061){Bang}
\rput(0.402,-1.066){Japa}
\rput(0.481,0.418){HoKo}
\rput(1.907,1.119){Chin}
\rput(-0.208,1.161){Taiw}
\rput(1.128,0.630){SoKo}
\rput(1.314,2.253){Thai}
\rput(0.521,1.008){Mala}
\rput(0.521,1.660){Indo}
\rput(1.112,1.119){Phil}
\rput(0.402,-0.518){Aust}
\rput(2.462,0.1){Moro}
\rput(1.658,-1.711){Ghan}
\rput(0.671,2.510){Keny}
\rput(-0.532,0.537){SoAf}
\rput(-0.688,-1.908){Maur}
\end{pspicture}

\begin{center}
Figure 9: minimum spanning tree for the second semester of 1997.
\end{center}

For 1998, we add Russia to the indices that were used in 1997. Figures 10 and 11 show the minimum spanning trees for both, respectively, the first and the second semesters of that year. The threshold for the first semester of 1998 is $0.70\pm 0.03$, what leaves below it the usual American and European clusters, a Pacific Asian cluster, and a new cluster centered around Russia (which is colored light blue for Eurasia), comprised of Hungary, the Czech Republic, Estonia, and Poland. Strange connections are those between Greece and Turkey with Canada, Indonesia-Chile-Switzerland, Venezuela-Norway, and Australia, South Korea and Pakistan with Poland. All connections above the threshold seem to be random.

The threshold for the second semester of 1998 is $0.71\pm 0.03$. The resulting graph (figure 10) shows a similar behavior to the first semester of the same year, except that Russia and Turkey are now more connected with Europe, and that now the Pacific Asian cluster is split into three parts, the main part being the cluster comprised of Japan, Hong Kong, Philipines, and Australia, although Australia belongs to Oceania. Chile is now connected with the American cluster via Mexico. Also to be noticed is that clusters clearly shrunk in the second semester of 1998, in particular the European cluster, whose indices agglomerated around the Netherlands.

\begin{pspicture}(-6,-5.2)(1,5.6)
\psset{xunit=3,yunit=3} \scriptsize
\psline(-1.170,0)(-1.385,0) 
\psline(-1.170,0)(-0.726,0) 
\psline(-1.385,0)(-1.385,0.475) 
\psline[linestyle=dashed](-0.726,0)(-1.463,-0.426) 
\psline(-0.726,0)(-0.279,0) 
\psline(-0.726,0)(-1.200,0.474) 
\psline(-0.726,0)(-1.034,-0.533) 
\psline(-1.385,0.915)(-1.385,1.325) 
\psline(-1.385,0.915)(-1.385,0.475) 
\psline[linestyle=dashed](3.163,0)(2.380,0) 
\psline[linestyle=dashed](3.163,0)(3.163,0.788) 
\psline[linestyle=dashed](0.800,-1.239)(0.800,-0.451) 
\psline(-0.279,0.584)(-0.279,0) 
\psline[linestyle=dashed](-0.279,0.584)(-0.811,1.116) 
\psline(-0.279,0.584)(-0.279,1.186) 
\psline(0,-1.625)(0,-0.977) 
\psline[linestyle=dashed](0,-1.625)(0.862,-1.625) 
\psline[linestyle=dashed](0,-1.625)(-0.812,-1.625) 
\psline(-0.593,-0.544)(-0.279,0) 
\psline(0.250,-0.433)(0.138,-0.239) 
\psline(0.216,-0.948)(0,-0.575) 
\psline(0.138,-0.239)(0,0) 
\psline(0,0.314)(0,0.735) 
\psline(0,0.314)(0,0) 
\psline(0,0.314)(0.473,0.587) 
\psline(0,0.314)(0.224,0.702) 
\psline(-0.279,0)(0,0) 
\psline(-0.279,0)(-0.536,-0.148) 
\psline[linestyle=dashed](-0.279,0)(-0.818,0.540) 
\psline(0.215,0)(0,0) 
\psline(0.215,0)(0.431,0.216) 
\psline(0.215,0)(0.593,0) 
\psline(0,0)(0,-0.292) 
\psline(0,0)(-0.278,-0.278) 
\psline(0,0)(0.518,-0.299) 
\psline(0,-0.292)(0,-0.575) 
\psline(0,-0.575)(0,-0.977) 
\psline[linestyle=dashed](0,-0.977)(-0.746,-0.977) 
\psline[linestyle=dashed](-1.200,0.474)(-1.200,1.266) 
\psline(1.630,-0.619)(1.630,0) 
\psline(1.061,0)(1.630,0) 
\psline(1.061,0)(0.593,0) 
\psline(1.630,0.536)(1.630,0) 
\psline(1.630,0.536)(2.091,0.997) 
\psline(1.630,0.536)(2.329,0.536) 
\psline(1.630,0.536)(1.630,1.099) 
\psline(2.085,-0.455)(1.630,0) 
\psline[linestyle=dashed](1.630,0)(2.380,0) 
\psline[linestyle=dashed](1.234,-1.126)(1.234,-0.370) 
\psline[linestyle=dashed](3.163,1.656)(3.163,0.788) 
\psline(1.240,0.777)(0.968,0.306) 
\psline(0.593,0)(0.800,-0.451) 
\psline(0.593,0)(0.968,0.306) 
\psline[linestyle=dashed](0.593,0)(1.234,-0.370) 
\psline(0.968,0.765)(0.968,0.306) 
\psline(0.968,0.765)(0.968,1.311) 
\psline[linestyle=dashed](-0.279,1.186)(0.495,1.633) 
\psdot[linecolor=orange,linewidth=1.2pt](-1.170,0) 
\psdot[linecolor=orange,linewidth=1.2pt](-1.385,0) 
\psdot[linecolor=orange,linewidth=1.2pt](-0.726,0) 
\psdot[linecolor=orange,linewidth=1.2pt](-1.385,0.915) 
\psdot[linecolor=darkgreen,linewidth=1.2pt](3.163,0) 
\psdot[linecolor=darkgreen,linewidth=1.2pt](0.800,-1.239) 
\psdot[linecolor=darkgreen,linewidth=1.2pt](-1.463,-0.426) 
\psdot[linecolor=green,linewidth=1.2pt](-1.385,1.325) 
\psdot[linecolor=green,linewidth=1.2pt](-1.385,0.475) 
\psdot[linecolor=green,linewidth=1.2pt](-0.279,0.584) 
\psdot[linecolor=green,linewidth=1.2pt](0,-1.625) 
\psdot[linecolor=green,linewidth=1.2pt](-0.593,-0.544) 
\psdot[linecolor=blue,linewidth=1.2pt](0.250,-0.433) 
\psdot[linecolor=blue,linewidth=1.2pt](0.216,-0.948) 
\psdot[linecolor=blue,linewidth=1.2pt](0.138,-0.239) 
\psdot[linecolor=blue,linewidth=1.2pt](0,0.314) 
\psdot[linecolor=blue,linewidth=1.2pt](-0.279,0) 
\psdot[linecolor=blue,linewidth=1.2pt](0,0.735) 
\psdot[linecolor=blue,linewidth=1.2pt](0.215,0) 
\psdot[linecolor=blue,linewidth=1.2pt](0,0) 
\psdot[linecolor=blue,linewidth=1.2pt](0,-0.292) 
\psdot[linecolor=blue,linewidth=1.2pt](0.431,0.216) 
\psdot[linecolor=blue,linewidth=1.2pt](0,-0.575) 
\psdot[linecolor=blue,linewidth=1.2pt](0,-0.977) 
\psdot[linecolor=blue,linewidth=1.2pt](0.862,-1.625) 
\psdot[linecolor=blue,linewidth=1.2pt](-0.536,-0.148) 
\psdot[linecolor=blue,linewidth=1.2pt](-0.278,-0.278) 
\psdot[linecolor=blue,linewidth=1.2pt](-1.200,0.474) 
\psdot[linecolor=blue,linewidth=1.2pt](1.630,-0.619) 
\psdot[linecolor=blue,linewidth=1.2pt](-0.818,0.540) 
\psdot[linecolor=blue,linewidth=1.2pt](1.061,0) 
\psdot[linecolor=blue,linewidth=1.2pt](1.630,0.536) 
\psdot[linecolor=blue,linewidth=1.2pt](2.085,-0.455) 
\psdot[linecolor=bluish,linewidth=1.2pt](1.630,0) 
\psdot[linecolor=bluish,linewidth=1.2pt](-1.034,-0.533) 
\psdot[linecolor=brown,linewidth=1.2pt](0.473,0.587) 
\psdot[linecolor=brown,linewidth=1.2pt](1.234,-1.126) 
\psdot[linecolor=brown,linewidth=1.2pt](2.380,0) 
\psdot[linecolor=brown,linewidth=1.2pt](3.163,1.656) 
\psdot[linecolor=brown,linewidth=1.2pt](2.091,0.997) 
\psdot[linecolor=brown,linewidth=1.2pt](0.518,-0.299) 
\psdot[linecolor=brown,linewidth=1.2pt](-0.811,1.116) 
\psdot[linecolor=brown,linewidth=1.2pt](3.163,0.788) 
\psdot[linecolor=red,linewidth=1.2pt](1.240,0.777) 
\psdot[linecolor=red,linewidth=1.2pt](0.593,0) 
\psdot[linecolor=red,linewidth=1.2pt](-1.200,1.266) 
\psdot[linecolor=red,linewidth=1.2pt](0.800,-0.451) 
\psdot[linecolor=red,linewidth=1.2pt](2.329,0.536) 
\psdot[linecolor=red,linewidth=1.2pt](0.968,0.765) 
\psdot[linecolor=red,linewidth=1.2pt](0.968,0.306) 
\psdot[linecolor=red,linewidth=1.2pt](-0.279,1.186) 
\psdot[linecolor=red,linewidth=1.2pt](0.968,1.311) 
\psdot[linecolor=black,linewidth=1.2pt](1.630,1.099) 
\psdot[linecolor=magenta,linewidth=1.2pt](-0.812,-1.625) 
\psdot[linecolor=magenta,linewidth=1.2pt](-0.746,-0.977) 
\psdot[linecolor=magenta,linewidth=1.2pt](0.495,1.633) 
\psdot[linecolor=magenta,linewidth=1.2pt](0.224,0.702) 
\psdot[linecolor=magenta,linewidth=1.2pt](1.234,-0.370) 
\rput(-1.170,0.1){S\&P}
\rput(-1.535,0){Nasd}
\rput(-0.726,0.1){Cana}
\rput(-1.535,0.915){Mexi}
\rput(3.333,0){CoRi}
\rput(0.800,-1.339){Berm}
\rput(-1.613,-0.426){Jama}
\rput(-1.385,1.425){Braz}
\rput(-1.535,0.475){Arge}
\rput(-0.429,0.584){Chil}
\rput(0,-1.725){Vene}
\rput(-0.593,-0.644){Peru}
\rput(0.250,-0.533){UK}
\rput(0.216,-1.048){Irel}
\rput(0.288,-0.239){Fran}
\rput(-0.15,0.314){Germ}
\rput(-0.279,0.1){Swit}
\rput(0,0.835){Autr}
\rput(0.215,0.1){Belg}
\rput(0,0.1){Neth}
\rput(-0.15,-0.192){Swed}
\rput(0.431,0.316){Denm}
\rput(-0.15,-0.575){Finl}
\rput(-0.15,-0.877){Norw}
\rput(1.012,-1.625){Icel}
\rput(-0.546,-0.248){Spai}
\rput(-0.278,-0.378){Port}
\rput(-1.050,0.474){Gree}
\rput(1.630,-0.719){CzRe}
\rput(-0.818,0.640){Slok}
\rput(1.061,0.1){Hung}
\rput(1.480,0.536){Pola}
\rput(2.085,-0.555){Esto}
\rput(1.630,0.1){Russ}
\rput(-1.184,-0.533){Turk}
\rput(0.473,0.687){Isra}
\rput(1.234,-1.226){Leba}
\rput(2.380,0.1){SaAr}
\rput(3.333,1.656){Ohma}
\rput(2.091,1.097){Paki}
\rput(0.518,-0.399){Indi}
\rput(-0.811,1.216){SrLa}
\rput(3.313,0.788){Bang}
\rput(1.240,0.877){Japa}
\rput(0.593,0.1){HoKo}
\rput(-1.200,1.366){Chin}
\rput(0.950,-0.451){Taiw}
\rput(2.329,0.636){SoKo}
\rput(0.818,0.765){Thai}
\rput(0.818,0.336){Mala}
\rput(-0.429,1.186){Indo}
\rput(0.968,1.411){Phil}
\rput(1.630,1.199){Aust}
\rput(-0.982,-1.625){Moro}
\rput(-0.896,-0.977){Ghan}
\rput(0.495,1.733){Keny}
\rput(0.244,0.802){SoAf}
\rput(1.384,-0.370){Maur}
\end{pspicture}

\begin{center}
Figure 10: minimum spanning tree for the first semester of 1998.
\end{center}

\begin{pspicture}(-6,-4.3)(1,6.5)
\psset{xunit=3,yunit=3} \scriptsize
\psline(-0.887,0)(-1.016,0) 
\psline(-0.887,0)(-0.575,0) 
\psline(-0.887,0)(-0.887,0.409) 
\psline(-0.575,0)(-0.142,0) 
\psline(-0.887,0.409)(-0.887,0.744) 
\psline[linestyle=dashed](3.318,0)(2.395,0) 
\psline[linestyle=dashed](-0.369,1.319)(-0.369,0.558) 
\psline[linestyle=dashed](-0.369,1.319)(-0.369,2.072) 
\psline[linestyle=dashed](2.395,0)(3.053,-0.380) 
\psline[linestyle=dashed](2.395,0)(1.620,0) 
\psline(-0.887,0.744)(-0.887,1.008) 
\psline(-0.887,1.008)(-1.227,1.008) 
\psline(-0.369,0.558)(0,0.189) 
\psline[linestyle=dashed](-0.369,0.558)(-0.806,1.316) 
\psline(0.766,-0.442)(0.272,-0.157) 
\psline[linestyle=dashed](0.766,-0.442)(0.766,-1.221) 
\psline(-0.147,0.085)(0,0) 
\psline(0.272,-0.157)(0,0) 
\psline(0.272,-0.157)(0.466,-0.492) 
\psline(0.143,0.082)(0,0) 
\psline(0.143,0.082)(0.305,0.176) 
\psline(-0.066,-0.114)(0,0) 
\psline(-0.066,-0.114)(-0.170,-0.293) 
\psline(-0.066,-0.114)(-0.066,-0.390) 
\psline(-0.142,0)(0,0) 
\psline(-0.142,0)(-0.381,-0.138) 
\psline(-0.267,-0.154)(0,0) 
\psline(-0.267,-0.154)(-0.267,-0.592) 
\psline(-0.077,0.133)(0,0) 
\psline(0,0)(0.098,-0.169) 
\psline(0,0)(0,0.189) 
\psline(0,0)(0.132,0.229) 
\psline(0,0)(0.280,0) 
\psline[linestyle=dashed](0.098,-0.169)(0.471,-0.815) 
\psline(0.098,-0.169)(0.098,-0.643) 
\psline[linestyle=dashed](0.766,-1.221)(1.492,-1.221) 
\psline(0,0.189)(0.215,0.404) 
\psline(-0.381,-0.138)(-0.755,-0.354) 
\psline(-0.381,-0.138)(-0.381,-0.830) 
\psline[linestyle=dashed](3.053,-0.380)(3.053,-1.195) 
\psline[linestyle=dashed](0.280,0)(0.916,0.368) 
\psline(0.280,0)(0.790,-0.294) 
\psline(0.280,0)(0.827,0) 
\psline[linestyle=dashed](-0.755,-0.354)(-0.755,-1.113) 
\psline(-0.755,-0.354)(-1.354,-0.354) 
\psline[linestyle=dashed](0.098,-0.643)(0.098,-1.391) 
\psline[linestyle=dashed](-0.369,2.072)(-1.194,2.072) 
\psline[linestyle=dashed](0.098,-1.391)(-0.764,-1.391) 
\psline[linestyle=dashed](0.774,0.963)(0.215,0.404) 
\psline(0.569,1.205)(0.215,0.851) 
\psline(-0.109,1.175)(0.215,0.851) 
\psline[linestyle=dashed](1.286,-1.075)(1.286,-0.265) 
\psline[linestyle=dashed](0.215,2.055)(0.215,1.331) 
\psline(1.286,-0.265)(0.827,0) 
\psline[linestyle=dashed](0.827,0)(1.620,0) 
\psline(0.215,1.331)(0.215,0.851) 
\psline(0.215,0.851)(0.215,0.404) 
\psdot[linecolor=orange,linewidth=1.2pt](-0.887,0) 
\psdot[linecolor=orange,linewidth=1.2pt](-1.016,0) 
\psdot[linecolor=orange,linewidth=1.2pt](-0.575,0) 
\psdot[linecolor=orange,linewidth=1.2pt](-0.887,0.409) 
\psdot[linecolor=darkgreen,linewidth=1.2pt](3.318,0) 
\psdot[linecolor=darkgreen,linewidth=1.2pt](-0.369,1.319) 
\psdot[linecolor=darkgreen,linewidth=1.2pt](2.395,0) 
\psdot[linecolor=green,linewidth=1.2pt](-0.887,0.744) 
\psdot[linecolor=green,linewidth=1.2pt](-0.887,1.008) 
\psdot[linecolor=green,linewidth=1.2pt](-1.227,1.008) 
\psdot[linecolor=green,linewidth=1.2pt](-0.369,0.558) 
\psdot[linecolor=green,linewidth=1.2pt](0.766,-0.442) 
\psdot[linecolor=blue,linewidth=1.2pt](-0.147,0.085) 
\psdot[linecolor=blue,linewidth=1.2pt](0.272,-0.157) 
\psdot[linecolor=blue,linewidth=1.2pt](0.143,0.082) 
\psdot[linecolor=blue,linewidth=1.2pt](-0.066,-0.114) 
\psdot[linecolor=blue,linewidth=1.2pt](-0.142,0) 
\psdot[linecolor=blue,linewidth=1.2pt](-0.267,-0.154) 
\psdot[linecolor=blue,linewidth=1.2pt](-0.077,0.133) 
\psdot[linecolor=blue,linewidth=1.2pt](0,0) 
\psdot[linecolor=blue,linewidth=1.2pt](0.305,0.176) 
\psdot[linecolor=blue,linewidth=1.2pt](-0.170,-0.293) 
\psdot[linecolor=blue,linewidth=1.2pt](0.098,-0.169) 
\psdot[linecolor=blue,linewidth=1.2pt](-0.066,-0.390) 
\psdot[linecolor=blue,linewidth=1.2pt](0.766,-1.221) 
\psdot[linecolor=blue,linewidth=1.2pt](0,0.189) 
\psdot[linecolor=blue,linewidth=1.2pt](0.132,0.229) 
\psdot[linecolor=blue,linewidth=1.2pt](0.466,-0.492) 
\psdot[linecolor=blue,linewidth=1.2pt](-0.381,-0.138) 
\psdot[linecolor=blue,linewidth=1.2pt](3.053,-0.380) 
\psdot[linecolor=blue,linewidth=1.2pt](0.280,0) 
\psdot[linecolor=blue,linewidth=1.2pt](-0.755,-0.354) 
\psdot[linecolor=blue,linewidth=1.2pt](0.471,-0.815) 
\psdot[linecolor=bluish,linewidth=1.2pt](-0.381,-0.830) 
\psdot[linecolor=bluish,linewidth=1.2pt](-0.267,-0.592) 
\psdot[linecolor=brown,linewidth=1.2pt](0.098,-0.643) 
\psdot[linecolor=brown,linewidth=1.2pt](-0.369,2.072) 
\psdot[linecolor=brown,linewidth=1.2pt](0.916,0.368) 
\psdot[linecolor=brown,linewidth=1.2pt](-0.806,1.316) 
\psdot[linecolor=brown,linewidth=1.2pt](0.098,-1.391) 
\psdot[linecolor=brown,linewidth=1.2pt](-0.755,-1.113) 
\psdot[linecolor=brown,linewidth=1.2pt](0.774,0.963) 
\psdot[linecolor=brown,linewidth=1.2pt](-0.764,-1.391) 
\psdot[linecolor=red,linewidth=1.2pt](0.569,1.205) 
\psdot[linecolor=red,linewidth=1.2pt](-0.109,1.175) 
\psdot[linecolor=red,linewidth=1.2pt](1.286,-1.075) 
\psdot[linecolor=red,linewidth=1.2pt](0.215,2.055) 
\psdot[linecolor=red,linewidth=1.2pt](0.790,-0.294) 
\psdot[linecolor=red,linewidth=1.2pt](1.286,-0.265) 
\psdot[linecolor=red,linewidth=1.2pt](-1.354,-0.354) 
\psdot[linecolor=red,linewidth=1.2pt](0.827,0) 
\psdot[linecolor=red,linewidth=1.2pt](0.215,1.331) 
\psdot[linecolor=black,linewidth=1.2pt](0.215,0.851) 
\psdot[linecolor=magenta,linewidth=1.2pt](3.053,-1.195) 
\psdot[linecolor=magenta,linewidth=1.2pt](1.620,0) 
\psdot[linecolor=magenta,linewidth=1.2pt](-1.194,2.072) 
\psdot[linecolor=magenta,linewidth=1.2pt](0.215,0.404) 
\psdot[linecolor=magenta,linewidth=1.2pt](1.492,-1.221) 
\rput(-0.887,-0.1){S\&P}
\rput(-1.166,0){Nasd}
\rput(-0.575,0.1){Cana}
\rput(-1.037,0.409){Mexi}
\rput(3.318,0.1){CoRi}
\rput(-0.519,1.319){Berm}
\rput(2.395,0.1){Jama}
\rput(-1.037,0.744){Braz}
\rput(-0.887,1.108){Arge}
\rput(-1.377,1.008){Chil}
\rput(-0.519,0.558){Vene}
\rput(0.956,-0.442){Peru}
\rput(-0.297,0.085){UK}
\rput(0.422,-0.157){Irel}
\rput(0.143,0.182){Fran}
\rput(-0.066,-0.214){Germ}
\rput(-0.142,-0.1){Swit}
\rput(-0.117,-0.154){Autr}
\rput(-0.227,0.183){Belg}
\rput(0,0.1){Neth}
\rput(0.455,0.176){Swed}
\rput(-0.170,-0.393){Denm}
\rput(0.098,-0.269){Finl}
\rput(-0.066,-0.490){Norw}
\rput(0.766,-1.321){Icel}
\rput(0,0.289){Spai}
\rput(0.282,0.229){Port}
\rput(0.466,-0.592){Gree}
\rput(-0.381,-0.038){CzRe}
\rput(3.203,-0.380){Slok}
\rput(0.380,0.1){Hung}
\rput(-0.605,-0.454){Pola}
\rput(0.471,-0.915){Esto}
\rput(-0.381,-0.930){Russ}
\rput(-0.267,-0.692){Turk}
\rput(-0.052,-0.643){Isra}
\rput(-0.369,2.172){Leba}
\rput(0.916,0.468){SaAr}
\rput(-0.806,1.416){Ohma}
\rput(0.098,-1.491){Paki}
\rput(-0.755,-1.213){Indi}
\rput(0.774,1.063){SrLa}
\rput(-0.914,-1.391){Bang}
\rput(0.569,1.305){Japa}
\rput(-0.109,1.275){HoKo}
\rput(1.436,-1.075){Chin}
\rput(0.215,2.155){Taiw}
\rput(0.790,-0.194){SoKo}
\rput(1.436,-0.265){Thai}
\rput(-1.504,-0.354){Mala}
\rput(0.827,0.1){Indo}
\rput(0.065,1.431){Phil}
\rput(0.065,0.851){Aust}
\rput(3.053,-1.295){Moro}
\rput(1.620,0.1){Ghan}
\rput(-1.344,2.072){Keny}
\rput(0.365,0.404){SoAf}
\rput(1.642,-1.221){Maur}
\end{pspicture}

\begin{center}
Figure 11: minimum spanning tree for the second semester of 1998.
\end{center}

\subsection{2001 - Burst of the Dot-Com Bubble and September 11}

The year 2001 saw two crises: the first one was internal, the result of excessive speculation on equities of internet-based companies, known as dot-com; the second, was purely exogenous, the result of the largest terrorist attack in history, perpetrated against the USA. Here, we analize the years 2000 and 2001, in order to check how the two crises influenced the minimum spanning trees resulting from data relative to many stock market exchanges in the world.

We now have 74 indices: S\&P, Nasdaq, Canada, and Mexico (North America, in orange), Panama, Costa Rica, Bermuda, and Jamaica (Central America and the Caribbean, in darkgreen), Brazil, Argentina, Chile, Venezuela, and Peru (South America, in green), UK, Ireland, France, Germany, Switzerland, Austria, Italy, Malta, Belgium, Netherlands, Luxembourg, Sweden, Denmark, Finland, Norway, Iceland, Spain, Portugal, Greece, Czech Republic, Slovakia, Hungary, Poland, Romania, Estonia, Latvia, Lithuania, and Ukraine (Europe, in blue), Russia and Turkey (Eurasia, in lightblue), Israel, Palestine, Lebanon, Jordan, Saudi Arabia, Qatar, Ohman, Pakistan, India, Sri Lanka, and Bangladesh (Western and Central Asia), Japan, Hong Kong, China, Mongolia, Taiwan, South Korea, Thailand, Malaysia, Singapore, Indonesia, and Philipines (Pacific Asia, in red), Australia (Oceania, in black), Morocco, Tunisia, Egypt, Ghana, Nigeria, Kenya, South Africa, and Mauritius (Africa, in magenta). Figures 12 and 13 show the minimum spanning trees for the first and second semesters of 2000.

\begin{pspicture}(-8,-6.5)(1,8)
\psset{xunit=3,yunit=3} \scriptsize
\psline(-1.220,0)(-1.028,0) 
\psline(-1.220,0)(-1.220,0.349) 
\psline[linestyle=dashed](-1.220,0)(-1.220,-0.813) 
\psline(-1.220,0)(-1.647,0) 
\psline(-1.028,0)(-0.698,0) 
\psline(-0.698,0)(-0.221,0) 
\psline(-1.220,0.349)(-1.745,0.349) 
\psline[linestyle=dashed](-0.373,-1.804)(-0.373,-0.970) 
\psline[linestyle=dashed](-0.373,-1.804)(-1.217,-1.804) 
\psline[linestyle=dashed](-0.446,1.330)(0,0.558) 
\psline[linestyle=dashed](2.200,-1.306)(1.789,-0.586) 
\psline(-1.647,0)(-2.097,0) 
\psline[linestyle=dashed](-1.647,0)(-1.647,-0.854) 
\psline[linestyle=dashed](-1.745,0.349)(-1.745,1.208) 
\psline[linestyle=dashed](-1.745,0.349)(-2.469,0.349) 
\psline(-0.844,-0.844)(-0.373,-0.373) 
\psline(-0.373,-0.970)(-0.373,-0.373) 
\psline(-0.214,0.214)(-0.806,0.214) 
\psline(-0.214,0.214)(0,0) 
\psline(-0.214,0.214)(-0.380,0.380) 
\psline(0,0)(-0.221,0) 
\psline(0,0)(0.168,-0.168) 
\psline(0,0)(0.241,0) 
\psline(0,0)(0,0.306) 
\psline(0,0)(-0.178,-0.178) 
\psline(0,0)(0,-0.523) 
\psline(0.497,-0.443)(0.241,0) 
\psline[linestyle=dashed](0.497,-0.443)(0.940,-1.211) 
\psline(0.781,-0.312)(0.241,0) 
\psline[linestyle=dashed](-0.795,1.609)(-0.795,0.795) 
\psline[linestyle=dashed](-0.795,1.609)(-0.376,2.335) 
\psline[linestyle=dashed](-0.795,1.609)(-1.200,2.310) 
\psline(0.994,0.698)(0.468,0.394) 
\psline[linestyle=dashed](0.994,0.698)(1.768,1.145) 
\psline(0.241,0)(0.643,0.232) 
\psline(0.241,0)(0.468,0.394) 
\psline(0.241,0)(0.773,0) 
\psline(0.241,0)(0.241,-0.511) 
\psline(-0.454,0.760)(0,0.306) 
\psline(0,0.306)(0,0.558) 
\psline[linestyle=dashed](0,0.558)(0,1.307) 
\psline(0.468,0.394)(0.724,0.837) 
\psline[linestyle=dashed](0.468,0.394)(0.468,1.274) 
\psline[linestyle=dashed](1.395,1.224)(0.724,0.837) 
\psline[linestyle=dashed](1.395,1.224)(1.395,2.067) 
\psline(-0.178,-0.178)(-0.373,-0.373) 
\psline[linestyle=dashed](2.391,-1.718)(2.391,-0.835) 
\psline(-0.380,0.380)(-0.795,0.795) 
\psline(1.277,0.291)(1.588,0.830) 
\psline(1.277,0.291)(0.773,0) 
\psline[linestyle=dashed](1.277,0.291)(1.999,0.708) 
\psline[linestyle=dashed](1.277,0.291)(2.004,0.291) 
\psline[linestyle=dashed](1.106,1.912)(0.468,1.274) 
\psline[linestyle=dashed](-1.350,1.350)(-0.795,0.795) 
\psline[linestyle=dashed](-1.350,1.350)(-2.120,1.794) 
\psline[linestyle=dashed](1.395,2.067)(2.231,2.067) 
\psline[linestyle=dashed](1.789,-1.358)(1.789,-0.586) 
\psline[linestyle=dashed](-2.469,1.210)(-2.469,0.349) 
\psline[linestyle=dashed](0.940,-2.073)(0.940,-1.211) 
\psline[linestyle=dashed](2.391,-0.835)(2.391,0) 
\psline[linestyle=dashed](0.618,-1.164)(0.241,-0.511) 
\psline(1.843,0)(1.304,0) 
\psline(1.843,0)(2.391,0) 
\psline(0.773,0)(1.304,0) 
\psline[linestyle=dashed](0.773,0)(1.136,-0.628) 
\psline(0.773,0)(1.217,-0.256) 
\psline[linestyle=dashed](-2.048,2.310)(-1.200,2.310) 
\psline[linestyle=dashed](0.468,2.155)(1.348,2.155) 
\psline[linestyle=dashed](0.468,2.155)(0.468,1.274) 
\psline[linestyle=dashed](1.956,-0.376)(1.304,0) 
\psline(1.789,-0.586)(1.217,-0.256) 
\psline[linestyle=dashed](1.136,-0.628)(1.565,-1.371) 
\psline[linestyle=dashed](-1.200,2.310)(-1.981,2.026) 
\psdot[linecolor=orange,linewidth=1.2pt](-1.220,0) 
\psdot[linecolor=orange,linewidth=1.2pt](-1.028,0) 
\psdot[linecolor=orange,linewidth=1.2pt](-0.698,0) 
\psdot[linecolor=orange,linewidth=1.2pt](-1.220,0.349) 
\psdot[linecolor=darkgreen,linewidth=1.2pt](-0.373,-1.804) 
\psdot[linecolor=darkgreen,linewidth=1.2pt](-0.446,1.330) 
\psdot[linecolor=darkgreen,linewidth=1.2pt](2.200,-1.306) 
\psdot[linecolor=darkgreen,linewidth=1.2pt](-1.220,-0.813) 
\psdot[linecolor=green,linewidth=1.2pt](-1.647,0) 
\psdot[linecolor=green,linewidth=1.2pt](-2.097,0) 
\psdot[linecolor=green,linewidth=1.2pt](-1.745,0.349) 
\psdot[linecolor=green,linewidth=1.2pt](-0.844,-0.844) 
\psdot[linecolor=green,linewidth=1.2pt](-0.373,-0.970) 
\psdot[linecolor=blue,linewidth=1.2pt](-0.214,0.214) 
\psdot[linecolor=blue,linewidth=1.2pt](-0.806,0.214) 
\psdot[linecolor=blue,linewidth=1.2pt](0,0) 
\psdot[linecolor=blue,linewidth=1.2pt](-0.221,0) 
\psdot[linecolor=blue,linewidth=1.2pt](0.497,-0.443) 
\psdot[linecolor=blue,linewidth=1.2pt](0.781,-0.312) 
\psdot[linecolor=blue,linewidth=1.2pt](0.168,-0.168) 
\psdot[linecolor=blue,linewidth=1.2pt](-0.795,1.609) 
\psdot[linecolor=blue,linewidth=1.2pt](0.994,0.698) 
\psdot[linecolor=blue,linewidth=1.2pt](0.241,0) 
\psdot[linecolor=blue,linewidth=1.2pt](-0.454,0.760) 
\psdot[linecolor=blue,linewidth=1.2pt](0,0.306) 
\psdot[linecolor=blue,linewidth=1.2pt](0.643,0.232) 
\psdot[linecolor=blue,linewidth=1.2pt](0,0.558) 
\psdot[linecolor=blue,linewidth=1.2pt](0.468,0.394) 
\psdot[linecolor=blue,linewidth=1.2pt](1.395,1.224) 
\psdot[linecolor=blue,linewidth=1.2pt](-0.178,-0.178) 
\psdot[linecolor=blue,linewidth=1.2pt](-0.373,-0.373) 
\psdot[linecolor=blue,linewidth=1.2pt](0,1.307) 
\psdot[linecolor=blue,linewidth=1.2pt](0.724,0.837) 
\psdot[linecolor=blue,linewidth=1.2pt](2.391,-1.718) 
\psdot[linecolor=blue,linewidth=1.2pt](-0.380,0.380) 
\psdot[linecolor=blue,linewidth=1.2pt](1.277,0.291) 
\psdot[linecolor=blue,linewidth=1.2pt](1.106,1.912) 
\psdot[linecolor=blue,linewidth=1.2pt](1.588,0.830) 
\psdot[linecolor=blue,linewidth=1.2pt](-1.745,1.208) 
\psdot[linecolor=blue,linewidth=1.2pt](-1.350,1.350) 
\psdot[linecolor=blue,linewidth=1.2pt](1.395,2.067) 
\psdot[linecolor=bluish,linewidth=1.2pt](-0.795,0.795) 
\psdot[linecolor=bluish,linewidth=1.2pt](1.789,-1.358) 
\psdot[linecolor=brown,linewidth=1.2pt](0,-0.523) 
\psdot[linecolor=brown,linewidth=1.2pt](-2.120,1.794) 
\psdot[linecolor=brown,linewidth=1.2pt](-2.469,1.210) 
\psdot[linecolor=brown,linewidth=1.2pt](0.940,-2.073) 
\psdot[linecolor=brown,linewidth=1.2pt](2.391,-0.835) 
\psdot[linecolor=brown,linewidth=1.2pt](0.940,-1.211) 
\psdot[linecolor=brown,linewidth=1.2pt](-1.217,-1.804) 
\psdot[linecolor=brown,linewidth=1.2pt](-0.376,2.335) 
\psdot[linecolor=brown,linewidth=1.2pt](0.618,-1.164) 
\psdot[linecolor=brown,linewidth=1.2pt](-1.647,-0.854) 
\psdot[linecolor=brown,linewidth=1.2pt](1.768,1.145) 
\psdot[linecolor=red,linewidth=1.2pt](1.843,0) 
\psdot[linecolor=red,linewidth=1.2pt](0.773,0) 
\psdot[linecolor=red,linewidth=1.2pt](-2.048,2.310) 
\psdot[linecolor=red,linewidth=1.2pt](0.468,2.155) 
\psdot[linecolor=red,linewidth=1.2pt](1.956,-0.376) 
\psdot[linecolor=red,linewidth=1.2pt](1.304,0) 
\psdot[linecolor=red,linewidth=1.2pt](1.789,-0.586) 
\psdot[linecolor=red,linewidth=1.2pt](1.136,-0.628) 
\psdot[linecolor=red,linewidth=1.2pt](1.217,-0.256) 
\psdot[linecolor=red,linewidth=1.2pt](1.999,0.708) 
\psdot[linecolor=red,linewidth=1.2pt](2.004,0.291) 
\psdot[linecolor=black,linewidth=1.2pt](2.391,0) 
\psdot[linecolor=magenta,linewidth=1.2pt](-2.469,0.349) 
\psdot[linecolor=magenta,linewidth=1.2pt](-1.200,2.310) 
\psdot[linecolor=magenta,linewidth=1.2pt](-1.981,2.026) 
\psdot[linecolor=magenta,linewidth=1.2pt](2.231,2.067) 
\psdot[linecolor=magenta,linewidth=1.2pt](1.565,-1.371) 
\psdot[linecolor=magenta,linewidth=1.2pt](1.348,2.155) 
\psdot[linecolor=magenta,linewidth=1.2pt](0.241,-0.511) 
\psdot[linecolor=magenta,linewidth=1.2pt](0.468,1.274) 
\rput(-1.370,0.1){S\&P}
\rput(-1.028,0.1){Nasd}
\rput(-0.698,0.1){Cana}
\rput(-1.220,0.449){Mexi}
\rput(-0.373,-1.904){Pana}
\rput(-0.446,1.430){CoRi}
\rput(2.200,-1.406){Berm}
\rput(-1.220,-0.913){Jama}
\rput(-1.647,0.1){Braz}
\rput(-2.097,0.1){Arge}
\rput(-1.895,0.449){Chil}
\rput(-0.844,-0.944){Vene}
\rput(-0.223,-0.970){Peru}
\rput(-0.214,0.314){UK}
\rput(-0.806,0.314){Irel}
\rput(0,0.1){Fran}
\rput(-0.371,0.1){Germ}
\rput(0.597,-0.343){Swit}
\rput(0.781,-0.412){Autr}
\rput(0.148,-0.268){Ital}
\rput(-0.945,1.609){Malt}
\rput(0.994,0.798){Belg}
\rput(0.241,0.1){Neth}
\rput(-0.454,0.860){Luxe}
\rput(0.15,0.306){Swed}
\rput(0.643,0.332){Denm}
\rput(0.15,0.558){Finl}
\rput(0.288,0.394){Norw}
\rput(1.545,1.224){Icel}
\rput(-0.118,-0.278){Spai}
\rput(-0.423,-0.273){Port}
\rput(0,1.407){Gree}
\rput(0.724,0.937){CzRe}
\rput(2.391,-1.818){Slok}
\rput(-0.380,0.480){Hung}
\rput(1.127,0.311){Pola}
\rput(1.206,1.812){Roma}
\rput(1.738,0.830){Esto}
\rput(-1.745,1.308){Latv}
\rput(-1.350,1.450){Lith}
\rput(1.545,1.967){Ukra}
\rput(-0.825,0.695){Russ}
\rput(1.789,-1.458){Turk}
\rput(0,-0.623){Isra}
\rput(-2.270,1.794){Pale}
\rput(-2.469,1.310){Leba}
\rput(0.940,-2.173){Jord}
\rput(2.241,-0.835){SaAr}
\rput(1.090,-1.211){Qata}
\rput(-1.407,-1.804){Ohma}
\rput(-0.376,2.435){Paki}
\rput(0.618,-1.264){Indi}
\rput(-1.647,-0.954){SrLa}
\rput(1.918,1.145){Bang}
\rput(1.843,0.1){Japa}
\rput(0.773,0.1){HoKo}
\rput(-2.198,2.310){Chin}
\rput(0.468,2.255){Mongo}
\rput(1.956,-0.276){Taiw}
\rput(1.304,0.1){SoKo}
\rput(1.789,-0.486){Thai}
\rput(1.136,-0.528){Mala}
\rput(1.217,-0.156){Sing}
\rput(2.149,0.708){Indo}
\rput(2.154,0.291){Phil}
\rput(2.391,0.1){Aust}
\rput(-2.469,0.249){Moro}
\rput(-1.200,2.410){Tuni}
\rput(-2.131,2.026){Egyp}
\rput(2.231,2.167){Ghan}
\rput(1.565,-1.471){Nige}
\rput(1.348,2.255){Keny}
\rput(0.391,-0.511){SoAf}
\rput(0.648,1.274){Maur}
\end{pspicture}

\begin{center}
Figure 12: minimum spanning tree for the first semester of 2000.
\end{center}

\begin{pspicture}(-8,-5.8)(1,7.4)
\psset{xunit=3,yunit=3} \scriptsize
\psline(-1.121,0)(-1.276,0) 
\psline[linestyle=dashed](-1.121,0)(-1.121,-0.822) 
\psline(-1.121,0)(-1.121,0.483) 
\psline(-1.121,0)(-0.743,0) 
\psline(-1.276,0)(-1.603,0) 
\psline(-1.276,0)(-1.547,0.271) 
\psline[linestyle=dashed](-1.547,0.271)(-1.547,1.112) 
\psline[linestyle=dashed](0,1.312)(0,0.558) 
\psline[linestyle=dashed](0,1.312)(0.400,2.004) 
\psline[linestyle=dashed](0,1.312)(0,2.113) 
\psline[linestyle=dashed](1.467,-1.084)(1.467,-0.276) 
\psline[linestyle=dashed](0,-1.270)(0,-0.452) 
\psline(-1.121,0.991)(-1.121,0.483) 
\psline(-0.874,0.602)(-0.462,0.190) 
\psline[linestyle=dashed](-0.272,1.290)(-0.272,0.570) 
\psline(-0.338,-0.584)(0,0) 
\psline(-0.462,0.190)(-0.272,0) 
\psline(-0.462,0.190)(-0.462,0.674) 
\psline(0.696,-0.509)(0.990,0) 
\psline(-0.272,0)(-0.500,0) 
\psline(-0.272,0)(0,0) 
\psline(-0.272,0)(-0.499,-0.131) 
\psline(-0.272,0)(-0.272,0.270) 
\psline(-0.743,0)(-0.500,0) 
\psline(0,-0.452)(0,0) 
\psline(-0.382,-0.220)(0,0) 
\psline[linestyle=dashed](-1.552,1.728)(-2.313,1.728) 
\psline[linestyle=dashed](-1.552,1.728)(-0.775,1.728) 
\psline(0,0.558)(0,0) 
\psline(0,0)(0.299,0.299) 
\psline(0,0)(0.444,0) 
\psline[linestyle=dashed](-0.848,-0.735)(-0.499,-0.131) 
\psline[linestyle=dashed](-0.848,-0.735)(-0.848,-1.511) 
\psline(-0.499,-0.131)(-0.753,-0.278) 
\psline(-0.499,-0.131)(-0.499,-0.687) 
\psline(0.555,0.742)(0.299,0.299) 
\psline[linestyle=dashed](0.555,0.742)(0.978,1.474) 
\psline(0.299,0.299)(0.299,0.819) 
\psline(0.299,0.299)(0.728,0.546) 
\psline[linestyle=dashed](0.688,1.493)(0.299,0.819) 
\psline(-0.272,0.270)(-0.272,0.570) 
\psline[linestyle=dashed](1.074,0.364)(0.444,0) 
\psline[linestyle=dashed](1.397,0.705)(1.397,1.524) 
\psline[linestyle=dashed](1.397,0.705)(0.990,0) 
\psline(1.169,0.800)(0.728,0.546) 
\psline(1.467,-0.276)(1.792,-0.838) 
\psline(1.467,-0.276)(0.990,0) 
\psline[linestyle=dashed](-1.728,-1.511)(-0.848,-1.511) 
\psline[linestyle=dashed](2.311,0)(2.311,0.780) 
\psline[linestyle=dashed](2.311,0)(1.556,0) 
\psline[linestyle=dashed](2.311,0)(2.311,-0.788) 
\psline[linestyle=dashed](0.990,-0.729)(0.573,-1.451) 
\psline[linestyle=dashed](0.990,-0.729)(0.990,-1.563) 
\psline[linestyle=dashed](0.990,-0.729)(0.990,0) 
\psline[linestyle=dashed](-0.499,-1.503)(-0.499,-0.687) 
\psline(0.728,0.546)(1.067,1.133) 
\psline[linestyle=dashed](1.397,2.342)(1.397,1.524) 
\psline[linestyle=dashed](1.397,2.342)(2.207,2.342) 
\psline[linestyle=dashed](-1.547,1.112)(-2.410,1.112) 
\psline(-0.775,1.728)(-0.462,1.186) 
\psline(1.556,0)(0.990,0) 
\psline[linestyle=dashed](0.990,0)(0.377,-0.354) 
\psline[linestyle=dashed](0.990,0)(1.657,0.385) 
\psline(0.990,0)(1.256,-0.461) 
\psline(0.990,0)(0.444,0) 
\psline[linestyle=dashed](-0.367,-1.949)(0.377,-1.949) 
\psline[linestyle=dashed](-0.462,1.932)(-0.462,1.186) 
\psline(-0.462,1.186)(-0.462,0.674) 
\psline[linestyle=dashed](0.377,-0.354)(0.728,-0.963) 
\psline[linestyle=dashed](0.377,-0.354)(0.377,-1.111) 
\psline[linestyle=dashed](0.377,-1.111)(0.377,-1.949) 
\psline[linestyle=dashed](0,2.113)(-0.791,2.113) 
\psline[linestyle=dashed](-2.410,0.248)(-2.410,1.112) 
\psdot[linecolor=orange,linewidth=1.2pt](-1.121,0) 
\psdot[linecolor=orange,linewidth=1.2pt](-1.276,0) 
\psdot[linecolor=orange,linewidth=1.2pt](-1.603,0) 
\psdot[linecolor=orange,linewidth=1.2pt](-1.547,0.271) 
\psdot[linecolor=darkgreen,linewidth=1.2pt](0,1.312) 
\psdot[linecolor=darkgreen,linewidth=1.2pt](1.467,-1.084) 
\psdot[linecolor=darkgreen,linewidth=1.2pt](0,-1.270) 
\psdot[linecolor=darkgreen,linewidth=1.2pt](-1.121,-0.822) 
\psdot[linecolor=green,linewidth=1.2pt](-1.121,0.991) 
\psdot[linecolor=green,linewidth=1.2pt](-1.121,0.483) 
\psdot[linecolor=green,linewidth=1.2pt](-0.874,0.602) 
\psdot[linecolor=green,linewidth=1.2pt](-0.272,1.290) 
\psdot[linecolor=green,linewidth=1.2pt](-0.338,-0.584) 
\psdot[linecolor=blue,linewidth=1.2pt](-0.462,0.190) 
\psdot[linecolor=blue,linewidth=1.2pt](0.696,-0.509) 
\psdot[linecolor=blue,linewidth=1.2pt](-0.272,0) 
\psdot[linecolor=blue,linewidth=1.2pt](-0.743,0) 
\psdot[linecolor=blue,linewidth=1.2pt](0,-0.452) 
\psdot[linecolor=blue,linewidth=1.2pt](-0.382,-0.220) 
\psdot[linecolor=blue,linewidth=1.2pt](-0.500,0) 
\psdot[linecolor=blue,linewidth=1.2pt](-1.552,1.728) 
\psdot[linecolor=blue,linewidth=1.2pt](0,0.558) 
\psdot[linecolor=blue,linewidth=1.2pt](0,0) 
\psdot[linecolor=blue,linewidth=1.2pt](-0.848,-0.735) 
\psdot[linecolor=blue,linewidth=1.2pt](-0.499,-0.131) 
\psdot[linecolor=blue,linewidth=1.2pt](0.555,0.742) 
\psdot[linecolor=blue,linewidth=1.2pt](-0.753,-0.278) 
\psdot[linecolor=blue,linewidth=1.2pt](0.299,0.299) 
\psdot[linecolor=blue,linewidth=1.2pt](0.688,1.493) 
\psdot[linecolor=blue,linewidth=1.2pt](-0.272,0.270) 
\psdot[linecolor=blue,linewidth=1.2pt](-0.272,0.570) 
\psdot[linecolor=blue,linewidth=1.2pt](1.074,0.364) 
\psdot[linecolor=blue,linewidth=1.2pt](0.299,0.819) 
\psdot[linecolor=blue,linewidth=1.2pt](1.397,0.705) 
\psdot[linecolor=blue,linewidth=1.2pt](1.169,0.800) 
\psdot[linecolor=blue,linewidth=1.2pt](1.467,-0.276) 
\psdot[linecolor=blue,linewidth=1.2pt](-1.728,-1.511) 
\psdot[linecolor=blue,linewidth=1.2pt](1.792,-0.838) 
\psdot[linecolor=blue,linewidth=1.2pt](2.311,0) 
\psdot[linecolor=blue,linewidth=1.2pt](0.990,-0.729) 
\psdot[linecolor=blue,linewidth=1.2pt](-0.499,-1.503) 
\psdot[linecolor=bluish,linewidth=1.2pt](0.728,0.546) 
\psdot[linecolor=bluish,linewidth=1.2pt](1.067,1.133) 
\psdot[linecolor=brown,linewidth=1.2pt](-0.499,-0.687) 
\psdot[linecolor=brown,linewidth=1.2pt](0.573,-1.451) 
\psdot[linecolor=brown,linewidth=1.2pt](1.397,2.342) 
\psdot[linecolor=brown,linewidth=1.2pt](-1.547,1.112) 
\psdot[linecolor=brown,linewidth=1.2pt](1.397,1.524) 
\psdot[linecolor=brown,linewidth=1.2pt](2.311,0.780) 
\psdot[linecolor=brown,linewidth=1.2pt](0.990,-1.563) 
\psdot[linecolor=brown,linewidth=1.2pt](-2.313,1.728) 
\psdot[linecolor=brown,linewidth=1.2pt](-0.775,1.728) 
\psdot[linecolor=brown,linewidth=1.2pt](0.978,1.474) 
\psdot[linecolor=brown,linewidth=1.2pt](0.400,2.004) 
\psdot[linecolor=red,linewidth=1.2pt](1.556,0) 
\psdot[linecolor=red,linewidth=1.2pt](0.990,0) 
\psdot[linecolor=red,linewidth=1.2pt](-0.367,-1.949) 
\psdot[linecolor=red,linewidth=1.2pt](2.207,2.342) 
\psdot[linecolor=red,linewidth=1.2pt](-0.462,1.932) 
\psdot[linecolor=red,linewidth=1.2pt](-0.462,1.186) 
\psdot[linecolor=red,linewidth=1.2pt](0.377,-0.354) 
\psdot[linecolor=red,linewidth=1.2pt](0.728,-0.963) 
\psdot[linecolor=red,linewidth=1.2pt](-0.462,0.674) 
\psdot[linecolor=red,linewidth=1.2pt](0.377,-1.111) 
\psdot[linecolor=red,linewidth=1.2pt](1.657,0.385) 
\psdot[linecolor=black,linewidth=1.2pt](1.256,-0.461) 
\psdot[linecolor=magenta,linewidth=1.2pt](0,2.113) 
\psdot[linecolor=magenta,linewidth=1.2pt](0.377,-1.949) 
\psdot[linecolor=magenta,linewidth=1.2pt](-0.848,-1.511) 
\psdot[linecolor=magenta,linewidth=1.2pt](-2.410,0.248) 
\psdot[linecolor=magenta,linewidth=1.2pt](-0.791,2.113) 
\psdot[linecolor=magenta,linewidth=1.2pt](-2.410,1.112) 
\psdot[linecolor=magenta,linewidth=1.2pt](0.444,0) 
\psdot[linecolor=magenta,linewidth=1.2pt](2.311,-0.788) 
\rput(-1.121,0.1){S\&P}
\rput(-1.276,-0.1){Nasd}
\rput(-1.603,0.1){Cana}
\rput(-1.697,0.271){Mexi}
\rput(0.15,1.312){Pana}
\rput(1.467,-1.184){CoRi}
\rput(0,-1.370){Berm}
\rput(-1.121,-0.922){Jama}
\rput(-1.121,1.091){Braz}
\rput(-1.271,0.483){Arge}
\rput(-0.874,0.702){Chil}
\rput(-0.272,1.390){Vene}
\rput(-0.338,-0.684){Peru}
\rput(-0.462,0.290){UK}
\rput(0.696,-0.609){Irel}
\rput(-0.272,0.1){Fran}
\rput(-0.743,0.1){Germ}
\rput(-0.15,-0.452){Swit}
\rput(-0.382,-0.320){Autr}
\rput(-0.500,0.1){Ital}
\rput(-1.552,1.828){Malt}
\rput(0.15,0.558){Belg}
\rput(0,0.1){Neth}
\rput(-0.938,-0.635){Luxe}
\rput(-0.679,-0.131){Swed}
\rput(0.375,0.742){Denm}
\rput(-0.903,-0.278){Finl}
\rput(0.149,0.299){Norw}
\rput(0.688,1.593){Icel}
\rput(-0.122,0.270){Spai}
\rput(-0.132,0.570){Port}
\rput(1.074,0.464){Gree}
\rput(0.149,0.819){CzRe}
\rput(1.247,0.705){Slok}
\rput(1.169,0.900){Hung}
\rput(1.617,-0.276){Pola}
\rput(-1.908,-1.511){Roma}
\rput(1.792,-0.938){Esto}
\rput(2.461,0){Latv}
\rput(0.840,-0.729){Lith}
\rput(-0.499,-1.603){Ukra}
\rput(0.728,0.446){Russ}
\rput(1.067,1.233){Turk}
\rput(-0.649,-0.687){Isra}
\rput(0.573,-1.551){Pale}
\rput(1.397,2.442){Leba}
\rput(-1.547,1.212){Jord}
\rput(1.247,1.524){SaAr}
\rput(2.311,0.880){Qata}
\rput(0.990,-1.663){Ohma}
\rput(-2.313,1.828){Paki}
\rput(-0.775,1.828){Indi}
\rput(0.978,1.574){SrLa}
\rput(0.400,2.104){Bang}
\rput(1.556,0.1){Japa}
\rput(0.990,0.1){HoKo}
\rput(-0.517,-1.949){Chin}
\rput(2.397,2.342){Mongo}
\rput(-0.462,2.032){Taiw}
\rput(-0.612,1.186){SoKo}
\rput(0.347,-0.254){Thai}
\rput(0.578,-0.963){Mala}
\rput(-0.612,0.674){Sing}
\rput(0.227,-1.111){Indo}
\rput(1.807,0.385){Phil}
\rput(1.256,-0.561){Aust}
\rput(0,2.213){Moro}
\rput(0.527,-1.949){Tuni}
\rput(-0.848,-1.611){Egyp}
\rput(-2.410,0.148){Ghan}
\rput(-0.941,2.113){Nige}
\rput(-2.410,1.212){Keny}
\rput(0.444,0.1){SoAf}
\rput(2.311,-0.888){Maur}
\end{pspicture}

\begin{center}
Figure 13: minimum spanning tree for the second semester of 2000.
\end{center}

Starting from the first semester of 2000, depicted in figure 12, we establish a threshold $0.70\pm 0.03$. The American cluster is almost unmodified, and the European cluster has now two main hubs: France and Netherlands. Israel and South Africa are still connected with the European cluster, and Australia is now definitely part of a Pacific Asian cluster, comprised of itself, Japan, Hong Kong, South Korea, Singapore, and Taiwan. Above the threshold, there is one connection that seems significant, the one between Hong Kong and Malaysia, and many other connections that seem random in nature.

Figure 13 shows the minimum spanning tree for the second semester of 2000, with the threshold $0.69\pm 0.03$. What one can see from it is that the American and the European clusters are not very altered, but the Pacific Asian cluster has been again pulverized into some looser connections. If we consider some of the connections above the threshold, then Hong Kong becomes a hub for some Pacific Asian indices and also for some European ones.

\newpage

For 2001, we now have 78 indices, for we added Bulgaria, Kazakhstan, Vietnam, New Zealand, and Botswana to the indices that were already being used. The minimum spanning tree for the first semester of 2001 is depicted in figure 14 with threshold $0.69\pm 0.03$. The South American cluster is detached from the North American one, and France slowly assumes the role of main hub for European indices, replacing Netherlands. We have some semblance of a cluster of Pacific Asian indices, together with those from Oceania (Australia and New Zealand). Other structures below the threshold remain, basically, unnaltered.

\begin{pspicture}(-9,-5.4)(1,7.7)
\psset{xunit=3,yunit=3} \scriptsize
\psline(-0.672,0)(-0.672,-0.144) 
\psline(-0.672,0)(-0.908,0) 
\psline(-0.672,0)(-0.225,0) 
\psline(-0.908,0)(-1.341,0) 
\psline[linestyle=dashed](-1.411,-1.531)(-0.595,-1.531) 
\psline[linestyle=dashed](2.467,1.625)(2.467,0.793) 
\psline[linestyle=dashed](-0.948,-1.370)(-0.595,-0.760) 
\psline[linestyle=dashed](0.175,-1.719)(0.175,-0.893) 
\psline(-1.030,0.469)(-0.598,0.469) 
\psline(-1.030,0.469)(-1.655,0.469) 
\psline(-0.598,0.469)(-0.108,0.186) 
\psline[linestyle=dashed](-0.381,0.659)(-0.381,1.389) 
\psline(-0.381,0.659)(-0.108,0.186) 
\psline[linestyle=dashed](-0.381,0.659)(-1.141,0.659) 
\psline[linestyle=dashed](-0.381,0.659)(-0.747,1.293) 
\psline(0,0.974)(0,0.342) 
\psline[linestyle=dashed](0,0.974)(0,1.703) 
\psline[linestyle=dashed](-0.381,1.389)(-0.795,2.106) 
\psline[linestyle=dashed](-0.381,1.389)(-0.381,2.197) 
\psline(-0.176,0.102)(0,0) 
\psline(0,-0.884)(0,-0.434) 
\psline(0,0)(-0.225,0) 
\psline(0,0)(0.471,-0.272) 
\psline(0,0)(-0.150,-0.086) 
\psline(0,0)(0.187,0) 
\psline(0,0)(-0.127,-0.220) 
\psline(0,0)(0,-0.434) 
\psline(0,0)(0.175,-0.303) 
\psline(0,0)(-0.108,0.186) 
\psline(0,0)(0,0.342) 
\psline(0,0)(0.228,0.394) 
\psline(0.449,0.152)(0.187,0) 
\psline[linestyle=dashed](-2.757,0.659)(-2.757,-0.148) 
\psline[linestyle=dashed](-2.757,0.659)(-2.757,1.454) 
\psline[linestyle=dashed](-2.757,0.659)(-1.939,0.659) 
\psline(0.476,-0.167)(0.187,0) 
\psline(0.187,0)(0.472,0.494) 
\psline(0.187,0)(0.774,0) 
\psline(0.472,0.494)(0.999,0.190) 
\psline[linestyle=dashed](0.472,0.494)(1.220,0.494) 
\psline[linestyle=dashed](0.472,0.494)(1.148,0.884) 
\psline(0.472,0.494)(0.784,1.034) 
\psline(-0.127,-0.220)(-0.362,-0.356) 
\psline(-0.127,-0.220)(-0.334,-0.578) 
\psline(-0.362,-0.356)(-0.811,-0.356) 
\psline(-0.362,-0.356)(-0.595,-0.760) 
\psline(0.175,-0.303)(0.175,-0.893) 
\psline(0.175,-0.303)(0.597,-0.771) 
\psline[linestyle=dashed](0.175,-0.303)(0.532,-0.921) 
\psline[linestyle=dashed](-1.969,-0.974)(-1.168,-0.974) 
\psline[linestyle=dashed](0.175,-0.893)(-0.244,-1.619) 
\psline(-0.811,-0.356)(-1.247,-0.608) 
\psline(-0.811,-0.356)(-1.328,-0.356) 
\psline[linestyle=dashed](-0.811,-0.356)(-1.168,-0.974) 
\psline[linestyle=dashed](-0.595,-0.760)(-0.595,-1.531) 
\psline[linestyle=dashed](-2.116,-0.356)(-1.328,-0.356) 
\psline[linestyle=dashed](-2.116,-0.356)(-2.116,0.486) 
\psline[linestyle=dashed](1.174,-0.693)(0.774,0) 
\psline[linestyle=dashed](0.228,0.394)(0.228,1.190) 
\psline[linestyle=dashed](0.228,0.394)(0.626,1.083) 
\psline[linestyle=dashed](-2.757,-1.000)(-2.757,-0.148) 
\psline[linestyle=dashed](-1.141,0.659)(-1.939,0.659) 
\psline[linestyle=dashed](-1.937,1.369)(-1.937,2.249) 
\psline[linestyle=dashed](1.910,0.356)(1.293,0) 
\psline(2.198,-0.244)(1.775,0) 
\psline(1.775,0)(1.293,0) 
\psline[linestyle=dashed](1.775,0)(2.467,0) 
\psline[linestyle=dashed](-2.135,0.832)(-2.757,1.454) 
\psline[linestyle=dashed](0.784,1.878)(0.784,1.034) 
\psline[linestyle=dashed](1.376,-0.348)(0.774,0) 
\psline(1.293,0)(0.774,0) 
\psline(1.293,0)(1.843,-0.318) 
\psline[linestyle=dashed](-2.757,2.249)(-1.937,2.249) 
\psline[linestyle=dashed](-2.757,2.249)(-2.757,1.454) 
\psline[linestyle=dashed](2.467,0)(2.467,0.793) 
\psline(0.774,0)(0.774,-0.575) 
\psline[linestyle=dashed](-1.937,2.249)(-1.106,2.249) 
\psline[linestyle=dashed](0.626,1.923)(0.626,1.083) 
\psdot[linecolor=orange,linewidth=1.2pt](-0.672,0) 
\psdot[linecolor=orange,linewidth=1.2pt](-0.672,-0.144) 
\psdot[linecolor=orange,linewidth=1.2pt](-0.908,0) 
\psdot[linecolor=orange,linewidth=1.2pt](-1.341,0) 
\psdot[linecolor=darkgreen,linewidth=1.2pt](-1.411,-1.531) 
\psdot[linecolor=darkgreen,linewidth=1.2pt](2.467,1.625) 
\psdot[linecolor=darkgreen,linewidth=1.2pt](-0.948,-1.370) 
\psdot[linecolor=darkgreen,linewidth=1.2pt](0.175,-1.719) 
\psdot[linecolor=green,linewidth=1.2pt](-1.030,0.469) 
\psdot[linecolor=green,linewidth=1.2pt](-0.598,0.469) 
\psdot[linecolor=green,linewidth=1.2pt](-0.381,0.659) 
\psdot[linecolor=green,linewidth=1.2pt](0,0.974) 
\psdot[linecolor=green,linewidth=1.2pt](-0.381,1.389) 
\psdot[linecolor=blue,linewidth=1.2pt](-0.176,0.102) 
\psdot[linecolor=blue,linewidth=1.2pt](0,-0.884) 
\psdot[linecolor=blue,linewidth=1.2pt](0,0) 
\psdot[linecolor=blue,linewidth=1.2pt](-0.225,0) 
\psdot[linecolor=blue,linewidth=1.2pt](0.449,0.152) 
\psdot[linecolor=blue,linewidth=1.2pt](0.471,-0.272) 
\psdot[linecolor=blue,linewidth=1.2pt](-0.150,-0.086) 
\psdot[linecolor=blue,linewidth=1.2pt](-2.757,0.659) 
\psdot[linecolor=blue,linewidth=1.2pt](0.476,-0.167) 
\psdot[linecolor=blue,linewidth=1.2pt](0.187,0) 
\psdot[linecolor=blue,linewidth=1.2pt](0.472,0.494) 
\psdot[linecolor=blue,linewidth=1.2pt](-0.127,-0.220) 
\psdot[linecolor=blue,linewidth=1.2pt](0,-0.434) 
\psdot[linecolor=blue,linewidth=1.2pt](-0.362,-0.356) 
\psdot[linecolor=blue,linewidth=1.2pt](0.175,-0.303) 
\psdot[linecolor=blue,linewidth=1.2pt](-1.969,-0.974) 
\psdot[linecolor=blue,linewidth=1.2pt](-0.108,0.186) 
\psdot[linecolor=blue,linewidth=1.2pt](0,0.342) 
\psdot[linecolor=blue,linewidth=1.2pt](0.175,-0.893) 
\psdot[linecolor=blue,linewidth=1.2pt](-0.811,-0.356) 
\psdot[linecolor=blue,linewidth=1.2pt](0.597,-0.771) 
\psdot[linecolor=blue,linewidth=1.2pt](-0.595,-0.760) 
\psdot[linecolor=blue,linewidth=1.2pt](-1.247,-0.608) 
\psdot[linecolor=blue,linewidth=1.2pt](-2.116,-0.356) 
\psdot[linecolor=blue,linewidth=1.2pt](-0.244,-1.619) 
\psdot[linecolor=blue,linewidth=1.2pt](0.999,0.190) 
\psdot[linecolor=blue,linewidth=1.2pt](-0.795,2.106) 
\psdot[linecolor=blue,linewidth=1.2pt](1.220,0.494) 
\psdot[linecolor=blue,linewidth=1.2pt](1.174,-0.693) 
\psdot[linecolor=bluish,linewidth=1.2pt](-1.328,-0.356) 
\psdot[linecolor=brown,linewidth=1.2pt](-0.595,-1.531) 
\psdot[linecolor=bluish,linewidth=1.2pt](-1.655,0.469) 
\psdot[linecolor=brown,linewidth=1.2pt](0.228,0.394) 
\psdot[linecolor=brown,linewidth=1.2pt](-2.757,-1.000) 
\psdot[linecolor=brown,linewidth=1.2pt](-1.168,-0.974) 
\psdot[linecolor=brown,linewidth=1.2pt](-0.381,2.197) 
\psdot[linecolor=brown,linewidth=1.2pt](-1.141,0.659) 
\psdot[linecolor=brown,linewidth=1.2pt](-0.747,1.293) 
\psdot[linecolor=brown,linewidth=1.2pt](-1.937,1.369) 
\psdot[linecolor=brown,linewidth=1.2pt](-2.116,0.486) 
\psdot[linecolor=brown,linewidth=1.2pt](1.910,0.356) 
\psdot[linecolor=brown,linewidth=1.2pt](0.532,-0.921) 
\psdot[linecolor=brown,linewidth=1.2pt](1.148,0.884) 
\psdot[linecolor=red,linewidth=1.2pt](2.198,-0.244) 
\psdot[linecolor=red,linewidth=1.2pt](1.775,0) 
\psdot[linecolor=red,linewidth=1.2pt](-2.135,0.832) 
\psdot[linecolor=red,linewidth=1.2pt](0.784,1.878) 
\psdot[linecolor=red,linewidth=1.2pt](1.376,-0.348) 
\psdot[linecolor=red,linewidth=1.2pt](1.293,0) 
\psdot[linecolor=red,linewidth=1.2pt](0.784,1.034) 
\psdot[linecolor=red,linewidth=1.2pt](-2.757,2.249) 
\psdot[linecolor=red,linewidth=1.2pt](2.467,0) 
\psdot[linecolor=red,linewidth=1.2pt](0.774,0) 
\psdot[linecolor=red,linewidth=1.2pt](2.467,0.793) 
\psdot[linecolor=red,linewidth=1.2pt](-1.937,2.249) 
\psdot[linecolor=black,linewidth=1.2pt](1.843,-0.318) 
\psdot[linecolor=black,linewidth=1.2pt](0.774,-0.575) 
\psdot[linecolor=magenta,linewidth=1.2pt](-2.757,-0.148) 
\psdot[linecolor=magenta,linewidth=1.2pt](-1.106,2.249) 
\psdot[linecolor=magenta,linewidth=1.2pt](0,1.703) 
\psdot[linecolor=magenta,linewidth=1.2pt](-2.757,1.454) 
\psdot[linecolor=magenta,linewidth=1.2pt](-1.939,0.659) 
\psdot[linecolor=magenta,linewidth=1.2pt](0.228,1.190) 
\psdot[linecolor=magenta,linewidth=1.2pt](0.626,1.923) 
\psdot[linecolor=magenta,linewidth=1.2pt](-0.334,-0.578) 
\psdot[linecolor=magenta,linewidth=1.2pt](0.626,1.083) 
\rput(-0.672,0.1){S\&P}
\rput(-0.672,-0.244){Nasd}
\rput(-0.908,0.1){Cana}
\rput(-1.341,0.1){Mexi}
\rput(-1.561,-1.531){Pana}
\rput(2.467,1.725){CoRi}
\rput(-1.098,-1.370){Berm}
\rput(0.175,-1.819){Jama}
\rput(-1.030,0.569){Braz}
\rput(-0.598,0.569){Arge}
\rput(-0.231,0.659){Chil}
\rput(-0.15,0.974){Vene}
\rput(-0.231,1.389){Peru}
\rput(-0.306,0.172){UK}
\rput(-0.15,-0.884){Irel}
\rput(0,0.1){Fran}
\rput(-0.345,0.07){Germ}
\rput(0.599,0.182){Swit}
\rput(0.471,-0.372){Autr}
\rput(-0.300,-0.1){Ital}
\rput(-2.907,0.659){Malt}
\rput(0.626,-0.167){Belg}
\rput(0.217,0.1){Neth}
\rput(0.622,0.594){Luxe}
\rput(-0.317,-0.220){Swed}
\rput(-0.18,-0.434){Denm}
\rput(-0.512,-0.406){Finl}
\rput(0.325,-0.303){Norw}
\rput(-2.119,-0.974){Icel}
\rput(-0.108,0.286){Spai}
\rput(-0.12,0.382){Port}
\rput(0.325,-0.893){Gree}
\rput(-0.811,-0.456){CzRe}
\rput(0.747,-0.771){Slok}
\rput(-0.745,-0.700){Hung}
\rput(-1.397,-0.608){Pola}
\rput(-2.116,-0.456){Roma}
\rput(-0.244,-1.719){Bulg}
\rput(0.999,0.290){Esto}
\rput(-0.795,2.206){Latv}
\rput(1.220,0.594){Lith}
\rput(1.174,-0.793){Ukra}
\rput(-1.328,-0.456){Russ}
\rput(-0.595,-1.631){Kaza}
\rput(-1.655,0.569){Turk}
\rput(0.108,0.454){Isra}
\rput(-2.757,-1.100){Pale}
\rput(-1.168,-1.074){Leba}
\rput(-0.381,2.297){Jord}
\rput(-1.141,0.759){SaAr}
\rput(-0.747,1.393){Qata}
\rput(-1.937,1.269){Ohma}
\rput(-2.116,0.586){Paki}
\rput(1.910,0.456){Indi}
\rput(0.532,-1.021){SrLa}
\rput(1.148,0.984){Bang}
\rput(2.198,-0.344){Japa}
\rput(1.775,0.1){HoKo}
\rput(-2.085,0.932){Chin}
\rput(0.854,1.978){Mongo}
\rput(1.376,-0.448){Taiw}
\rput(1.273,0.1){SoKo}
\rput(0.934,1.034){Thai}
\rput(-2.907,2.249){Viet}
\rput(2.317,0.1){Mala}
\rput(0.774,0.1){Sing}
\rput(2.317,0.793){Indo}
\rput(-1.937,2.349){Phil}
\rput(1.843,-0.418){Aust}
\rput(0.924,-0.575){NeZe}
\rput(-2.927,-0.148){Moro}
\rput(-1.106,2.349){Tuni}
\rput(0,1.803){Egyp}
\rput(-2.907,1.454){Ghan}
\rput(-1.939,0.759){Nige}
\rput(0.228,1.290){Keny}
\rput(0.586,2.023){Bots}
\rput(-0.334,-0.678){SoAf}
\rput(0.466,1.083){Maur}
\end{pspicture}

\begin{center}
Figure 14: minimum spanning tree for the first semester of 2001.
\end{center}

\newpage

For the second semester of 2001 (figure 15), the threshold is $0.69\pm 1$. France is now the undisputable hub for Europe, and both the American and the Pacific Asian clusters are clearly defined. The newcomer index, New Zealand, connects with Australia, as it would be expected, but a good number of indices are still loosely connected, and the connections shown in dashed lines are probably all due to random noise. The exception are the loose connections between some Arab indices (Jordan-Qatar-Ohman-Egypt).

\begin{pspicture}(-4.5,-6.5)(1,5.3)
\psset{xunit=3,yunit=3} \scriptsize
\psline(-0.752,0)(-0.874,0) 
\psline(-0.752,0)(-0.507,0) 
\psline(-0.874,0)(-1.354,0) 
\psline(-0.507,0)(-0.875,0.368) 
\psline(-0.507,0)(-0.117,0) 
\psline[linestyle=dashed](2.917,-0.544)(3.461,0) 
\psline[linestyle=dashed](1.872,0.720)(1.456,0) 
\psline[linestyle=dashed](0.410,-1.420)(0.410,-0.641) 
\psline[linestyle=dashed](0,-2.056)(0.604,-1.707) 
\psline[linestyle=dashed](0,-2.056)(0.758,-2.056) 
\psline[linestyle=dashed](0,-2.056)(0,-1.422) 
\psline[linestyle=dashed](0,-2.056)(-0.797,-2.056) 
\psline[linestyle=dashed](0,-2.056)(-0.664,-1.672) 
\psline(-1.023,-0.184)(-0.491,-0.491) 
\psline(-0.491,-0.491)(-0.081,-0.081) 
\psline(-0.491,-0.491)(-1.098,-0.491) 
\psline[linestyle=dashed](-0.540,-0.881)(0,-1.422) 
\psline(-0.579,0.459)(-0.158,0.216) 
\psline(-0.080,0.080)(-0.433,0.284) 
\psline(-0.080,0.080)(0,0) 
\psline(-0.080,0.080)(-0.158,0.216) 
\psline(0,0)(-0.117,0) 
\psline(0,0)(0.158,-0.091) 
\psline(0,0)(0,-0.160) 
\psline(0,0)(0,0.086) 
\psline(0,0)(0.410,0) 
\psline(0,0)(0.208,-0.360) 
\psline(0,0)(-0.081,-0.081) 
\psline(0,0)(0.402,0.402) 
\psline(0,0.939)(0,0.325) 
\psline(0,-0.160)(0,-0.680) 
\psline[linestyle=dashed](3.461,-1.507)(3.461,-0.753) 
\psline(0,0.325)(0,0.086) 
\psline[linestyle=dashed](0,-0.680)(0,-1.422) 
\psline(0,-0.680)(-0.522,-0.680) 
\psline(-0.158,0.216)(-0.268,0.407) 
\psline(0.410,0)(0.934,0) 
\psline(0.410,0)(0.410,-0.641) 
\psline(-0.081,-0.081)(-0.270,-0.407) 
\psline(0.934,1.006)(0.934,0.511) 
\psline(0.934,1.006)(1.514,1.006) 
\psline(0.934,0.511)(0.934,0) 
\psline(0.934,0)(1.456,0) 
\psline[linestyle=dashed](0.934,0)(0.934,-0.824) 
\psline[linestyle=dashed](0.934,0)(1.340,-0.703) 
\psline[linestyle=dashed](1.456,0)(1.456,0.779) 
\psline(1.456,0)(2.010,0) 
\psline[linestyle=dashed](1.946,-0.370)(1.584,-0.997) 
\psline[linestyle=dashed](1.946,-0.370)(2.587,0) 
\psline(3.192,0.351)(2.989,0) 
\psline(3.192,0.351)(3.192,0.845) 
\psline[linestyle=dashed](4.021,1.512)(3.192,1.512) 
\psline[linestyle=dashed](2.622,0.635)(3.031,1.343) 
\psline[linestyle=dashed](2.622,0.635)(2.989,0) 
\psline[linestyle=dashed](2.622,0.635)(2.622,1.468) 
\psline(2.010,0)(2.587,0) 
\psline[linestyle=dashed](2.010,0)(1.350,-0.381) 
\psline[linestyle=dashed](2.398,-1.701)(1.584,-1.701) 
\psline[linestyle=dashed](1.584,-0.997)(1.222,-1.625) 
\psline[linestyle=dashed](1.584,-0.997)(1.584,-1.701) 
\psline[linestyle=dashed](1.584,-1.701)(2.272,-2.098) 
\psline[linestyle=dashed](3.752,1.243)(3.752,0.505) 
\psline(3.752,0.505)(3.461,0) 
\psline[linestyle=dashed](0,-1.422)(-0.849,-1.422) 
\psline(2.549,-0.254)(2.989,0) 
\psline[linestyle=dashed](2.549,-0.254)(2.194,-0.868) 
\psline(2.989,0)(2.587,0) 
\psline(2.989,0)(3.461,0) 
\psline[linestyle=dashed](3.461,-0.753)(3.461,0) 
\psline[linestyle=dashed](1.475,1.227)(2.338,1.227) 
\psline(2.338,0.431)(2.587,0) 
\psline[linestyle=dashed](2.338,0.431)(1.952,1.100) 
\psline[linestyle=dashed](2.338,0.431)(2.338,1.227) 
\psline(0.410,-0.641)(0.182,-1.267) 
\psline(2.194,-1.545)(2.194,-0.868) 
\psline[linestyle=dashed](2.194,-0.868)(2.772,-1.446) 
\psline(3.192,0.845)(3.192,1.512) 
\psline[linestyle=dashed](1.772,1.468)(2.622,1.468) 
\psdot[linecolor=orange,linewidth=1.2pt](-0.752,0) 
\psdot[linecolor=orange,linewidth=1.2pt](-0.874,0) 
\psdot[linecolor=orange,linewidth=1.2pt](-0.507,0) 
\psdot[linecolor=orange,linewidth=1.2pt](-1.354,0) 
\psdot[linecolor=darkgreen,linewidth=1.2pt](2.917,-0.544) 
\psdot[linecolor=darkgreen,linewidth=1.2pt](1.872,0.720) 
\psdot[linecolor=darkgreen,linewidth=1.2pt](0.410,-1.420) 
\psdot[linecolor=darkgreen,linewidth=1.2pt](0,-2.056) 
\psdot[linecolor=green,linewidth=1.2pt](-0.875,0.368) 
\psdot[linecolor=green,linewidth=1.2pt](-1.023,-0.184) 
\psdot[linecolor=green,linewidth=1.2pt](-0.491,-0.491) 
\psdot[linecolor=green,linewidth=1.2pt](-0.540,-0.881) 
\psdot[linecolor=green,linewidth=1.2pt](-0.579,0.459) 
\psdot[linecolor=blue,linewidth=1.2pt](-0.080,0.080) 
\psdot[linecolor=blue,linewidth=1.2pt](-0.433,0.284) 
\psdot[linecolor=blue,linewidth=1.2pt](0,0) 
\psdot[linecolor=blue,linewidth=1.2pt](-0.117,0) 
\psdot[linecolor=blue,linewidth=1.2pt](0.158,-0.091) 
\psdot[linecolor=blue,linewidth=1.2pt](0,0.939) 
\psdot[linecolor=blue,linewidth=1.2pt](0,-0.160) 
\psdot[linecolor=blue,linewidth=1.2pt](3.461,-1.507) 
\psdot[linecolor=blue,linewidth=1.2pt](0,0.325) 
\psdot[linecolor=blue,linewidth=1.2pt](0,0.086) 
\psdot[linecolor=blue,linewidth=1.2pt](0,-0.680) 
\psdot[linecolor=blue,linewidth=1.2pt](-0.158,0.216) 
\psdot[linecolor=blue,linewidth=1.2pt](0.410,0) 
\psdot[linecolor=blue,linewidth=1.2pt](-0.268,0.407) 
\psdot[linecolor=blue,linewidth=1.2pt](0.208,-0.360) 
\psdot[linecolor=blue,linewidth=1.2pt](0.604,-1.707) 
\psdot[linecolor=blue,linewidth=1.2pt](-0.081,-0.081) 
\psdot[linecolor=blue,linewidth=1.2pt](-0.270,-0.407) 
\psdot[linecolor=blue,linewidth=1.2pt](0.934,1.006) 
\psdot[linecolor=blue,linewidth=1.2pt](0.934,0.511) 
\psdot[linecolor=blue,linewidth=1.2pt](0.758,-2.056) 
\psdot[linecolor=blue,linewidth=1.2pt](0.934,0) 
\psdot[linecolor=blue,linewidth=1.2pt](1.456,0) 
\psdot[linecolor=blue,linewidth=1.2pt](1.456,0.779) 
\psdot[linecolor=blue,linewidth=1.2pt](1.946,-0.370) 
\psdot[linecolor=blue,linewidth=1.2pt](3.192,0.351) 
\psdot[linecolor=blue,linewidth=1.2pt](4.021,1.512) 
\psdot[linecolor=blue,linewidth=1.2pt](2.622,0.635) 
\psdot[linecolor=blue,linewidth=1.2pt](3.031,1.343) 
\psdot[linecolor=bluish,linewidth=1.2pt](2.010,0) 
\psdot[linecolor=brown,linewidth=1.2pt](2.398,-1.701) 
\psdot[linecolor=bluish,linewidth=1.2pt](1.514,1.006) 
\psdot[linecolor=brown,linewidth=1.2pt](0.402,0.402) 
\psdot[linecolor=brown,linewidth=1.2pt](0.934,-0.824) 
\psdot[linecolor=brown,linewidth=1.2pt](1.340,-0.703) 
\psdot[linecolor=brown,linewidth=1.2pt](1.584,-0.997) 
\psdot[linecolor=brown,linewidth=1.2pt](-1.098,-0.491) 
\psdot[linecolor=brown,linewidth=1.2pt](1.222,-1.625) 
\psdot[linecolor=brown,linewidth=1.2pt](1.584,-1.701) 
\psdot[linecolor=brown,linewidth=1.2pt](3.752,1.243) 
\psdot[linecolor=brown,linewidth=1.2pt](3.752,0.505) 
\psdot[linecolor=brown,linewidth=1.2pt](0,-1.422) 
\psdot[linecolor=brown,linewidth=1.2pt](-0.797,-2.056) 
\psdot[linecolor=red,linewidth=1.2pt](2.549,-0.254) 
\psdot[linecolor=red,linewidth=1.2pt](2.989,0) 
\psdot[linecolor=red,linewidth=1.2pt](3.461,-0.753) 
\psdot[linecolor=red,linewidth=1.2pt](1.475,1.227) 
\psdot[linecolor=red,linewidth=1.2pt](2.338,0.431) 
\psdot[linecolor=red,linewidth=1.2pt](2.587,0) 
\psdot[linecolor=red,linewidth=1.2pt](0.410,-0.641) 
\psdot[linecolor=red,linewidth=1.2pt](1.350,-0.381) 
\psdot[linecolor=red,linewidth=1.2pt](0.182,-1.267) 
\psdot[linecolor=red,linewidth=1.2pt](3.461,0) 
\psdot[linecolor=red,linewidth=1.2pt](2.194,-1.545) 
\psdot[linecolor=red,linewidth=1.2pt](2.194,-0.868) 
\psdot[linecolor=black,linewidth=1.2pt](3.192,0.845) 
\psdot[linecolor=black,linewidth=1.2pt](3.192,1.512) 
\psdot[linecolor=magenta,linewidth=1.2pt](-0.664,-1.672) 
\psdot[linecolor=magenta,linewidth=1.2pt](-0.849,-1.422) 
\psdot[linecolor=magenta,linewidth=1.2pt](2.272,-2.098) 
\psdot[linecolor=magenta,linewidth=1.2pt](1.772,1.468) 
\psdot[linecolor=magenta,linewidth=1.2pt](2.622,1.468) 
\psdot[linecolor=magenta,linewidth=1.2pt](1.952,1.100) 
\psdot[linecolor=magenta,linewidth=1.2pt](2.338,1.227) 
\psdot[linecolor=magenta,linewidth=1.2pt](-0.522,-0.680) 
\psdot[linecolor=magenta,linewidth=1.2pt](2.772,-1.446) 
\rput(-0.752,0.1){S\&P}
\rput(-0.874,-0.1){Nasd}
\rput(-0.507,-0.1){Cana}
\rput(-1.504,0){Mexi}
\rput(2.917,-0.644){Pana}
\rput(1.872,0.820){CoRi}
\rput(0.410,-1.520){Berm}
\rput(0,-2.156){Jama}
\rput(-1.025,0.368){Braz}
\rput(-1.173,-0.184){Arge}
\rput(-0.531,-0.391){Chil}
\rput(-0.690,-0.881){Vene}
\rput(-0.729,0.459){Peru}
\rput(-0.080,0.180){UK}
\rput(-0.583,0.284){Irel}
\rput(0.14,0.06){Fran}
\rput(-0.237,0.08){Germ}
\rput(0.288,-0.091){Swit}
\rput(0,1.039){Autr}
\rput(0.12,-0.230){Ital}
\rput(3.461,-1.607){Malt}
\rput(0.15,0.325){Belg}
\rput(0.14,0.13){Neth}
\rput(0.15,-0.680){Luxe}
\rput(-0.158,0.316){Swed}
\rput(0.410,0.1){Denm}
\rput(-0.268,0.507){Finl}
\rput(0.208,-0.460){Norw}
\rput(0.754,-1.707){Icel}
\rput(-0.231,-0.081){Spai}
\rput(-0.270,-0.507){Port}
\rput(0.784,1.006){Gree}
\rput(0.784,0.511){CzRe}
\rput(0.908,-2.056){Slok}
\rput(0.934,0.1){Hung}
\rput(1.306,0.1){Pola}
\rput(1.456,0.879){Roma}
\rput(1.796,-0.370){Bulg}
\rput(3.342,0.351){Esto}
\rput(4.021,1.612){Latv}
\rput(2.472,0.635){Lith}
\rput(3.031,1.443){Ukra}
\rput(2.010,0.1){Russ}
\rput(2.548,-1.701){Kaza}
\rput(1.514,1.106){Turk}
\rput(0.552,0.402){Isra}
\rput(0.934,-0.924){Pale}
\rput(1.340,-0.803){Leba}
\rput(1.434,-0.997){Jord}
\rput(-1.248,-0.491){SaAr}
\rput(1.222,-1.725){Qata}
\rput(1.744,-1.601){Ohma}
\rput(3.602,1.243){Paki}
\rput(3.602,0.505){Indi}
\rput(-0.15,-1.522){SrLa}
\rput(-0.947,-2.056){Bang}
\rput(2.399,-0.254){Japa}
\rput(2.989,-0.1){HoKo}
\rput(3.311,-0.753){Chin}
\rput(1.475,1.327){Mongo}
\rput(2.188,0.431){Taiw}
\rput(2.587,0.1){SoKo}
\rput(0.560,-0.641){Thai}
\rput(1.350,-0.481){Viet}
\rput(0.182,-1.367){Mala}
\rput(3.611,0){Sing}
\rput(2.044,-1.545){Indo}
\rput(2.044,-0.868){Phil}
\rput(3.342,0.845){Aust}
\rput(3.192,1.612){NeZe}
\rput(-0.814,-1.672){Moro}
\rput(-0.999,-1.422){Tuni}
\rput(2.422,-2.098){Egyp}
\rput(1.772,1.568){Ghan}
\rput(2.622,1.568){Nige}
\rput(1.802,1.100){Keny}
\rput(2.338,1.327){Bots}
\rput(-0.672,-0.680){SoAf}
\rput(2.772,-1.546){Maur}
\end{pspicture}

\begin{center}
Figure 15: minimum spanning tree for the second semester of 2001.
\end{center}

\newpage

\subsection{2008 - Subprime Mortgage Crisis}

The last crisis to be analized is the so called Subprime Mortage Crisis, which had its origins in the USA and then spread to the world as a Credit Crisis. It started at the end of 1997 and reached its peak in 2008, with the banckrupcy of major investment banks, and has its effects up to now. We here analyze the years 2007 and 2008, using 92 indices, namely S\&P and Nasdaq from the USA, and indices from Canada, Mexico, Panama, Costa Rica, Bermuda, Jamaica, Brazil, Argentina, Chile, Colombia, Venezuela, Peru, UK, Ireland, France, Germany, Switzerland, Austria, Italy, Malta, Belgium, Netherlands, Luxembourg, Sweden, Denmark, Finland, Norway, Iceland, Spain, Portugal, Greece, Czech Republic, Slovakia, Hungary, Serbia, Croatia, Slovenia, Bosnia and Herzegovina, Montenegro, Macedonia, Poland, Romania, Bulgaria, Estonia, Latvia, Lithuania, Ukraine, Russia, Kazakhstan, Turkey, Cyprus, Israel, Palestine, Lebanon, Jordan, Saudi Arabia, Kuwait, Bahrein, Qatar, United Arab Emirates, Ohman, Pakistan, India, Sri Lanka, Bangladesh, Japan, Hong Kong, China, Mongolia, Taiwan, South Korea, Thailand, Vietnam, Malaysia, Singapore, Indonesia, Philipines, Australia, New Zealand, Morocco, Tunisia, Egypt, Ghana, Nigeria, Kenya, Tanzania, Namibia, Botswana, South Africa, and Mauritius. The number of indices grew due to the emergence of new financial markets and also of new countries, like those that became independent after the end of the former Yugoslavia (although Yugoslavia began disintegrating in 1991, it took some time for the new countries to develope solid stock exchanges).

We start from figure 16, which shows the minimum spanning tree for the first semester of 2007 with threshold $0.68\pm 0.03$. Again, one can see a large aglomeration of indices around France, and a very compact European cluster of indices. The American cluster remains stable, now with the addition of Colombia. The presence of a Pacific Asian cluster (together with the indices from Oceania) is now much clearer. Namibia is tightly connected with South Africa, what is to be expected, for Namibia has been under the economic and political influence of South Africa for the past six decades. Cyprus is connected with Greece, what is to be expected given their common cultures and ethnicities (Cyprus' stock exchange is located at the Greek part of the island). There is also the appearance of a small cluster of some of the former members of the extinct Yugoslavia: Croatia, Bosnia and Herzegovina, and Macedonia. Among the weaker links, one can devise an Arab cluster, comprised of Bahrain, Qatar, Kuwait, United Arab Emirates, Lebanon, and Saudi Arabia. These connections, although weak, do not seem random. One may also notice that Singapore is the main hub for Pacific Asian indices, probably due to the country's expertise in finance.

\begin{pspicture}(-7,-8.7)(1,11)
\psset{xunit=3,yunit=3} \scriptsize
\psline(-1.392,0)(-1.490,0) 
\psline(-1.392,0)(-1.392,0.285) 
\psline(-1.392,0)(-1.538,-0.146) 
\psline(-1.392,0)(-0.952,0) 
\psline(-1.392,0)(-1.108,-0.426) 
\psline(-1.538,-0.399)(-1.538,-0.146) 
\psline[linestyle=dashed](-1.404,1.451)(-0.585,1.451) 
\psline[linestyle=dashed](1.630,2.464)(1.630,3.254) 
\psline[linestyle=dashed](-1.904,-2.876)(-1.072,-2.876) 
\psline[linestyle=dashed](-0.625,2.886)(0,2.510) 
\psline[linestyle=dashed](-0.625,2.886)(-1.285,3.267) 
\psline(-1.538,-0.146)(-1.751,-0.359) 
\psline(-0.952,0)(-0.170,0) 
\psline[linestyle=dashed](-0.952,0)(-0.952,-0.802) 
\psline[linestyle=dashed](0.815,3.254)(0,3.254) 
\psline[linestyle=dashed](0.815,3.254)(1.630,3.254) 
\psline[linestyle=dashed](0.945,-0.455)(0.945,-1.168) 
\psline(0.945,-0.455)(1.613,-0.455) 
\psline(0.945,-0.455)(0.472,-0.182) 
\psline[linestyle=dashed](0.945,-0.455)(0.395,-1.005) 
\psline(0.157,-0.052)(0.327,-0.222) 
\psline(0.157,-0.052)(0.067,0) 
\psline(0.157,-0.052)(0.472,-0.182) 
\psline(0.067,0)(0,0) 
\psline(0.067,0)(0.148,0.081) 
\psline(0.067,0)(0.067,0.137) 
\psline(0.067,0)(0.182,0) 
\psline(0.067,0)(0.067,-0.141) 
\psline(0.067,0)(0.278,-0.366) 
\psline(0,0)(-0.170,0) 
\psline(0,0)(0,0.246) 
\psline(0,0)(-0.157,-0.272) 
\psline(0,0)(0,-0.396) 
\psline[linestyle=dashed](0,0)(-0.486,0.486) 
\psline(0,0)(-0.453,-0.262) 
\psline(-0.170,0)(-0.441,0) 
\psline(0,0.536)(0,0.866) 
\psline(0,0.536)(0,0.246) 
\psline(0,0.536)(0.372,0.536) 
\psline(0,0.536)(0.247,0.289) 
\psline(0,0.536)(-0.280,0.816) 
\psline[linestyle=dashed](-1.162,1.628)(-0.477,1.628) 
\psline(0.182,0)(0.546,0) 
\psline(0,0.866)(0,1.250) 
\psline[linestyle=dashed](0,0.866)(-0.585,1.451) 
\psline(-0.182,0.428)(0,0.246) 
\psline[linestyle=dashed](-0.182,0.428)(-0.719,0.965) 
\psline(0.372,1.142)(0.372,0.536) 
\psline[linestyle=dashed](0,-0.396)(-0.508,-0.904) 
\psline(0.372,0.536)(0.839,0.536) 
\psline(0.247,0.289)(0.655,0.289) 
\psline[linestyle=dashed](0,3.254)(0,2.510) 
\psline(0.980,0)(0.546,0) 
\psline(0.945,-1.816)(0.945,-1.168) 
\psline(1.166,1.628)(1.166,2.274) 
\psline(1.166,1.628)(0.552,1.628) 
\psline[linestyle=dashed](-1.457,2.584)(-0.827,2.220) 
\psline[linestyle=dashed](1.166,2.274)(1.166,3.079) 
\psline(0,2.510)(0,1.943) 
\psline[linestyle=dashed](-0.114,-2.281)(-0.114,-1.514) 
\psline(-0.384,2.220)(-0.827,2.220) 
\psline(-0.384,2.220)(0,1.943) 
\psline[linestyle=dashed](-0.384,2.220)(-1.009,2.580) 
\psline(-0.846,2.117)(-0.259,1.778) 
\psline[linestyle=dashed](-0.827,2.220)(-1.619,2.220) 
\psline[linestyle=dashed](0.655,0.289)(1.354,0.289) 
\psline[linestyle=dashed](-1.619,2.220)(-1.619,2.996) 
\psline[linestyle=dashed](-1.900,-2.143)(-1.072,-2.143) 
\psline[linestyle=dashed](1.354,1.011)(1.354,0.289) 
\psline[linestyle=dashed](1.354,1.011)(1.354,1.805) 
\psline[linestyle=dashed](-1.072,-2.876)(-1.072,-2.143) 
\psline[linestyle=dashed](-1.914,-1.468)(-1.072,-1.468) 
\psline[linestyle=dashed](-0.508,-0.904)(-1.072,-1.468) 
\psline(-1.072,-1.468)(-1.072,-2.143) 
\psline[linestyle=dashed](2.101,0.289)(1.354,0.289) 
\psline(-0.477,1.628)(0,1.628) 
\psline(0.213,1.464)(0,1.250) 
\psline[linestyle=dashed](-0.259,1.778)(-1.072,1.778) 
\psline(-0.259,1.778)(0,1.628) 
\psline[linestyle=dashed](0.528,2.542)(0.150,1.887) 
\psline(0.735,1.250)(0.314,1.250) 
\psline(0.314,1.250)(0,1.250) 
\psline[linestyle=dashed](1.354,1.805)(1.354,2.511) 
\psline(0,1.943)(0,1.628) 
\psline[linestyle=dashed](0,1.943)(0.418,2.666) 
\psline(0,1.628)(0.255,1.776) 
\psline(0,1.628)(0.552,1.628) 
\psline(0,1.628)(0,1.250) 
\psline(0,1.628)(0.150,1.887) 
\psline(0.552,1.628)(1.166,1.628) 
\psline[linestyle=dashed](-0.114,-1.514)(0.395,-1.005) 
\psline(0.592,-0.182)(0.472,-0.182) 
\psdot[linecolor=orange,linewidth=1.2pt](-1.392,0) 
\psdot[linecolor=orange,linewidth=1.2pt](-1.490,0) 
\psdot[linecolor=orange,linewidth=1.2pt](-1.392,0.285) 
\psdot[linecolor=orange,linewidth=1.2pt](-1.538,-0.399) 
\psdot[linecolor=darkgreen,linewidth=1.2pt](-1.404,1.451) 
\psdot[linecolor=darkgreen,linewidth=1.2pt](1.630,2.464) 
\psdot[linecolor=darkgreen,linewidth=1.2pt](-1.904,-2.876) 
\psdot[linecolor=darkgreen,linewidth=1.2pt](-0.625,2.886) 
\psdot[linecolor=green,linewidth=1.2pt](-1.538,-0.146) 
\psdot[linecolor=green,linewidth=1.2pt](-1.751,-0.359) 
\psdot[linecolor=green,linewidth=1.2pt](-0.952,0) 
\psdot[linecolor=green,linewidth=1.2pt](-1.108,-0.426) 
\psdot[linecolor=green,linewidth=1.2pt](0.815,3.254) 
\psdot[linecolor=green,linewidth=1.2pt](0.945,-0.455) 
\psdot[linecolor=blue,linewidth=1.2pt](0.157,-0.052) 
\psdot[linecolor=blue,linewidth=1.2pt](0.327,-0.222) 
\psdot[linecolor=blue,linewidth=1.2pt](0.067,0) 
\psdot[linecolor=blue,linewidth=1.2pt](0,0) 
\psdot[linecolor=blue,linewidth=1.2pt](-0.170,0) 
\psdot[linecolor=blue,linewidth=1.2pt](0,0.536) 
\psdot[linecolor=blue,linewidth=1.2pt](0.148,0.081) 
\psdot[linecolor=blue,linewidth=1.2pt](-1.162,1.628) 
\psdot[linecolor=blue,linewidth=1.2pt](0.067,0.137) 
\psdot[linecolor=blue,linewidth=1.2pt](0.182,0) 
\psdot[linecolor=blue,linewidth=1.2pt](0,0.866) 
\psdot[linecolor=blue,linewidth=1.2pt](-0.182,0.428) 
\psdot[linecolor=blue,linewidth=1.2pt](-0.441,0) 
\psdot[linecolor=blue,linewidth=1.2pt](0,0.246) 
\psdot[linecolor=blue,linewidth=1.2pt](-0.157,-0.272) 
\psdot[linecolor=blue,linewidth=1.2pt](0.372,1.142) 
\psdot[linecolor=blue,linewidth=1.2pt](0.067,-0.141) 
\psdot[linecolor=blue,linewidth=1.2pt](0,-0.396) 
\psdot[linecolor=blue,linewidth=1.2pt](0.372,0.536) 
\psdot[linecolor=blue,linewidth=1.2pt](0.247,0.289) 
\psdot[linecolor=blue,linewidth=1.2pt](0,3.254) 
\psdot[linecolor=blue,linewidth=1.2pt](0.980,0) 
\psdot[linecolor=blue,linewidth=1.2pt](0.945,-1.816) 
\psdot[linecolor=blue,linewidth=1.2pt](1.166,1.628) 
\psdot[linecolor=blue,linewidth=1.2pt](-1.457,2.584) 
\psdot[linecolor=blue,linewidth=1.2pt](1.166,2.274) 
\psdot[linecolor=blue,linewidth=1.2pt](0.945,-1.168) 
\psdot[linecolor=blue,linewidth=1.2pt](1.166,3.079) 
\psdot[linecolor=blue,linewidth=1.2pt](0.546,0) 
\psdot[linecolor=blue,linewidth=1.2pt](0,2.510) 
\psdot[linecolor=blue,linewidth=1.2pt](-0.114,-2.281) 
\psdot[linecolor=blue,linewidth=1.2pt](-0.384,2.220) 
\psdot[linecolor=blue,linewidth=1.2pt](-0.846,2.117) 
\psdot[linecolor=blue,linewidth=1.2pt](-0.827,2.220) 
\psdot[linecolor=blue,linewidth=1.2pt](1.613,-0.455) 
\psdot[linecolor=bluish,linewidth=1.2pt](0.655,0.289) 
\psdot[linecolor=brown,linewidth=1.2pt](-1.619,2.220) 
\psdot[linecolor=bluish,linewidth=1.2pt](-0.280,0.816) 
\psdot[linecolor=brown,linewidth=1.2pt](0.839,0.536) 
\psdot[linecolor=brown,linewidth=1.2pt](0.278,-0.366) 
\psdot[linecolor=brown,linewidth=1.2pt](-0.952,-0.802) 
\psdot[linecolor=brown,linewidth=1.2pt](-1.900,-2.143) 
\psdot[linecolor=brown,linewidth=1.2pt](1.354,1.011) 
\psdot[linecolor=brown,linewidth=1.2pt](-1.072,-2.876) 
\psdot[linecolor=brown,linewidth=1.2pt](-1.914,-1.468) 
\psdot[linecolor=brown,linewidth=1.2pt](-0.508,-0.904) 
\psdot[linecolor=brown,linewidth=1.2pt](-1.072,-1.468) 
\psdot[linecolor=brown,linewidth=1.2pt](-1.072,-2.143) 
\psdot[linecolor=brown,linewidth=1.2pt](2.101,0.289) 
\psdot[linecolor=brown,linewidth=1.2pt](-1.285,3.267) 
\psdot[linecolor=brown,linewidth=1.2pt](-0.477,1.628) 
\psdot[linecolor=brown,linewidth=1.2pt](1.354,0.289) 
\psdot[linecolor=brown,linewidth=1.2pt](-1.619,2.996) 
\psdot[linecolor=red,linewidth=1.2pt](0.213,1.464) 
\psdot[linecolor=red,linewidth=1.2pt](-0.259,1.778) 
\psdot[linecolor=red,linewidth=1.2pt](0.528,2.542) 
\psdot[linecolor=red,linewidth=1.2pt](-1.072,1.778) 
\psdot[linecolor=red,linewidth=1.2pt](0.735,1.250) 
\psdot[linecolor=red,linewidth=1.2pt](0.314,1.250) 
\psdot[linecolor=red,linewidth=1.2pt](-0.486,0.486) 
\psdot[linecolor=red,linewidth=1.2pt](1.354,1.805) 
\psdot[linecolor=red,linewidth=1.2pt](0,1.943) 
\psdot[linecolor=red,linewidth=1.2pt](0,1.628) 
\psdot[linecolor=red,linewidth=1.2pt](0.255,1.776) 
\psdot[linecolor=red,linewidth=1.2pt](0.552,1.628) 
\psdot[linecolor=black,linewidth=1.2pt](0,1.250) 
\psdot[linecolor=black,linewidth=1.2pt](0.150,1.887) 
\psdot[linecolor=magenta,linewidth=1.2pt](1.630,3.254) 
\psdot[linecolor=magenta,linewidth=1.2pt](0.418,2.666) 
\psdot[linecolor=magenta,linewidth=1.2pt](-0.453,-0.262) 
\psdot[linecolor=magenta,linewidth=1.2pt](-0.719,0.965) 
\psdot[linecolor=magenta,linewidth=1.2pt](-1.009,2.580) 
\psdot[linecolor=magenta,linewidth=1.2pt](-0.114,-1.514) 
\psdot[linecolor=magenta,linewidth=1.2pt](1.354,2.511) 
\psdot[linecolor=magenta,linewidth=1.2pt](0.592,-0.182) 
\psdot[linecolor=magenta,linewidth=1.2pt](-0.585,1.451) 
\psdot[linecolor=magenta,linewidth=1.2pt](0.472,-0.182) 
\psdot[linecolor=magenta,linewidth=1.2pt](0.395,-1.005) 
\rput(-1.242,0.1){S\&P}
\rput(-1.660,0){Nasd}
\rput(-1.392,0.385){Cana}
\rput(-1.538,-0.499){Mexi}
\rput(-1.554,1.451){Pana}
\rput(1.630,2.364){CoRi}
\rput(-2.054,-2.876){Berm}
\rput(-0.445,2.886){Jama}
\rput(-1.698,-0.146){Braz}
\rput(-1.771,-0.459){Arge}
\rput(-0.952,0.1){Chil}
\rput(-1.108,-0.526){Colo}
\rput(0.815,3.354){Vene}
\rput(0.795,-0.455){Peru}
\rput(0.157,-0.152){UK}
\rput(0.327,-0.302){Irel}
\rput(0.067,0.1){Fran}
\rput(-0.03,-0.1){Germ}
\rput(-0.170,0.1){Swit}
\rput(-0.15,0.536){Autr}
\rput(0.148,0.181){Ital}
\rput(-1.312,1.628){Malt}
\rput(0.117,0.237){Belg}
\rput(0.292,0.1){Neth}
\rput(-0.15,0.866){Luxe}
\rput(-0.332,0.428){Swed}
\rput(-0.441,0.1){Denm}
\rput(-0.15,0.246){Finl}
\rput(-0.157,-0.372){Norw}
\rput(0.522,1.142){Icel}
\rput(0.097,-0.241){Spai}
\rput(0.15,-0.396){Port}
\rput(0.222,0.636){Gree}
\rput(0.247,0.389){CzRe}
\rput(-0.15,3.254){Slok}
\rput(0.980,0.1){Hung}
\rput(0.945,-1.916){Serb}
\rput(1.016,1.728){Croa}
\rput(-1.457,2.684){Slov}
\rput(0.996,2.274){BoHe}
\rput(1.095,-1.168){Mont}
\rput(1.166,3.179){Mace}
\rput(0.546,0.1){Pola}
\rput(-0.18,2.510){Roma}
\rput(-0.114,-2.381){Bulg}
\rput(-0.254,2.240){Esto}
\rput(-0.996,2.117){Latv}
\rput(-0.807,2.320){Lith}
\rput(1.763,-0.455){Ukra}
\rput(0.655,0.389){Russ}
\rput(-1.769,2.220){Kaza}
\rput(-0.400,0.916){Turk}
\rput(0.839,0.636){Cypr}
\rput(0.278,-0.466){Isra}
\rput(-0.952,-0.902){Pale}
\rput(-2.050,-2.143){Leba}
\rput(1.504,1.011){Jord}
\rput(-1.222,-2.776){SaAr}
\rput(-2.064,-1.468){Kuwa}
\rput(-0.658,-0.904){Bahr}
\rput(-1.222,-1.368){Qata}
\rput(-1.222,-2.043){UAE}
\rput(2.271,0.289){Ohma}
\rput(-1.435,3.267){Paki}
\rput(-0.477,1.728){Indi}
\rput(1.504,0.389){SrLa}
\rput(-1.619,3.096){Bang}
\rput(0.213,1.564){Japa}
\rput(-0.109,1.798){HoKo}
\rput(0.578,2.642){Chin}
\rput(-1.232,1.778){Mong}
\rput(0.735,1.350){Taiw}
\rput(0.314,1.350){SoKo}
\rput(-0.486,0.586){Thai}
\rput(1.504,1.805){Viet}
\rput(-0.15,1.943){Mala}
\rput(-0.15,1.528){Sing}
\rput(0.405,1.826){Indo}
\rput(0.552,1.728){Phil}
\rput(-0.15,1.250){Aust}
\rput(0.320,1.927){NeZe}
\rput(1.630,3.354){Moro}
\rput(0.418,2.766){Tuni}
\rput(-0.453,-0.362){Egyp}
\rput(-0.869,0.965){Ghan}
\rput(-1.009,2.680){Nige}
\rput(-0.264,-1.514){Keny}
\rput(1.354,2.611){Tanz}
\rput(0.742,-0.182){Nami}
\rput(-0.585,1.551){Bots}
\rput(0.472,-0.082){SoAf}
\rput(0.225,-1.005){Maur}
\end{pspicture}

\vskip 0.3 cm

\noindent Figure 16: minimum spanning tree for the first semester of 2007. The inset shows an amplification of the connections around France and Germany.

\vskip -6.5 cm
\hskip 11 cm
\begin{minipage}{4 cm }
\begin{pspicture}(-3,-1)(1,1)
\psline(-2.3,-4)(3.3,-4)(3.3,2.7)(-2.3,2.7)(-2.3,-4)
\psset{xunit=9,yunit=9} \scriptsize
\psline(0.157,-0.052)(0.327,-0.222) 
\psline(0.157,-0.052)(0.067,0) 
\psline(0.157,-0.052)(0.367,-0.141) 
\psline(0.067,0)(0,0) 
\psline(0.067,0)(0.148,0.081) 
\psline(0.067,0)(0.067,0.137) 
\psline(0.067,0)(0.182,0) 
\psline(0.067,0)(0.067,-0.141) 
\psline(0.067,0)(0.278,-0.366) 
\psline(0,0)(-0.170,0) 
\psline(0,0)(0,0.246) 
\psline(0,0)(-0.157,-0.272) 
\psline(0,0)(0,-0.396) 
\psline(0,0)(-0.256,0.256) 
\psline(0,0)(-0.256,-0.148) 
\psline(-0.170,0)(-0.256,0) 
\psline(0,0.3)(0,0.246) 
\psline(0.182,0)(0.367,0) 
\psline(-0.054,0.3)(0,0.246) 
\psline[linestyle=dashed](0,-0.396)(-0.048,-0.444) 
\psdot[linecolor=blue,linewidth=1.2pt](0.157,-0.052) 
\psdot[linecolor=blue,linewidth=1.2pt](0.327,-0.222) 
\psdot[linecolor=blue,linewidth=1.2pt](0.067,0) 
\psdot[linecolor=blue,linewidth=1.2pt](0,0) 
\psdot[linecolor=blue,linewidth=1.2pt](-0.170,0) 
\psdot[linecolor=blue,linewidth=1.2pt](0.148,0.081) 
\psdot[linecolor=blue,linewidth=1.2pt](0.067,0.137) 
\psdot[linecolor=blue,linewidth=1.2pt](0.182,0) 
\psdot[linecolor=blue,linewidth=1.2pt](0,0.246) 
\psdot[linecolor=blue,linewidth=1.2pt](-0.157,-0.272) 
\psdot[linecolor=blue,linewidth=1.2pt](0.067,-0.141) 
\psdot[linecolor=blue,linewidth=1.2pt](0,-0.396) 
\psdot[linecolor=brown,linewidth=1.2pt](0.278,-0.366) 
\rput(0.147,-0.082){UK}
\rput(0.327,-0.252){Irel}
\rput(0.030,0.03){Fran}
\rput(-0.07,0.03){Germ}
\rput(-0.170,0.03){Swit}
\rput(0.148,0.111){Ital}
\rput(0.067,0.167){Belg}
\rput(0.232,0.03){Neth}
\rput(0.05,0.246){Finl}
\rput(-0.157,-0.302){Norw}
\rput(0.067,-0.171){Spai}
\rput(0.05,-0.396){Port}
\rput(0.278,-0.396){Isra}
\end{pspicture}
\end{minipage}

\newpage

Figure 17 shows the minimum spanning tree for the second semester of 2007, also with threshold $0.68\pm 0.03$. Notice that the graph is more compact, with most of the European indices aglomerating around France. The basic American cluster and the Pacific Asian cluster are still solid, although smaller economies, like Colombia, Peru, Vietnam, and Mongolia are not attached to their geographical clusters. Namibia and South Africa keep strongly connected, and South Africa and Israel keep strongly connected with Europe. Some links which were previously weak are now reinforced, like the ones between the former members of Yugoslavia, and the links between some Arab countries. One can also see two pairs of African indices, Tanzania-Ghana and Nigeria-Tunisia that may or may not be the resut of random noise.

\begin{pspicture}(-7.5,-8.7)(1,5)
\psset{xunit=3,yunit=3} \scriptsize
\psline(-0.642,0)(-0.578,0) 
\psline(-0.578,0)(-0.734,0.156) 
\psline(-0.578,0)(-0.578,0.247) 
\psline(-0.578,0)(-0.352,0) 
\psline(-0.578,0.247)(-0.578,0.595) 
\psline[linestyle=dashed](-0.883,-2.798)(-0.883,-2.042) 
\psline[linestyle=dashed](1.790,-1.427)(1.790,-0.713) 
\psline[linestyle=dashed](1.790,-1.427)(1.790,-2.220) 
\psline[linestyle=dashed](-1.836,-0.948)(-1.056,-0.739) 
\psline[linestyle=dashed](-0.009,-2.005)(-0.009,-1.194) 
\psline(-0.352,0)(-0.352,0.268) 
\psline(-0.352,0)(0,0) 
\psline(0.643,0.334)(0.207,0.082) 
\psline[linestyle=dashed](-1.292,-0.173)(-0.662,-0.537) 
\psline(-0.211,0.789)(0,0.203) 
\psline(0,-0.054)(0,0) 
\psline(0,-0.054)(0,-0.140) 
\psline(0,-0.054)(0.096,-0.413) 
\psline(0,-0.054)(-0.125,-0.270) 
\psline(-0.213,-0.123)(0,0) 
\psline(-0.213,-0.123)(-0.820,-0.123) 
\psline(0,0)(-0.114,0.066) 
\psline(0,0)(-0.039,-0.068) 
\psline(0,0)(0.056,0.056) 
\psline(0,0)(0.066,0) 
\psline(0,0)(-0.064,0.111) 
\psline(0,0)(0,0.203) 
\psline(0,0)(0.087,0.151) 
\psline(0,0)(0.245,-0.142) 
\psline[linestyle=dashed](0,0)(0.358,-0.620) 
\psline(0,-0.140)(0,-0.320) 
\psline(0,-0.140)(-0.088,-0.470) 
\psline(0,-0.320)(0,-0.606) 
\psline[linestyle=dashed](-1.014,-2.172)(-0.883,-1.430) 
\psline(0.056,0.056)(0.223,0.223) 
\psline(0.066,0)(0.207,0.082) 
\psline(0.066,0)(0.261,0) 
\psline(0,0.203)(0,0.578) 
\psline(0.261,0)(0.530,0) 
\psline[linestyle=dashed](0,0.578)(0,1.403) 
\psline(0.087,0.151)(0.256,0.443) 
\psline(0.530,0)(0.753,0) 
\psline(0.530,0)(0.530,-0.262) 
\psline[linestyle=dashed](-1.987,-0.469)(-1.209,-0.537) 
\psline[linestyle=dashed](0.482,-1.685)(-0.009,-1.194) 
\psline[linestyle=dashed](0.482,-1.685)(0.482,-2.372) 
\psline[linestyle=dashed](0.482,-1.685)(1.016,-2.220) 
\psline[linestyle=dashed](0.482,-1.685)(1.300,-1.685) 
\psline(-0.009,-1.194)(-0.464,-0.739) 
\psline[linestyle=dashed](1.777,-2.419)(1.016,-2.220) 
\psline[linestyle=dashed](0.482,-2.372)(1.226,-2.571) 
\psline[linestyle=dashed](1.016,-2.220)(1.790,-2.220) 
\psline(-1.308,-1.456)(-0.883,-1.158) 
\psline(-0.748,-0.270)(-0.125,-0.270) 
\psline(-0.883,-2.042)(-0.883,-1.430) 
\psline(-1.797,-0.620)(-1.209,-0.537) 
\psline(-1.209,-0.537)(-0.662,-0.537) 
\psline(-1.128,-1.581)(-0.883,-1.158) 
\psline(1.015,0)(0.753,0) 
\psline(-0.213,-1.943)(-0.326,-1.303) 
\psline[linestyle=dashed](0.725,-1.256)(0.358,-0.620) 
\psline[linestyle=dashed](-0.443,-2.763)(-0.443,-1.965) 
\psline[linestyle=dashed](-1.961,0.311)(-1.167,0.311) 
\psline[linestyle=dashed](-1.961,0.311)(-1.961,-0.372) 
\psline[linestyle=dashed](-1.961,0.311)(-1.961,1.103) 
\psline[linestyle=dashed](-1.682,-1.228)(-0.883,-1.158) 
\psline[linestyle=dashed](-1.278,-0.372)(-1.961,-0.372) 
\psline(-1.278,-0.372)(-0.662,-0.537) 
\psline[linestyle=dashed](-1.961,-0.372)(-2.158,0.365) 
\psline(-0.976,-0.223)(-0.662,-0.537) 
\psline[linestyle=dashed](-1.539,-1.027)(-0.762,-0.819) 
\psline[linestyle=dashed](-1.539,-1.027)(-2.242,-1.215) 
\psline[linestyle=dashed](-0.029,-2.682)(-0.443,-1.965) 
\psline(-0.677,-1.211)(-0.677,-0.952) 
\psline(-0.464,-0.739)(-1.056,-0.739) 
\psline(-0.464,-0.739)(-0.662,-0.537) 
\psline(-0.464,-0.739)(-0.304,-0.579) 
\psline(-0.464,-0.739)(-0.762,-0.819) 
\psline(-0.464,-0.739)(-0.677,-0.952) 
\psline[linestyle=dashed](-1.056,-0.739)(-1.896,-0.739) 
\psline(-0.883,-1.430)(-0.883,-1.158) 
\psline(-0.883,-1.158)(-0.677,-0.952) 
\psline(-0.030,-0.853)(-0.304,-0.579) 
\psline(-0.443,-1.965)(-0.326,-1.303) 
\psline(-0.304,-0.579)(-0.125,-0.270) 
\psline(-0.461,-1.326)(-0.677,-0.952) 
\psline[linestyle=dashed](-0.461,-1.326)(-0.647,-2.020) 
\psline(-0.677,-0.952)(-0.326,-1.303) 
\psline[linestyle=dashed](-0.820,-0.123)(-1.615,0.090) 
\psline[linestyle=dashed](-1.137,1.103)(-1.961,1.103) 
\psline(-0.203,-0.315)(-0.125,-0.270) 
\psdot[linecolor=orange,linewidth=1.2pt](-0.642,0) 
\psdot[linecolor=orange,linewidth=1.2pt](-0.578,0) 
\psdot[linecolor=orange,linewidth=1.2pt](-0.734,0.156) 
\psdot[linecolor=orange,linewidth=1.2pt](-0.578,0.247) 
\psdot[linecolor=darkgreen,linewidth=1.2pt](-0.883,-2.798) 
\psdot[linecolor=darkgreen,linewidth=1.2pt](1.790,-1.427) 
\psdot[linecolor=darkgreen,linewidth=1.2pt](-1.836,-0.948) 
\psdot[linecolor=darkgreen,linewidth=1.2pt](-0.009,-2.005) 
\psdot[linecolor=green,linewidth=1.2pt](-0.352,0) 
\psdot[linecolor=green,linewidth=1.2pt](-0.352,0.268) 
\psdot[linecolor=green,linewidth=1.2pt](-0.578,0.595) 
\psdot[linecolor=green,linewidth=1.2pt](0.643,0.334) 
\psdot[linecolor=green,linewidth=1.2pt](-1.292,-0.173) 
\psdot[linecolor=green,linewidth=1.2pt](-0.211,0.789) 
\psdot[linecolor=blue,linewidth=1.2pt](0,-0.054) 
\psdot[linecolor=blue,linewidth=1.2pt](-0.213,-0.123) 
\psdot[linecolor=blue,linewidth=1.2pt](0,0) 
\psdot[linecolor=blue,linewidth=1.2pt](0,-0.140) 
\psdot[linecolor=blue,linewidth=1.2pt](-0.114,0.066) 
\psdot[linecolor=blue,linewidth=1.2pt](0,-0.320) 
\psdot[linecolor=blue,linewidth=1.2pt](-0.039,-0.068) 
\psdot[linecolor=blue,linewidth=1.2pt](-1.014,-2.172) 
\psdot[linecolor=blue,linewidth=1.2pt](0.056,0.056) 
\psdot[linecolor=blue,linewidth=1.2pt](0.066,0) 
\psdot[linecolor=blue,linewidth=1.2pt](0.223,0.223) 
\psdot[linecolor=blue,linewidth=1.2pt](-0.064,0.111) 
\psdot[linecolor=blue,linewidth=1.2pt](0,0.203) 
\psdot[linecolor=blue,linewidth=1.2pt](0.207,0.082) 
\psdot[linecolor=blue,linewidth=1.2pt](0.261,0) 
\psdot[linecolor=blue,linewidth=1.2pt](0,0.578) 
\psdot[linecolor=blue,linewidth=1.2pt](0.087,0.151) 
\psdot[linecolor=blue,linewidth=1.2pt](0.256,0.443) 
\psdot[linecolor=blue,linewidth=1.2pt](0.530,0) 
\psdot[linecolor=blue,linewidth=1.2pt](0,-0.606) 
\psdot[linecolor=blue,linewidth=1.2pt](-1.987,-0.469) 
\psdot[linecolor=blue,linewidth=1.2pt](-0.088,-0.470) 
\psdot[linecolor=blue,linewidth=1.2pt](0.482,-1.685) 
\psdot[linecolor=blue,linewidth=1.2pt](-0.009,-1.194) 
\psdot[linecolor=blue,linewidth=1.2pt](1.777,-2.419) 
\psdot[linecolor=blue,linewidth=1.2pt](0.482,-2.372) 
\psdot[linecolor=blue,linewidth=1.2pt](1.226,-2.571) 
\psdot[linecolor=blue,linewidth=1.2pt](1.016,-2.220) 
\psdot[linecolor=blue,linewidth=1.2pt](0.245,-0.142) 
\psdot[linecolor=blue,linewidth=1.2pt](-1.308,-1.456) 
\psdot[linecolor=blue,linewidth=1.2pt](-0.748,-0.270) 
\psdot[linecolor=blue,linewidth=1.2pt](-0.883,-2.042) 
\psdot[linecolor=blue,linewidth=1.2pt](-1.797,-0.620) 
\psdot[linecolor=blue,linewidth=1.2pt](-1.209,-0.537) 
\psdot[linecolor=blue,linewidth=1.2pt](-1.128,-1.581) 
\psdot[linecolor=bluish,linewidth=1.2pt](1.015,0) 
\psdot[linecolor=brown,linewidth=1.2pt](-0.213,-1.943) 
\psdot[linecolor=bluish,linewidth=1.2pt](0.753,0) 
\psdot[linecolor=brown,linewidth=1.2pt](0.530,-0.262) 
\psdot[linecolor=brown,linewidth=1.2pt](0.096,-0.413) 
\psdot[linecolor=brown,linewidth=1.2pt](0.725,-1.256) 
\psdot[linecolor=brown,linewidth=1.2pt](-0.443,-2.763) 
\psdot[linecolor=brown,linewidth=1.2pt](-1.961,0.311) 
\psdot[linecolor=brown,linewidth=1.2pt](-1.682,-1.228) 
\psdot[linecolor=brown,linewidth=1.2pt](-1.167,0.311) 
\psdot[linecolor=brown,linewidth=1.2pt](1.790,-0.713) 
\psdot[linecolor=brown,linewidth=1.2pt](-1.278,-0.372) 
\psdot[linecolor=brown,linewidth=1.2pt](-1.961,-0.372) 
\psdot[linecolor=brown,linewidth=1.2pt](-2.158,0.365) 
\psdot[linecolor=brown,linewidth=1.2pt](0.358,-0.620) 
\psdot[linecolor=brown,linewidth=1.2pt](-0.976,-0.223) 
\psdot[linecolor=brown,linewidth=1.2pt](-1.539,-1.027) 
\psdot[linecolor=brown,linewidth=1.2pt](-0.029,-2.682) 
\psdot[linecolor=red,linewidth=1.2pt](-0.677,-1.211) 
\psdot[linecolor=red,linewidth=1.2pt](-0.464,-0.739) 
\psdot[linecolor=red,linewidth=1.2pt](-1.056,-0.739) 
\psdot[linecolor=red,linewidth=1.2pt](1.790,-2.220) 
\psdot[linecolor=red,linewidth=1.2pt](-0.883,-1.430) 
\psdot[linecolor=red,linewidth=1.2pt](-0.883,-1.158) 
\psdot[linecolor=red,linewidth=1.2pt](-0.030,-0.853) 
\psdot[linecolor=red,linewidth=1.2pt](-0.443,-1.965) 
\psdot[linecolor=red,linewidth=1.2pt](-0.662,-0.537) 
\psdot[linecolor=red,linewidth=1.2pt](-0.304,-0.579) 
\psdot[linecolor=red,linewidth=1.2pt](-0.762,-0.819) 
\psdot[linecolor=red,linewidth=1.2pt](-0.461,-1.326) 
\psdot[linecolor=black,linewidth=1.2pt](-0.677,-0.952) 
\psdot[linecolor=black,linewidth=1.2pt](-0.326,-1.303) 
\psdot[linecolor=magenta,linewidth=1.2pt](-1.896,-0.739) 
\psdot[linecolor=magenta,linewidth=1.2pt](-0.820,-0.123) 
\psdot[linecolor=magenta,linewidth=1.2pt](-0.647,-2.020) 
\psdot[linecolor=magenta,linewidth=1.2pt](-1.137,1.103) 
\psdot[linecolor=magenta,linewidth=1.2pt](-1.615,0.090) 
\psdot[linecolor=magenta,linewidth=1.2pt](-2.242,-1.215) 
\psdot[linecolor=magenta,linewidth=1.2pt](-1.961,1.103) 
\psdot[linecolor=magenta,linewidth=1.2pt](-0.203,-0.315) 
\psdot[linecolor=magenta,linewidth=1.2pt](1.300,-1.685) 
\psdot[linecolor=magenta,linewidth=1.2pt](-0.125,-0.270) 
\psdot[linecolor=magenta,linewidth=1.2pt](0,1.403) 
\rput(-0.792,0){S\&P}
\rput(-0.578,-0.054){Nasd}
\rput(-0.884,0.156){Cana}
\rput(-0.728,0.247){Mexi}
\rput(-0.883,-2.898){Pana}
\rput(1.940,-1.427){CoRi}
\rput(-1.986,-0.948){Berm}
\rput(-0.009,-2.105){Jama}
\rput(-0.352,-0.054){Braz}
\rput(-0.352,0.368){Arge}
\rput(-0.578,0.695){Chil}
\rput(0.793,0.334){Colo}
\rput(-1.442,-0.173){Vene}
\rput(-0.211,0.889){Peru}
\rput(-0.13,-0.054){UK}
\rput(-0.253,-0.173){Irel}
\rput(0.07,0.054){Fran}
\rput(0.15,-0.140){Germ}
\rput(-0.244,0.066){Swit}
\rput(0.15,-0.320){Autr}
\rput(-0.099,-0.118){Ital}
\rput(-1.014,-2.272){Malt}
\rput(0.056,0.106){Belg}
\rput(0.126,-0.054){Neth}
\rput(0.223,0.273){Luxe}
\rput(-0.214,0.141){Swed}
\rput(-0.17,0.223){Denm}
\rput(0.257,0.162){Finl}
\rput(0.351,-0.054){Norw}
\rput(0.15,0.578){Icel}
\rput(0.127,0.211){Spai}
\rput(0.406,0.493){Port}
\rput(0.530,0.1){Gree}
\rput(0,-0.706){CzRe}
\rput(-2.137,-0.469){Slok}
\rput(-0.148,-0.570){Hung}
\rput(0.632,-1.585){Serb}
\rput(0.141,-1.194){Croa}
\rput(1.927,-2.419){Slov}
\rput(0.312,-2.372){BoHe}
\rput(1.406,-2.571){Monte}
\rput(1.166,-2.120){Mace}
\rput(0.395,-0.142){Pola}
\rput(-1.458,-1.456){Roma}
\rput(-0.748,-0.370){Bulg}
\rput(-0.763,-1.962){Esto}
\rput(-1.947,-0.620){Latv}
\rput(-1.209,-0.460){Lith}
\rput(-1.278,-1.581){Ukra}
\rput(1.165,0){Russ}
\rput(-0.213,-2.023){Kaza}
\rput(0.753,0.1){Turk}
\rput(0.530,-0.362){Cypr}
\rput(0.096,-0.513){Isra}
\rput(0.725,-1.356){Pale}
\rput(-0.443,-2.863){Leba}
\rput(-1.811,0.411){Jord}
\rput(-1.832,-1.228){SaAr}
\rput(-1.017,0.311){Kuwa}
\rput(1.790,-0.613){Bahr}
\rput(-1.278,-0.272){Qata}
\rput(-2.111,-0.372){UAE}
\rput(-2.158,0.465){Ohma}
\rput(0.508,-0.620){Paki}
\rput(-1.096,-0.183){Indi}
\rput(-1.689,-1.007){SrLa}
\rput(-0.029,-2.782){Bang}
\rput(-0.677,-1.311){Japa}
\rput(-0.634,-0.639){HoKo}
\rput(-1.056,-0.639){Chin}
\rput(1.650,-2.120){Mong}
\rput(-0.733,-1.430){Taiw}
\rput(-1.033,-1.118){SoKo}
\rput(-0.030,-0.953){Thai}
\rput(-0.563,-1.905){Viet}
\rput(-0.662,-0.437){Mala}
\rput(-0.454,-0.579){Sing}
\rput(-0.912,-0.789){Indo}
\rput(-0.471,-1.426){Phil}
\rput(-0.827,-0.952){Aust}
\rput(-0.176,-1.303){NeZe}
\rput(-2.046,-0.739){Moro}
\rput(-0.820,-0.193){Tuni}
\rput(-0.647,-2.120){Egyp}
\rput(-0.987,1.103){Ghan}
\rput(-1.765,0.090){Nige}
\rput(-2.392,-1.215){Keny}
\rput(-1.961,1.203){Tanz}
\rput(-0.353,-0.315){Nami}
\rput(1.450,-1.685){Bots}
\rput(-0.125,-0.350){SoAf}
\rput(0,1.503){Maur}
\end{pspicture}

\vskip 0.4 cm

\noindent Figure 17: minimum spanning tree for the second semester of 2007. The inset shows an amplification of the connections closest to France.

\vskip -12.1 cm
\hskip 10.5 cm
\begin{minipage}{4 cm }
\begin{pspicture}(-3,-1)(-1,1)
\psline(-1.778,-2.222)(3.556,-2.222)(3.556,3)(-1.778,3)(-1.778,-2.222)
\psset{xunit=10,yunit=10} \scriptsize
\psline(-0.18,0)(0,0) 
\psline(0.356,0.185)(0.207,0.082) 
\psline(-0.080,0.3)(0,0.203) 
\psline(0,-0.054)(0,0) 
\psline(0,-0.054)(0,-0.140) 
\psline(-0.178,-0.103)(0,0) 
\psline(0,0)(-0.114,0.066) 
\psline(0,0)(-0.039,-0.068) 
\psline(0,0)(0.056,0.056) 
\psline(0,0)(0.066,0) 
\psline(0,0)(-0.064,0.111) 
\psline(0,0)(0,0.203) 
\psline(0,0)(0.087,0.151) 
\psline(0,0)(0.245,-0.142) 
\psline[linestyle=dashed](0,0)(0.125,-0.217) 
\psline(0,-0.140)(0,-0.222) 
\psline(0,-0.140)(-0.042,-0.222) 
\psline(0.056,0.056)(0.223,0.223) 
\psline(0.066,0)(0.207,0.082) 
\psline(0.066,0)(0.261,0) 
\psline(0,0.203)(0,0.3) 
\psline(0.261,0)(0.356,0) 
\psline(0.087,0.151)(0.173,0.3) 
\psdot[linecolor=blue,linewidth=1.2pt](0,-0.054) 
\psdot[linecolor=blue,linewidth=1.2pt](0,0) 
\psdot[linecolor=blue,linewidth=1.2pt](0,-0.140) 
\psdot[linecolor=blue,linewidth=1.2pt](-0.114,0.066) 
\psdot[linecolor=blue,linewidth=1.2pt](-0.039,-0.068) 
\psdot[linecolor=blue,linewidth=1.2pt](0.056,0.056) 
\psdot[linecolor=blue,linewidth=1.2pt](0.066,0) 
\psdot[linecolor=blue,linewidth=1.2pt](0.223,0.223) 
\psdot[linecolor=blue,linewidth=1.2pt](-0.064,0.111) 
\psdot[linecolor=blue,linewidth=1.2pt](0,0.203) 
\psdot[linecolor=blue,linewidth=1.2pt](0.207,0.082) 
\psdot[linecolor=blue,linewidth=1.2pt](0.261,0) 
\psdot[linecolor=blue,linewidth=1.2pt](0.087,0.151) 
\psdot[linecolor=blue,linewidth=1.2pt](0.245,-0.142) 
\rput(0.04,-0.074){UK}
\rput(-0.05,0.030){Fran}
\rput(-0.05,-0.140){Germ}
\rput(-0.114,0.096){Swit}
\rput(-0.039,-0.098){Ital}
\rput(0.106,0.056){Belg}
\rput(0.116,-0.03){Neth}
\rput(0.273,0.223){Luxe}
\rput(-0.064,0.141){Swed}
\rput(-0.05,0.203){Denm}
\rput(0.207,0.052){Finl}
\rput(0.261,-0.030){Norw}
\rput(0.037,0.151){Spai}
\rput(0.295,-0.142){Pola}
\end{pspicture}
\end{minipage}

\newpage

Figure 18 shows the minimum spanning tree for the first semester of 2008, with threshold $0.68\pm 0.03$. Here, again, one can clearly see the presence of the American, European, Pacific Asian, and Arab clusters. Connections between African indices, with the exception of the pair Namibia - South Africa, haven't lasted. Some of the former members of Yugoslavia, Croatia and Macedonia, seem now more integrated with Europe, and a weakly connected cluster is made of Serbia, Montenegro, and Bosnia and Herzegovina. We also have a strong connection between Ukraine and Russia.

\begin{pspicture}(-3.7,-8.3)(1,5.5)
\psset{xunit=3,yunit=3} \scriptsize
\psline(-1.038,0)(-1.129,0) 
\psline(-1.038,0)(-0.721,0) 
\psline(-1.038,0)(-1.038,-0.365) 
\psline(-0.721,0)(-0.398,0) 
\psline[linestyle=dashed](-1.038,-0.365)(-1.038,-1.055) 
\psline[linestyle=dashed](2.421,-0.816)(2.061,-0.192) 
\psline[linestyle=dashed](2.489,-2.370)(1.932,-1.813) 
\psline[linestyle=dashed](0.716,-0.733)(0.520,0) 
\psline[linestyle=dashed](0.716,-0.733)(0.897,-1.409) 
\psline[linestyle=dashed](-1.038,-1.868)(-1.038,-1.055) 
\psline[linestyle=dashed](-1.038,-1.868)(-1.038,-2.638) 
\psline[linestyle=dashed](-1.038,-1.868)(-0.188,-1.868) 
\psline(-0.398,0)(-0.681,-0.206) 
\psline(-0.398,0)(0,0) 
\psline(0.059,0.512)(-0.020,0.063) 
\psline(0.428,-0.774)(0.339,-0.268) 
\psline[linestyle=dashed](0.428,-0.774)(0.428,-1.572) 
\psline[linestyle=dashed](0.814,-2.240)(0.428,-1.572) 
\psline(-0.374,-0.576)(-0.131,-0.242) 
\psline(-0.088,0.064)(-0.293,0.213) 
\psline(-0.088,0.064)(0,0) 
\psline(0,0)(-0.020,0.063) 
\psline(0,0)(0.051,0.157) 
\psline(0,0)(0.184,0) 
\psline(0,0)(0.102,0.074) 
\psline(0,0)(0.042,-0.128) 
\psline(0,0)(0.058,-0.042) 
\psline(0,0)(-0.036,-0.111) 
\psline(0,0)(-0.377,-0.274) 
\psline(-0.020,0.063)(-0.068,0.147) 
\psline(-0.020,0.063)(-0.054,0.387) 
\psline(0.184,0)(0.242,0.331) 
\psline(0.184,0)(0.520,0) 
\psline(0.184,0)(0.339,-0.268) 
\psline[linestyle=dashed](0.006,-2.303)(0.428,-1.572) 
\psline(0.042,-0.128)(0.113,-0.346) 
\psline(0.113,-0.346)(0.291,-0.895) 
\psline(-0.036,-0.111)(-0.110,-0.338) 
\psline(-0.036,-0.111)(-0.131,-0.242) 
\psline(-0.036,-0.111)(-0.036,-0.348) 
\psline(-0.110,-0.338)(-0.190,-0.585) 
\psline[linestyle=dashed](-0.190,-0.585)(-0.166,-1.201) 
\psline(-0.190,-0.585)(-0.399,-0.947) 
\psline(0.242,0.331)(0.142,0.716) 
\psline(0.242,0.331)(0.286,0.494) 
\psline(0.520,0)(0.396,0.461) 
\psline(0.520,0)(0.931,0) 
\psline(0.520,0)(0.601,0.303) 
\psline[linestyle=dashed](2.632,-1.938)(2.104,-1.410) 
\psline[linestyle=dashed](2.632,-1.938)(3.214,-2.520) 
\psline[linestyle=dashed](1.389,-1.270)(1.932,-1.813) 
\psline[linestyle=dashed](1.389,-1.270)(1.389,-0.511) 
\psline(1.206,0.477)(0.931,0) 
\psline[linestyle=dashed](1.206,0.477)(1.206,1.199) 
\psline[linestyle=dashed](1.206,0.477)(1.562,1.094) 
\psline[linestyle=dashed](1.334,-2.410)(1.932,-1.813) 
\psline[linestyle=dashed](1.561,0.364)(0.931,0) 
\psline[linestyle=dashed](1.561,0.364)(1.939,1.019) 
\psline[linestyle=dashed](1.561,0.364)(2.179,0.721) 
\psline[linestyle=dashed](0.931,0)(0.931,0.696) 
\psline(0.931,0)(1.389,0) 
\psline(2.104,-1.410)(1.624,-0.930) 
\psline(2.511,0.311)(1.972,0) 
\psline[linestyle=dashed](2.511,0.311)(2.511,1.197) 
\psline(1.624,-0.930)(1.624,-0.406) 
\psline[linestyle=dashed](0.931,0.696)(0.931,1.420) 
\psline(-0.441,0.716)(0.142,0.716) 
\psline(0.601,0.303)(0.601,0.939) 
\psline[linestyle=dashed](0.286,0.494)(0.286,1.240) 
\psline(3.231,-0.657)(3.231,0) 
\psline(2.802,-0.428)(3.231,0) 
\psline(3.231,0)(3.826,0) 
\psline(3.231,0)(2.624,0) 
\psline(3.826,0)(3.826,-0.655) 
\psline[linestyle=dashed](3.826,-0.655)(3.826,-1.357) 
\psline(1.624,-0.406)(1.389,0) 
\psline[linestyle=dashed](1.104,-2.180)(0.897,-1.409) 
\psline(1.972,0)(1.728,0) 
\psline(1.972,0)(2.450,-0.276) 
\psline(1.972,0)(2.624,0) 
\psline(1.178,-0.211)(0.783,-0.606) 
\psline(1.178,-0.211)(1.389,0) 
\psline[linestyle=dashed](1.178,-0.211)(1.178,-0.895) 
\psline[linestyle=dashed](2.179,1.439)(2.179,0.721) 
\psline(2.060,0.192)(1.728,0) 
\psline(1.728,0)(1.389,0) 
\psline(1.728,0)(2.061,-0.192) 
\psline(1.389,-0.511)(1.389,0) 
\psline(2.061,-0.746)(2.061,-0.192) 
\psline(2.061,-0.192)(1.796,-0.651) 
\psline(-0.474,-1.076)(-0.399,-0.947) 
\psdot[linecolor=orange,linewidth=1.2pt](-1.038,0) 
\psdot[linecolor=orange,linewidth=1.2pt](-1.129,0) 
\psdot[linecolor=orange,linewidth=1.2pt](-0.721,0) 
\psdot[linecolor=orange,linewidth=1.2pt](-1.038,-0.365) 
\psdot[linecolor=darkgreen,linewidth=1.2pt](2.421,-0.816) 
\psdot[linecolor=darkgreen,linewidth=1.2pt](2.489,-2.370) 
\psdot[linecolor=darkgreen,linewidth=1.2pt](0.716,-0.733) 
\psdot[linecolor=darkgreen,linewidth=1.2pt](-1.038,-1.868) 
\psdot[linecolor=green,linewidth=1.2pt](-0.398,0) 
\psdot[linecolor=green,linewidth=1.2pt](-0.681,-0.206) 
\psdot[linecolor=green,linewidth=1.2pt](0.059,0.512) 
\psdot[linecolor=green,linewidth=1.2pt](0.428,-0.774) 
\psdot[linecolor=green,linewidth=1.2pt](0.814,-2.240) 
\psdot[linecolor=green,linewidth=1.2pt](-0.374,-0.576) 
\psdot[linecolor=blue,linewidth=1.2pt](-0.088,0.064) 
\psdot[linecolor=blue,linewidth=1.2pt](-0.293,0.213) 
\psdot[linecolor=blue,linewidth=1.2pt](0,0) 
\psdot[linecolor=blue,linewidth=1.2pt](-0.020,0.063) 
\psdot[linecolor=blue,linewidth=1.2pt](0.051,0.157) 
\psdot[linecolor=blue,linewidth=1.2pt](0.184,0) 
\psdot[linecolor=blue,linewidth=1.2pt](0.102,0.074) 
\psdot[linecolor=blue,linewidth=1.2pt](0.006,-2.303) 
\psdot[linecolor=blue,linewidth=1.2pt](0.042,-0.128) 
\psdot[linecolor=blue,linewidth=1.2pt](0.058,-0.042) 
\psdot[linecolor=blue,linewidth=1.2pt](0.113,-0.346) 
\psdot[linecolor=blue,linewidth=1.2pt](-0.036,-0.111) 
\psdot[linecolor=blue,linewidth=1.2pt](-0.110,-0.338) 
\psdot[linecolor=blue,linewidth=1.2pt](-0.131,-0.242) 
\psdot[linecolor=blue,linewidth=1.2pt](-0.190,-0.585) 
\psdot[linecolor=blue,linewidth=1.2pt](0.291,-0.895) 
\psdot[linecolor=blue,linewidth=1.2pt](-0.068,0.147) 
\psdot[linecolor=blue,linewidth=1.2pt](-0.036,-0.348) 
\psdot[linecolor=blue,linewidth=1.2pt](0.242,0.331) 
\psdot[linecolor=blue,linewidth=1.2pt](0.520,0) 
\psdot[linecolor=blue,linewidth=1.2pt](2.632,-1.938) 
\psdot[linecolor=blue,linewidth=1.2pt](-0.054,0.387) 
\psdot[linecolor=blue,linewidth=1.2pt](1.389,-1.270) 
\psdot[linecolor=blue,linewidth=1.2pt](0.396,0.461) 
\psdot[linecolor=blue,linewidth=1.2pt](1.206,0.477) 
\psdot[linecolor=blue,linewidth=1.2pt](1.334,-2.410) 
\psdot[linecolor=blue,linewidth=1.2pt](1.932,-1.813) 
\psdot[linecolor=blue,linewidth=1.2pt](1.561,0.364) 
\psdot[linecolor=blue,linewidth=1.2pt](0.339,-0.268) 
\psdot[linecolor=blue,linewidth=1.2pt](0.931,0) 
\psdot[linecolor=blue,linewidth=1.2pt](2.104,-1.410) 
\psdot[linecolor=blue,linewidth=1.2pt](2.511,0.311) 
\psdot[linecolor=blue,linewidth=1.2pt](1.624,-0.930) 
\psdot[linecolor=blue,linewidth=1.2pt](0.931,0.696) 
\psdot[linecolor=blue,linewidth=1.2pt](-0.441,0.716) 
\psdot[linecolor=bluish,linewidth=1.2pt](0.142,0.716) 
\psdot[linecolor=brown,linewidth=1.2pt](1.206,1.199) 
\psdot[linecolor=bluish,linewidth=1.2pt](0.601,0.303) 
\psdot[linecolor=brown,linewidth=1.2pt](0.286,0.494) 
\psdot[linecolor=brown,linewidth=1.2pt](-0.377,-0.274) 
\psdot[linecolor=brown,linewidth=1.2pt](1.562,1.094) 
\psdot[linecolor=brown,linewidth=1.2pt](-0.166,-1.201) 
\psdot[linecolor=brown,linewidth=1.2pt](0.931,1.420) 
\psdot[linecolor=brown,linewidth=1.2pt](0.601,0.939) 
\psdot[linecolor=brown,linewidth=1.2pt](3.231,-0.657) 
\psdot[linecolor=brown,linewidth=1.2pt](1.939,1.019) 
\psdot[linecolor=brown,linewidth=1.2pt](2.802,-0.428) 
\psdot[linecolor=brown,linewidth=1.2pt](3.231,0) 
\psdot[linecolor=brown,linewidth=1.2pt](3.826,0) 
\psdot[linecolor=brown,linewidth=1.2pt](3.826,-0.655) 
\psdot[linecolor=brown,linewidth=1.2pt](1.624,-0.406) 
\psdot[linecolor=brown,linewidth=1.2pt](-1.038,-1.055) 
\psdot[linecolor=brown,linewidth=1.2pt](1.104,-2.180) 
\psdot[linecolor=red,linewidth=1.2pt](1.972,0) 
\psdot[linecolor=red,linewidth=1.2pt](1.178,-0.211) 
\psdot[linecolor=red,linewidth=1.2pt](0.783,-0.606) 
\psdot[linecolor=red,linewidth=1.2pt](2.179,1.439) 
\psdot[linecolor=red,linewidth=1.2pt](2.060,0.192) 
\psdot[linecolor=red,linewidth=1.2pt](1.728,0) 
\psdot[linecolor=red,linewidth=1.2pt](1.389,-0.511) 
\psdot[linecolor=red,linewidth=1.2pt](-1.038,-2.638) 
\psdot[linecolor=red,linewidth=1.2pt](2.450,-0.276) 
\psdot[linecolor=red,linewidth=1.2pt](1.389,0) 
\psdot[linecolor=red,linewidth=1.2pt](1.178,-0.895) 
\psdot[linecolor=red,linewidth=1.2pt](2.061,-0.746) 
\psdot[linecolor=black,linewidth=1.2pt](2.061,-0.192) 
\psdot[linecolor=black,linewidth=1.2pt](1.796,-0.651) 
\psdot[linecolor=magenta,linewidth=1.2pt](2.179,0.721) 
\psdot[linecolor=magenta,linewidth=1.2pt](0.428,-1.572) 
\psdot[linecolor=magenta,linewidth=1.2pt](2.624,0) 
\psdot[linecolor=magenta,linewidth=1.2pt](-0.188,-1.868) 
\psdot[linecolor=magenta,linewidth=1.2pt](3.214,-2.520) 
\psdot[linecolor=magenta,linewidth=1.2pt](0.897,-1.409) 
\psdot[linecolor=magenta,linewidth=1.2pt](3.826,-1.357) 
\psdot[linecolor=magenta,linewidth=1.2pt](-0.474,-1.076) 
\psdot[linecolor=magenta,linewidth=1.2pt](2.511,1.197) 
\psdot[linecolor=magenta,linewidth=1.2pt](-0.399,-0.947) 
\psdot[linecolor=magenta,linewidth=1.2pt](0.286,1.240) 
\rput(-1.038,0.08){S\&P}
\rput(-1.279,0){Nasd}
\rput(-0.721,0.08){Cana}
\rput(-1.188,-0.365){Mexi}
\rput(2.421,-0.916){Pana}
\rput(2.489,-2.470){CoRi}
\rput(0.866,-0.733){Berm}
\rput(-1.188,-1.868){Jama}
\rput(-0.398,0.08){Braz}
\rput(-0.681,-0.276){Arge}
\rput(0.059,0.592){Chil}
\rput(0.578,-0.774){Colo}
\rput(0.814,-2.340){Vene}
\rput(-0.374,-0.656){Peru}
\rput(-0.088,0.144){UK}
\rput(-0.293,0.293){Irel}
\rput(0,0.08){Fran}
\rput(-0.020,0.143){Germ}
\rput(0.051,0.237){Swit}
\rput(0.184,0.08){Autr}
\rput(0.102,0.154){Ital}
\rput(0.006,-2.403){Malt}
\rput(0.042,-0.178){Belg}
\rput(0.058,-0.092){Neth}
\rput(0.263,-0.346){Luxe}
\rput(-0.186,-0.111){Swed}
\rput(-0.260,-0.338){Denm}
\rput(-0.281,-0.242){Finl}
\rput(-0.040,-0.585){Norw}
\rput(0.291,-0.985){Icel}
\rput(-0.068,0.227){Spai}
\rput(-0.036,-0.438){Port}
\rput(0.092,0.331){Gree}
\rput(0.520,0.08){CzRe}
\rput(2.782,-1.938){Slok}
\rput(-0.054,0.467){Hung}
\rput(1.239,-1.270){Serb}
\rput(0.396,0.541){Croa}
\rput(1.056,0.477){Slov}
\rput(1.334,-2.510){BoHe}
\rput(2.082,-1.813){Mont}
\rput(1.711,0.364){Mace}
\rput(0.489,-0.268){Pola}
\rput(0.781,0.08){Roma}
\rput(2.254,-1.410){Bulg}
\rput(2.661,0.311){Esto}
\rput(1.774,-0.930){Latv}
\rput(0.781,0.696){Lith}
\rput(-0.441,0.806){Ukra}
\rput(0.142,0.816){Russ}
\rput(1.206,1.299){Kaza}
\rput(0.751,0.303){Turk}
\rput(0.286,0.574){Cypr}
\rput(-0.487,-0.334){Isra}
\rput(1.562,1.194){Pale}
\rput(-0.166,-1.301){Leba}
\rput(0.931,1.520){Jord}
\rput(0.601,1.039){SaAr}
\rput(3.231,-0.757){Kuwa}
\rput(1.939,1.119){Bahr}
\rput(2.802,-0.528){Qata}
\rput(3.081,0.08){UAE}
\rput(3.826,0.08){Ohma}
\rput(3.676,-0.655){Paki}
\rput(1.774,-0.406){Indi}
\rput(-1.188,-1.055){SrLa}
\rput(1.104,-2.280){Bang}
\rput(1.972,0.08){Japa}
\rput(1.028,-0.211){HoKo}
\rput(0.933,-0.606){Chin}
\rput(2.179,1.539){Mong}
\rput(2.060,0.292){Taiw}
\rput(1.728,0.08){SoKo}
\rput(1.509,-0.511){Thai}
\rput(-1.038,-2.718){Viet}
\rput(2.450,-0.376){Mala}
\rput(1.389,0.08){Sing}
\rput(1.178,-0.995){Indo}
\rput(2.061,-0.846){Phil}
\rput(2.211,-0.192){Aust}
\rput(1.796,-0.751){NeZe}
\rput(2.329,0.721){Moro}
\rput(0.578,-1.572){Tuni}
\rput(2.624,0.08){Egyp}
\rput(-0.0388,-1.868){Ghan}
\rput(3.214,-2.620){Nige}
\rput(1.047,-1.409){Keny}
\rput(3.826,-1.457){Tanz}
\rput(-0.474,-1.176){Nami}
\rput(2.511,1.297){Bots}
\rput(-0.549,-0.947){SoAf}
\rput(0.286,1.340){Maur}
\end{pspicture}

\vskip 0.15 cm

\noindent Figure 18: minimum spanning tree for the first semester of 2008. The inset shows an amplification of the connections closest to France.

\vskip -13 cm
\hskip 12 cm
\begin{minipage}{4 cm }
\begin{pspicture}(-2,-1)(-1,1)
\psline(-1.5,-2)(3.3,-2)(3.3,2.3)(-1.5,2.3)(-1.5,-2)
\psset{xunit=10,yunit=10} \scriptsize
\psline(-0.150,0)(0,0) 
\psline(0.026,0.23)(-0.020,0.063) 
\psline(-0.088,0.064)(-0.150,0.109) 
\psline(-0.088,0.064)(0,0) 
\psline(0,0)(-0.020,0.063) 
\psline(0,0)(0.051,0.157) 
\psline(0,0)(0.184,0) 
\psline(0,0)(0.102,0.074) 
\psline(0,0)(0.042,-0.128) 
\psline(0,0)(0.058,-0.042) 
\psline(0,0)(-0.036,-0.111) 
\psline(0,0)(-0.150,-0.109) 
\psline(-0.020,0.063)(-0.068,0.147) 
\psline(-0.020,0.063)(-0.032,0.230) 
\psline(0.184,0)(0.168,0.230) 
\psline(0.184,0)(0.330,0) 
\psline(0.184,0)(0.253,-0.200) 
\psline(0.042,-0.128)(0.065,-0.200) 
\psline(-0.036,-0.111)(-0.065,-0.200) 
\psline(-0.036,-0.111)(-0.108,-0.200) 
\psline(-0.036,-0.111)(-0.036,-0.200) 
\psdot[linecolor=blue,linewidth=1.2pt](-0.088,0.064) 
\psdot[linecolor=blue,linewidth=1.2pt](0,0) 
\psdot[linecolor=blue,linewidth=1.2pt](-0.020,0.063) 
\psdot[linecolor=blue,linewidth=1.2pt](0.051,0.157) 
\psdot[linecolor=blue,linewidth=1.2pt](0.184,0) 
\psdot[linecolor=blue,linewidth=1.2pt](0.102,0.074) 
\psdot[linecolor=blue,linewidth=1.2pt](0.042,-0.128) 
\psdot[linecolor=blue,linewidth=1.2pt](0.058,-0.042) 
\psdot[linecolor=blue,linewidth=1.2pt](-0.036,-0.111) 
\psdot[linecolor=blue,linewidth=1.2pt](-0.068,0.147) 
\rput(-0.088,0.034){UK}
\rput(-0.05,-0.02){Fran}
\rput(0.030,0.063){Germ}
\rput(0.051,0.187){Swit}
\rput(0.134,0.03){Autr}
\rput(0.102,0.104){Ital}
\rput(0.092,-0.128){Belg}
\rput(0.108,-0.042){Neth}
\rput(-0.086,-0.111){Swed}
\rput(-0.068,0.177){Spai}
\end{pspicture}
\end{minipage}

\newpage

The second semester of 2008 is represented in figure 19 for a threshold $0.68\pm 0.03$. This is a high volatility period, especially after the main crisis, which occurred in September of that same year. The network has shrunk due to the increase in correlations between markets during and after the peak of the crisis: S\&P and Nasdaq have nearly joined, and France also nearly merges with other European indices. The Czech Republic now works as a hub for some Eastern European indices, and Australia and Singapore are the main hubs for the Pacific Asian cluster. There is a clear cluster of other Eastern European indices, namely Slovakia, Serbia, Bulgaria, Estonia, Lithuania, and Latvia, and a well formed cluster of Arab indices (United Arab Emirates, Qatar, Palestine, Jordan, Ohman, Bahrein, and Kuwait). Saudi Arabia is still isolated, though. There are also connections between Botswana and Ghana, and between Egypt, Morocco, and Tanzania.

\begin{pspicture}(-3,-6.7)(1,5.5)
\psset{xunit=3,yunit=3} \scriptsize
\psline(-0.724,0)(-0.677,0) 
\psline(-0.677,0)(-0.443,0) 
\psline(-0.763,0.265)(-0.443,0.265) 
\psline(-0.443,0)(-0.443,0.265) 
\psline(-0.443,0)(-0.083,0) 
\psline[linestyle=dashed](-0.306,-1.469)(-0.103,-0.712) 
\psline[linestyle=dashed](0,-1.479)(0,-0.755) 
\psline[linestyle=dashed](1.956,-1.708)(1.956,-0.890) 
\psline[linestyle=dashed](0.184,-1.015)(0,-0.328) 
\psline[linestyle=dashed](0.184,-1.015)(0.770,-1.601) 
\psline[linestyle=dashed](0.184,-1.015)(0.184,-1.777) 
\psline(-0.443,0.265)(-0.443,0.608) 
\psline(-0.265,0.315)(-0.083,0) 
\psline(-0.265,0.315)(-0.265,0.895) 
\psline(0.297,0.375)(0.297,0) 
\psline[linestyle=dashed](0.984,-1.115)(0.614,-0.473) 
\psline(-0.103,-0.712)(0,-0.328) 
\psline(-0.061,-0.016)(0,0) 
\psline(-0.061,-0.016)(-0.249,-0.066) 
\psline(-0.061,-0.016)(-0.506,-0.273) 
\psline(-0.196,-0.371)(-0.067,-0.147) 
\psline(0,0)(-0.083,0) 
\psline(0,0)(-0.062,-0.062) 
\psline(0,0)(0,0.090) 
\psline(0,0)(-0.016,-0.059) 
\psline(0,0)(0.106,0) 
\psline(0,0)(0,-0.092) 
\psline(-0.062,-0.062)(-0.346,-0.346) 
\psline(0.297,0)(0.106,0) 
\psline(0.297,0)(0.522,0) 
\psline(0,0.090)(0,0.178) 
\psline[linestyle=dashed](2.672,0)(1.967,0) 
\psline[linestyle=dashed](2.672,0)(3.228,-0.556) 
\psline(-0.067,-0.147)(-0.016,-0.059) 
\psline(-0.016,-0.059)(-0.075,-0.278) 
\psline(0.106,0)(0.106,0.169) 
\psline(0.106,0.169)(0.106,0.701) 
\psline(0.106,0.169)(0.011,0.525) 
\psline(0,-0.092)(0,-0.328) 
\psline(0,-0.328)(0,-0.755) 
\psline(0.614,-0.282)(0.522,0) 
\psline(0.614,-0.282)(0.442,-0.579) 
\psline(0.614,-0.282)(0.614,-0.473) 
\psline(0.522,0)(0.384,-0.238) 
\psline(0.522,0)(0.802,-0.280) 
\psline(0.522,0)(0.857,0) 
\psline(0.522,0)(0.075,-0.163) 
\psline(0.522,0)(0.310,-0.177) 
\psline(1.967,0)(1.442,0) 
\psline(1.442,0)(1.805,0.210) 
\psline(1.442,0)(1.643,0.348) 
\psline(1.442,0)(1.696,-0.441) 
\psline[linestyle=dashed](1.442,0)(2.146,-0.406) 
\psline(1.442,0)(1.127,0) 
\psline(1.840,0.650)(1.333,0.357) 
\psline(1.840,0.650)(2.368,0.955) 
\psline(1.640,0.889)(1.333,0.357) 
\psline(0.384,-0.238)(0.188,-0.578) 
\psline(0.614,-0.839)(0.614,-0.473) 
\psline(1.958,0.348)(1.643,0.348) 
\psline(1.958,0.348)(2.530,0.348) 
\psline(2.054,0.586)(1.643,0.348) 
\psline(1.127,0.679)(1.127,0) 
\psline(1.956,-0.890)(1.696,-0.441) 
\psline(0.857,-0.545)(0.857,0) 
\psline(2.364,-0.587)(1.949,-0.587) 
\psline(2.364,-0.587)(3.031,-0.587) 
\psline(0.857,1.516)(0.857,0.927) 
\psline(2.140,-1.331)(2.140,-0.906) 
\psline(2.140,-0.906)(1.949,-0.587) 
\psline(1.949,-0.587)(1.696,-0.441) 
\psline(1.949,-0.587)(2.281,-0.778) 
\psline[linestyle=dashed](3.031,-0.587)(3.031,-1.279) 
\psline(0.971,0.270)(1.127,0) 
\psline(0.857,0.229)(0.485,0.601) 
\psline(0.857,0.229)(0.857,0.569) 
\psline(0.857,0.229)(0.857,0) 
\psline[linestyle=dashed](1.379,-2.133)(0.540,-2.133) 
\psline[linestyle=dashed](1.379,-2.133)(2.255,-2.133) 
\psline(0.857,0.927)(0.857,0.569) 
\psline(0.539,0.318)(0.857,0) 
\psline(1.354,-0.392)(1.127,0) 
\psline(1.354,-0.392)(1.354,-1.026) 
\psline(0.857,0)(1.127,0) 
\psline(1.333,0.357)(1.127,0) 
\psline(1.127,0)(0.971,-0.270) 
\psline(1.127,0)(1.127,-0.525) 
\psline(1.354,-1.685)(1.354,-1.026) 
\psline[linestyle=dashed](1.354,-1.026)(1.565,-1.813) 
\psline(0.540,-2.133)(0.184,-1.777) 
\psline(0.253,-0.224)(0.310,-0.177) 
\psdot[linecolor=orange,linewidth=1.2pt](-0.724,0) 
\psdot[linecolor=orange,linewidth=1.2pt](-0.677,0) 
\psdot[linecolor=orange,linewidth=1.2pt](-0.763,0.265) 
\psdot[linecolor=orange,linewidth=1.2pt](-0.443,0) 
\psdot[linecolor=darkgreen,linewidth=1.2pt](-0.306,-1.469) 
\psdot[linecolor=darkgreen,linewidth=1.2pt](0,-1.479) 
\psdot[linecolor=darkgreen,linewidth=1.2pt](1.956,-1.708) 
\psdot[linecolor=darkgreen,linewidth=1.2pt](0.184,-1.015) 
\psdot[linecolor=green,linewidth=1.2pt](-0.443,0.265) 
\psdot[linecolor=green,linewidth=1.2pt](-0.443,0.608) 
\psdot[linecolor=green,linewidth=1.2pt](-0.265,0.315) 
\psdot[linecolor=green,linewidth=1.2pt](0.297,0.375) 
\psdot[linecolor=green,linewidth=1.2pt](0.984,-1.115) 
\psdot[linecolor=green,linewidth=1.2pt](-0.103,-0.712) 
\psdot[linecolor=blue,linewidth=1.2pt](-0.061,-0.016) 
\psdot[linecolor=blue,linewidth=1.2pt](-0.196,-0.371) 
\psdot[linecolor=blue,linewidth=1.2pt](0,0) 
\psdot[linecolor=blue,linewidth=1.2pt](-0.083,0) 
\psdot[linecolor=blue,linewidth=1.2pt](-0.062,-0.062) 
\psdot[linecolor=blue,linewidth=1.2pt](0.297,0) 
\psdot[linecolor=blue,linewidth=1.2pt](0,0.090) 
\psdot[linecolor=blue,linewidth=1.2pt](2.672,0) 
\psdot[linecolor=blue,linewidth=1.2pt](-0.067,-0.147) 
\psdot[linecolor=blue,linewidth=1.2pt](-0.016,-0.059) 
\psdot[linecolor=blue,linewidth=1.2pt](-0.075,-0.278) 
\psdot[linecolor=blue,linewidth=1.2pt](0.106,0) 
\psdot[linecolor=blue,linewidth=1.2pt](0.106,0.169) 
\psdot[linecolor=blue,linewidth=1.2pt](0,-0.092) 
\psdot[linecolor=blue,linewidth=1.2pt](0,-0.328) 
\psdot[linecolor=blue,linewidth=1.2pt](0.106,0.701) 
\psdot[linecolor=blue,linewidth=1.2pt](0,0.178) 
\psdot[linecolor=blue,linewidth=1.2pt](-0.249,-0.066) 
\psdot[linecolor=blue,linewidth=1.2pt](0.614,-0.282) 
\psdot[linecolor=blue,linewidth=1.2pt](0.522,0) 
\psdot[linecolor=blue,linewidth=1.2pt](0.770,-1.601) 
\psdot[linecolor=blue,linewidth=1.2pt](-0.346,-0.346) 
\psdot[linecolor=blue,linewidth=1.2pt](1.967,0) 
\psdot[linecolor=blue,linewidth=1.2pt](0.011,0.525) 
\psdot[linecolor=blue,linewidth=1.2pt](1.442,0) 
\psdot[linecolor=blue,linewidth=1.2pt](-0.265,0.895) 
\psdot[linecolor=blue,linewidth=1.2pt](1.840,0.650) 
\psdot[linecolor=blue,linewidth=1.2pt](1.640,0.889) 
\psdot[linecolor=blue,linewidth=1.2pt](0.384,-0.238) 
\psdot[linecolor=blue,linewidth=1.2pt](0.614,-0.839) 
\psdot[linecolor=blue,linewidth=1.2pt](1.805,0.210) 
\psdot[linecolor=blue,linewidth=1.2pt](1.958,0.348) 
\psdot[linecolor=blue,linewidth=1.2pt](2.054,0.586) 
\psdot[linecolor=blue,linewidth=1.2pt](1.643,0.348) 
\psdot[linecolor=blue,linewidth=1.2pt](0.802,-0.280) 
\psdot[linecolor=bluish,linewidth=1.2pt](0,-0.755) 
\psdot[linecolor=brown,linewidth=1.2pt](1.127,0.679) 
\psdot[linecolor=bluish,linewidth=1.2pt](0.442,-0.579) 
\psdot[linecolor=brown,linewidth=1.2pt](0.614,-0.473) 
\psdot[linecolor=brown,linewidth=1.2pt](-0.506,-0.273) 
\psdot[linecolor=brown,linewidth=1.2pt](1.956,-0.890) 
\psdot[linecolor=brown,linewidth=1.2pt](0.857,-0.545) 
\psdot[linecolor=brown,linewidth=1.2pt](2.364,-0.587) 
\psdot[linecolor=brown,linewidth=1.2pt](0.857,1.516) 
\psdot[linecolor=brown,linewidth=1.2pt](2.140,-1.331) 
\psdot[linecolor=brown,linewidth=1.2pt](2.140,-0.906) 
\psdot[linecolor=brown,linewidth=1.2pt](1.949,-0.587) 
\psdot[linecolor=brown,linewidth=1.2pt](1.696,-0.441) 
\psdot[linecolor=brown,linewidth=1.2pt](2.281,-0.778) 
\psdot[linecolor=brown,linewidth=1.2pt](2.146,-0.406) 
\psdot[linecolor=brown,linewidth=1.2pt](0.188,-0.578) 
\psdot[linecolor=brown,linewidth=1.2pt](3.031,-0.587) 
\psdot[linecolor=brown,linewidth=1.2pt](3.228,-0.556) 
\psdot[linecolor=red,linewidth=1.2pt](0.971,0.270) 
\psdot[linecolor=red,linewidth=1.2pt](0.857,0.229) 
\psdot[linecolor=red,linewidth=1.2pt](0.485,0.601) 
\psdot[linecolor=red,linewidth=1.2pt](1.379,-2.133) 
\psdot[linecolor=red,linewidth=1.2pt](0.857,0.927) 
\psdot[linecolor=red,linewidth=1.2pt](0.857,0.569) 
\psdot[linecolor=red,linewidth=1.2pt](0.539,0.318) 
\psdot[linecolor=red,linewidth=1.2pt](2.530,0.348) 
\psdot[linecolor=red,linewidth=1.2pt](1.354,-0.392) 
\psdot[linecolor=red,linewidth=1.2pt](0.857,0) 
\psdot[linecolor=red,linewidth=1.2pt](0.075,-0.163) 
\psdot[linecolor=red,linewidth=1.2pt](1.333,0.357) 
\psdot[linecolor=black,linewidth=1.2pt](1.127,0) 
\psdot[linecolor=black,linewidth=1.2pt](0.971,-0.270) 
\psdot[linecolor=magenta,linewidth=1.2pt](1.354,-1.685) 
\psdot[linecolor=magenta,linewidth=1.2pt](3.031,-1.279) 
\psdot[linecolor=magenta,linewidth=1.2pt](1.354,-1.026) 
\psdot[linecolor=magenta,linewidth=1.2pt](0.540,-2.133) 
\psdot[linecolor=magenta,linewidth=1.2pt](2.255,-2.133) 
\psdot[linecolor=magenta,linewidth=1.2pt](2.368,0.955) 
\psdot[linecolor=magenta,linewidth=1.2pt](1.565,-1.813) 
\psdot[linecolor=magenta,linewidth=1.2pt](0.253,-0.224) 
\psdot[linecolor=magenta,linewidth=1.2pt](0.184,-1.777) 
\psdot[linecolor=magenta,linewidth=1.2pt](0.310,-0.177) 
\psdot[linecolor=magenta,linewidth=1.2pt](1.127,-0.525) 
\rput(-0.874,0){S\&P}
\rput(-0.677,0.1){Nasd}
\rput(-0.913,0.265){Cana}
\rput(-0.443,-0.08){Mexi}
\rput(-0.306,-1.569){Pana}
\rput(0,-1.579){CoRi}
\rput(1.956,-1.808){Berm}
\rput(0.334,-1.015){Jama}
\rput(-0.593,0.365){Braz}
\rput(-0.443,0.708){Arge}
\rput(-0.135,0.385){Chil}
\rput(0.297,0.475){Colo}
\rput(0.984,-1.215){Vene}
\rput(-0.253,-0.712){Peru}
\rput(-0.061,-0.076){UK}
\rput(-0.196,-0.431){Irel}
\rput(0,0.06){Fran}
\rput(-0.263,0.06){Germ}
\rput(-0.062,-0.122){Swit}
\rput(0.297,0.1){Autr}
\rput(-0.1,0.150){Ital}
\rput(2.672,0.1){Malt}
\rput(-0.187,-0.147){Belg}
\rput(-0.166,-0.059){Neth}
\rput(-0.075,-0.338){Luxe}
\rput(0.106,-0.07){Swed}
\rput(0.256,0.209){Denm}
\rput(0,-0.132){Finl}
\rput(0,-0.388){Norw}
\rput(0.106,0.801){Icel}
\rput(-0.03,0.258){Spai}
\rput(-0.249,-0.126){Port}
\rput(0.614,-0.382){Gree}
\rput(0.522,0.1){CzRe}
\rput(0.770,-1.701){Slok}
\rput(-0.346,-0.406){Hung}
\rput(1.967,0.1){Serb}
\rput(-0.019,0.625){Croa}
\rput(1.442,0.1){Slov}
\rput(-0.265,0.995){BoHe}
\rput(1.660,0.650){Mont}
\rput(1.790,0.889){Mace}
\rput(0.504,-0.238){Pola}
\rput(0.614,-0.939){Roma}
\rput(1.955,0.210){Bulg}
\rput(1.988,0.448){Esto}
\rput(2.204,0.586){Latv}
\rput(1.643,0.448){Lith}
\rput(0.732,-0.180){Ukra}
\rput(-0.15,-0.805){Russ}
\rput(1.127,0.779){Kaza}
\rput(0.442,-0.679){Turk}
\rput(0.614,-0.573){Cypr}
\rput(-0.656,-0.273){Isra}
\rput(1.806,-0.890){Pale}
\rput(0.857,-0.645){Leba}
\rput(2.514,-0.487){Jord}
\rput(0.857,1.616){SaAr}
\rput(2.140,-1.431){Kuwa}
\rput(2.290,-0.906){Bahr}
\rput(1.949,-0.487){Qata}
\rput(1.546,-0.441){UAE}
\rput(2.461,-0.778){Ohma}
\rput(2.296,-0.406){Paki}
\rput(0.188,-0.678){Indi}
\rput(3.031,-0.487){SrLa}
\rput(3.378,-0.556){Bang}
\rput(0.971,0.370){Japa}
\rput(0.677,0.229){HoKo}
\rput(0.485,0.701){Chin}
\rput(1.379,-2.233){Mong}
\rput(0.687,0.927){Taiw}
\rput(0.677,0.569){SoKo}
\rput(0.539,0.418){Thai}
\rput(2.680,0.348){Viet}
\rput(1.464,-0.292){Mala}
\rput(0.857,0.1){Sing}
\rput(0.075,-0.263){Indo}
\rput(1.233,0.447){Phil}
\rput(1.127,0.1){Aust}
\rput(0.971,-0.370){NeZe}
\rput(1.354,-1.785){Moro}
\rput(3.031,-1.379){Tuni}
\rput(1.204,-1.026){Egyp}
\rput(0.540,-2.233){Ghan}
\rput(2.255,-2.233){Nige}
\rput(2.518,0.955){Keny}
\rput(1.565,-1.913){Tanz}
\rput(0.273,-0.334){Nami}
\rput(0.154,-1.877){Bots}
\rput(0.440,-0.127){SoAf}
\rput(1.127,-0.625){Maur}
\end{pspicture}

\vskip 0.3 cm

\noindent Figure 19: minimum spanning tree for the second semester of 2008. The inset shows an amplification of the connections closest to France.

\vskip -11.4 cm
\hskip 13.3 cm
\begin{minipage}{4 cm }
\begin{pspicture}(-2,-1)(-1,1)
\psline(-3,-3)(2.3,-3)(2.3,2.3)(-3,2.3)(-3,-3)
\psset{xunit=10,yunit=10} \scriptsize
\psline(-0.3,0)(-0.083,0) 
\psline(-0.061,-0.016)(0,0) 
\psline(-0.061,-0.016)(-0.249,-0.066) 
\psline(-0.061,-0.016)(-0.3,-0.154) 
\psline(-0.155,-0.300)(-0.067,-0.147) 
\psline(0,0)(-0.083,0) 
\psline(0,0)(-0.062,-0.062) 
\psline(0,0)(0,0.090) 
\psline(0,0)(-0.016,-0.059) 
\psline(0,0)(0.106,0) 
\psline(0,0)(0,-0.092) 
\psline(-0.062,-0.062)(-0.3,-0.3) 
\psline(0.230,0)(0.106,0) 
\psline(0,0.090)(0,0.178) 
\psline[linestyle=dashed](2.672,0)(1.967,0) 
\psline[linestyle=dashed](2.672,0)(3.459,0) 
\psline(-0.067,-0.147)(-0.016,-0.059) 
\psline(-0.016,-0.059)(-0.075,-0.278) 
\psline(0.106,0)(0.106,0.169) 
\psline(0.106,0.169)(0.106,0.230) 
\psline(0.106,0.169)(0.090,0.230) 
\psline(0,-0.092)(0,-0.300) 
\psline(0.230,-0.056)(0.075,-0.163) 
\psline(-0.216,0.23)(-0.083,0) 
\psdot[linecolor=blue,linewidth=1.2pt](-0.061,-0.016) 
\psdot[linecolor=blue,linewidth=1.2pt](0,0) 
\psdot[linecolor=blue,linewidth=1.2pt](-0.083,0) 
\psdot[linecolor=blue,linewidth=1.2pt](-0.062,-0.062) 
\psdot[linecolor=blue,linewidth=1.2pt](0,0.090) 
\psdot[linecolor=blue,linewidth=1.2pt](-0.067,-0.147) 
\psdot[linecolor=blue,linewidth=1.2pt](-0.016,-0.059) 
\psdot[linecolor=blue,linewidth=1.2pt](-0.075,-0.278) 
\psdot[linecolor=blue,linewidth=1.2pt](0.106,0) 
\psdot[linecolor=blue,linewidth=1.2pt](0.106,0.169) 
\psdot[linecolor=blue,linewidth=1.2pt](0,-0.092) 
\psdot[linecolor=blue,linewidth=1.2pt](0,0.178) 
\psdot[linecolor=blue,linewidth=1.2pt](-0.249,-0.066) 
\psdot[linecolor=red,linewidth=1.2pt](0.075,-0.163) 
\rput(-0.061,-0.036){UK}
\rput(0.04,0.02){Fran}
\rput(-0.143,0.03){Germ}
\rput(-0.062,-0.082){Swit}
\rput(-0.04,0.090){Ital}
\rput(-0.097,-0.177){Belg}
\rput(0.034,-0.059){Neth}
\rput(-0.125,-0.278){Luxe}
\rput(0.106,-0.03){Swed}
\rput(0.156,0.169){Denm}
\rput(0.05,-0.092){Finl}
\rput(-0.05,0.178){Spai}
\rput(-0.249,-0.036){Port}
\rput(0.075,-0.193){Indo}
\end{pspicture}
\end{minipage}

\vskip 10.2 cm

\section{Pruning}

As we could see from the previous section, the connections between indices in a minimum spanning tree, which in our case is based on distances which are themselves based on the correlations between them, is plagued by random noise, which may place suspicion on the weakly connected indices. By performing simulations on shuffled data, which preserve the frequency distributions of the times series used but destroys correlations between them, we could see that, in average, distances above a threshold of aproximately 0.7 are typical of random interactions. So, any connection above this threshold may be due to random noise. When those connections are removed, the resulting network is called a k-minimum spanning tree, which is applied to the study of some networks \cite{kmst1}-\cite{kmst3}. Nevertheless, some weaker connections may fall beyond the threshold, and using the threshold as a cut-off will probably eliminate some real and important connections from our results.

One way to separate true connections from false ones is to hypothesize that if two indices are truly connected, then that connection must be enduring in time. How long the connection must endure is open to debate, but the longer a connection lasts, the more reliable it is. So, in this section we develop a method for testing the survivability of connections in a mininum spanning tree and use it to determine which connections are resilient in time and which are not.

Robustness of connections of a network have already been studied using stock returns in \cite{robus1}-\cite{robus2}, and pruning was used in \cite{pruned1} based on the survival ratio of connections \cite{surv1}-\cite{surv2}, although the sort of pruning is different from the one adopted in the present article.

We begin by considering an array, called the {\sl adjacency matrix} $A$, whose elements $A_{ij}$ are 1 when there is a connection between indices $i$ and $j$ and zero otherwise. We shall consider each element of this matrix as a function of time and consider sequences $A(t_0),A(t_1),\cdots ,A(t_n)$ of snapshots of such matrix at different times. We consider matrices calculated over periods of 60 days in a window that moves one day at a time. Such a choice offers a compromise between eliminating some statistical noise for calculating correlation matrices using a small amount of data and not extending our results over a long period of time, when the structure of the global market may change by itself. We shall use the year 2008 as an example.

Our first attempt at testing survivability follows the approach of other researchers \cite{robus1}-\cite{surv2}, which is to define a survival matrix $R$ as the result of the product of a sequence of adjency matrices in time. So, we may define the elements of such a survival matrix as
\begin{equation}
\label{survival 1}
R_{ij}=\prod_{k=1}^nA_{ij}(t_k)\ ,
\end{equation}
where $n$ is the number of sliding windows used.

If the connection is strong, then it shall be present at all times, but if it is zero in a single period of time, then the entire product will be zero as well. Applying this method to the data concerning 2008, then we obtain the following results: since a minimum spanning tree of $n$ nodes has $n-1$ connections, then we start with 91 connections in the first calculation, done over the first 60 days of that year; moving the window five days, 53 of those connections survive; for 20 days, the connections that survive are 18; for forty days, we have 11 connections that survive, and so on. Table 1 shows the results obtained, and the graph in figure 20 shows the evolution of surviving connections in time (solid line). Also in table 1 and figure 20 are the average number of surviving connections in time resulting from 1000 simulations with randomized data based on the same time series as the real data (dashed line). Dotted lines represent the average simulated values plus or minus their standard deviation.

\vskip 0.4 cm

\noindent
\begin{minipage}{8 cm }
\[ \begin{array}{|c|c|c|} \hline \text{Timespan} & \text{Surviving} & \text{Surviving random} \\ \text{(days)} & \text{connections} & \text{connections} \\ \hline 0 & 91 & 91 \\ 5 & 53 & 40.5 \\ 10 & 38 & 25.0 \\ 20 & 18 & 10.9 \\ 40 & 11 & 2.0 \\ 60 & 9 & 0.3 \\ 80 & 8 & 0.0 \\ 100 & 8 & 0.0 \\ 120 & 8 & 0 \\ 140 & 5 & 0 \\ 160 & 3 & 0 \\ 180 & 3 & 0 \\ 190 & 3 & 0 \\ \hline \end{array} \]
Table 1: surviving connections for matrix $A$ over periods of time.
\end{minipage}
\begin{minipage}{9 cm }
\begin{pspicture}(-1,2.5)(5.5,2.5)
\psset{xunit=0.04,yunit=0.06} \small \psline{->}(0,0)(200,0) \psline{->}(0,0)(0,100) \rput(213,0){days} \rput(47,100){surviving connections}
\psline(5,-1.7)(5,1.7) \psline(20,-1.7)(20,1.7) \psline(40,-1.7)(40,1.7) \psline(60,-1.7)(60,1.7) \psline(80,-1.7)(80,1.7) \psline(100,-1.7)(100,1.7) \psline(120,-1.7)(120,1.7) \psline(140,-1.7)(140,1.7) \psline(160,-1.7)(160,1.7) \psline(180,-1.7)(180,1.7) \rput(5,-5){5} \rput(20,-5){20} \rput(40,-5){40} \rput(60,-5){60} \rput(80,-5){80} \rput(100,-5){100} \rput(120,-5){120} \rput(140,-5){140} \rput(160,-5){160} \rput(180,-5){180}
\psline(-2.5,20)(2.5,20) \psline(-2.5,40)(2.5,40) \psline(-2.5,60)(2.5,60) \psline(-2.5,80)(2.5,80) \rput(-9,20){20} \rput(-9,40){40} \rput(-9,60){60} \rput(-9,80){80}
\psline(0,91) (1,69) (2,65) (3,61) (4,53) (5,48) (6,46) (7,44) (8,42) (9,38) (10,35) (11,32) (12,32) (13,30) (14,27) (15,25) (16,24) (17,24) (18,22) (19,18) (20,17) (21,17) (22,16) (23,16) (24,15) (25,15) (26,15) (27,15) (28,15) (29,14) (30,13) (31,12) (32,12) (33,12) (34,11) (35,11) (36,11) (37,11) (38,11) (39,11) (40,11) (41,10) (42,10) (43,10) (44,10) (45,10) (46,10) (47,10) (48,10) (49,9) (50,9) (51,9) (52,9) (53,9) (54,9) (55,9) (56,9) (57,9) (58,9) (59,9) (60,9) (61,9) (62,9) (63,9) (64,9) (65,9) (66,9) (67,9) (68,9) (69,9) (70,9) (71,9) (72,9) (73,9) (74,9) (75,9) (76,9) (77,9) (78,9) (79,8) (80,8) (81,8) (82,8) (83,8) (84,8) (85,8) (86,8) (87,8) (88,8) (89,8) (90,8) (91,8) (92,8) (93,8) (94,8) (95,8) (96,8) (97,8) (98,8) (99,8) (100,8) (101,8) (102,8) (103,8) (104,8) (105,8) (106,8) (107,8) (108,8) (109,8) (110,8) (111,8) (112,8) (113,8) (114,8) (115,8) (116,8) (117,8) (118,8) (119,8) (120,8) (121,8) (122,7) (123,7) (124,7) (125,7) (126,7) (127,7) (128,7) (129,7) (130,7) (131,7) (132,7) (133,7) (134,7) (135,7) (136,7) (137,7) (138,5) (139,5) (140,5) (141,4) (142,4) (143,4) (144,4) (145,4) (146,4) (147,4) (148,4) (149,4) (150,4) (151,4) (152,4) (153,4) (154,3) (155,3) (156,3) (157,3) (158,3) (159,3) (160,3) (161,3) (162,3) (163,3) (164,3) (165,3) (166,3) (167,3) (168,3) (169,3) (170,3) (171,3) (172,3) (173,3) (174,3) (175,3) (176,3) (177,3) (178,3) (179,3) (180,3) (181,3) (182,3) (183,3) (184,3) (185,3) (186,3) (187,3) (188,3) (189,3) (190,3)
\psline[linestyle=dashed](0,91) (1,68.653) (2,58.566) (3,51.123) (4,45.272) (5,40.456) (6,36.486) (7,33.054) (8,30.071) (9,27.399) (10,24.955) (11,22.777) (12,20.945) (13,19.227) (14,17.742) (15,16.351) (16,15.031) (17,13.915) (18,12.774) (19,11.851) (20,10.912) (21,10.047) (22,9.276) (23,8.541) (24,7.882) (25,7.29) (26,6.712) (27,6.169) (28,5.677) (29,5.245) (30,4.8) (31,4.433) (32,4.072) (33,3.741) (34,3.427) (35,3.139) (36,2.897) (37,2.659) (38,2.415) (39,2.196) (40,2.011) (41,1.85) (42,1.677) (43,1.525) (44,1.382) (45,1.258) (46,1.142) (47,1.033) (48,0.922) (49,0.827) (50,0.749) (51,0.678) (52,0.605) (53,0.544) (54,0.48) (55,0.447) (56,0.401) (57,0.355) (58,0.326) (59,0.296) (60,0.26) (61,0.228) (62,0.201) (63,0.176) (64,0.157) (65,0.135) (66,0.122) (67,0.11) (68,0.099) (69,0.095) (70,0.084) (71,0.074) (72,0.062) (73,0.051) (74,0.048) (75,0.045) (76,0.04) (77,0.036) (78,0.032) (79,0.031) (80,0.028) (81,0.025) (82,0.02) (83,0.019) (84,0.017) (85,0.017) (86,0.015) (87,0.014) (88,0.013) (89,0.012) (90,0.011) (91,0.01) (92,0.01) (93,0.009) (94,0.009) (95,0.009) (96,0.008) (97,0.007) (98,0.007) (99,0.007) (100,0.006) (101,0.004) (102,0.004) (103,0.004) (104,0.004) (105,0.003) (106,0.001) (107,0.001) (108,0.001) (109,0.001) (110,0.001) (111,0.001) (112,0.001) (113,0.001) (114,0.001) (115,0.001) (116,0.001) (117,0) (190,0)
\psline[linestyle=dotted](0,91.000) (1,64.983) (2,54.444) (3,46.701) (4,40.807) (5,36.095) (6,32.258) (7,28.807) (8,25.932) (9,23.377) (10,21.032) (11,18.908) (12,17.188) (13,15.600) (14,14.250) (15,12.956) (16,11.787) (17,10.820) (18,9.717) (19,8.897) (20,8.028) (21,7.253) (22,6.571) (23,5.921) (24,5.344) (25,4.849) (26,4.381) (27,3.910) (28,3.508) (29,3.129) (30,2.788) (31,2.494) (32,2.176) (33,1.917) (34,1.711) (35,1.483) (36,1.298) (37,1.106) (38,0.932) (39,0.778) (40,0.659) (41,0.541) (42,0.419) (43,0.327) (44,0.248) (45,0.159) (46,0.091) (47,0.020) (48,-0.047) (49,-0.095) (50,-0.117) (51,-0.154) (52,-0.177) (53,-0.191) (54,-0.207) (55,-0.216) (56,-0.230) (57,-0.238) (58,-0.238) (59,-0.236) (60,-0.237) (61,-0.228) (62,-0.229) (63,-0.230) (64,-0.231) (65,-0.227) (66,-0.226) (67,-0.225) (68,-0.219) (69,-0.218) (70,-0.214) (71,-0.210) (72,-0.187) (73,-0.174) (74,-0.171) (75,-0.167) (76,-0.161) (77,-0.150) (78,-0.144) (79,-0.142) (80,-0.137) (81,-0.131) (82,-0.120) (83,-0.118) (84,-0.112) (85,-0.112) (86,-0.107) (87,-0.104) (88,-0.100) (89,-0.097) (90,-0.093) (91,-0.090) (92,-0.090) (93,-0.085) (94,-0.085) (95,-0.085) (96,-0.081) (97,-0.076) (98,-0.076) (99,-0.076) (100,-0.071) (101,-0.059) (102,-0.059) (103,-0.059) (104,-0.059) (105,-0.052) (106,-0.031) (107,-0.031) (108,-0.031) (109,-0.031) (110,-0.031) (111,-0.031) (112,-0.031) (113,-0.031) (114,-0.031) (115,-0.031) (116,-0.031) (117,0.000) (190,0.000)
\psline[linestyle=dotted](0,91.000) (1,72.323) (2,62.688) (3,55.545) (4,49.737) (5,44.817) (6,40.714) (7,37.301) (8,34.210) (9,31.421) (10,28.878) (11,26.646) (12,24.702) (13,22.854) (14,21.234) (15,19.746) (16,18.275) (17,17.010) (18,15.831) (19,14.805) (20,13.796) (21,12.841) (22,11.981) (23,11.161) (24,10.420) (25,9.731) (26,9.043) (27,8.428) (28,7.846) (29,7.361) (30,6.812) (31,6.372) (32,5.968) (33,5.565) (34,5.143) (35,4.795) (36,4.496) (37,4.212) (38,3.898) (39,3.614) (40,3.363) (41,3.159) (42,2.935) (43,2.723) (44,2.516) (45,2.357) (46,2.193) (47,2.046) (48,1.891) (49,1.749) (50,1.615) (51,1.510) (52,1.387) (53,1.279) (54,1.167) (55,1.110) (56,1.032) (57,0.948) (58,0.890) (59,0.828) (60,0.757) (61,0.684) (62,0.631) (63,0.582) (64,0.545) (65,0.497) (66,0.470) (67,0.445) (68,0.417) (69,0.408) (70,0.382) (71,0.358) (72,0.311) (73,0.276) (74,0.267) (75,0.257) (76,0.241) (77,0.222) (78,0.208) (79,0.204) (80,0.193) (81,0.181) (82,0.160) (83,0.156) (84,0.146) (85,0.146) (86,0.137) (87,0.132) (88,0.126) (89,0.121) (90,0.115) (91,0.110) (92,0.110) (93,0.103) (94,0.103) (95,0.103) (96,0.097) (97,0.090) (98,0.090) (99,0.090) (100,0.083) (101,0.067) (102,0.067) (103,0.067) (104,0.067) (105,0.058) (106,0.033) (107,0.033) (108,0.033) (109,0.033) (110,0.033) (111,0.033) (112,0.033) (113,0.033) (114,0.033) (115,0.033) (116,0.033) (117,0.000) (190,0.000)
\rput(100,-17){\normalsize Figure 20: surviving connections as a function}
\rput(32,-26){\normalsize of time for $A$.}
\end{pspicture}
\end{minipage}

\vskip 0.5 cm

As one can see, the number of connections decreases exponentially in time, a consequence of the fact that one sinlge zero value eliminates the survival rate of a connection. The three connections that survive after 160 days are the ones between S\&P and Nasdaq, Greece and Cyprus, and Namibia and South Africa. These are all reasonable connections, and their survival in time is not a surprise. The five connections that survive up to 140 days are, besides those already cited, the ones between Belgium and Luxembourg, and between Japan and South Korea. The eight connections that survive after 120 days include the connections between France and Netherlands, Japan and Singapore, and between Taiwan and South Korea. For 60 days of survival, we also have a connection between Austria and the Czech Republic, and for 40 days of survival, connections between Germany and Spain, and between Latvia and India, this last one indicating that probably connections made by random noise may survive for some days, although for randomized data, obtained by shuffling the log returns for 2008, the survivability falls much sharper than with real data, and survivability for over 50 days is virtually zero.

Even though this seems to be a good method for pruning a minimum spanning tree, it is a little too radical, for it eliminates any connections that are disrupted for a brief period of time, but that overall are stable in time. An example would be some connections among European indices, or among members of the Arab cluster. One index may be connected to one element of its cluster and at a later time connect with a different member of the same cluster. This does not indicate that the connections of that index with the others of its own cluster is inexistent, as the multiplication of the coefficients of the adjency matrix would lead us to believe, so an alternative method should be devised in order to try to distinguish enduring connections from those that are feeble.

One such alternative is to also consider connections of the second order. A node $i$ may be connected with another node $j$ via a third node $k$. Such connections can be computed if we consider the square of the adjency matrix $A$. By taking $A+A^2$, one is considering all first and second order connections between indices, so that an index may be connected to another at certain times, and then be connected to another index that is connected with the previous one. We may then define a new survival matrix by
\begin{equation}
\label{survival 2}
S_{ij}=\prod_{k=1}^nA^2_{ij}(t_k)\ ,
\end{equation}
where $n$ is the number of sliding windows used and $A_{ij}^2$ are the elements of matrix $A+A^2$.

Starting from 152 connections in the first calculation, done over the first 60 days of that year, and moving the window five days, 85 of the original connections survive; sliding the window for 20 days, the connections that survive are 26; for forty days, 14 connections that survive, going down to three surviving connections after 160 days. Table 2 shows the results obtained, and the graph in figure 21 shows the evolution of surviving connections in time (solid line). Table 2 and figure 22 also exhibit the results of the average values obtained from 1000 simulations with randomized data (dashed line for the average and dotted lines for mean plus or minus standard deviation).

\vskip 0.4 cm

\noindent
\begin{minipage}{8 cm }
\[ \begin{array}{|c|c|c|} \hline \text{Timespan} & \text{Surviving} & \text{Surviving random} \\ \text{(days)} & \text{connections} & \text{connections} \\ \hline 0 & 152 & 152 \\ 5 & 85 & 52.3 \\ 10 & 55 & 30.0 \\ 20 & 26 & 12.1 \\ 40 & 14 & 2.2 \\ 60 & 10 & 0.3 \\ 80 & 9 & 0.0 \\ 100 & 9 & 0.0 \\ 120 & 9 & 0.0 \\ 140 & 6 & 0 \\ 160 & 3 & 0 \\ 180 & 3 & 0 \\ 190 & 3 & 0 \\ \hline \end{array} \]
Table 2: surviving connections for $A+A^2$ over periods of time.
\end{minipage}
\begin{minipage}{9 cm }
\begin{pspicture}(-1,2.5)(5.5,2.5)
\psset{xunit=0.04,yunit=0.03} \small \psline{->}(0,0)(200,0) \psline{->}(0,0)(0,200) \rput(213,0){days} \rput(47,200){surviving connections}
\psline(5,-3.4)(5,3.4) \psline(20,-3.4)(20,3.4) \psline(40,-3.4)(40,3.4) \psline(60,-3.4)(60,3.4) \psline(80,-3.4)(80,3.4) \psline(100,-3.4)(100,3.4) \psline(120,-3.4)(120,3.4) \psline(140,-3.4)(140,3.4) \psline(160,-3.4)(160,3.4) \psline(180,-3.4)(180,3.4) \rput(5,-10){5} \rput(20,-10){20} \rput(40,-10){40} \rput(60,-10){60} \rput(80,-10){80} \rput(100,-10){100} \rput(120,-10){120} \rput(140,-10){140} \rput(160,-10){160} \rput(180,-10){180} \psline(-2.5,20)(2.5,20) \psline(-2.5,40)(2.5,40) \psline(-2.5,60)(2.5,60) \psline(-2.5,80)(2.5,80) \psline(-2.5,100)(2.5,100) \psline(-2.5,120)(2.5,120) \psline(-2.5,140)(2.5,140) \psline(-2.5,160)(2.5,160) \psline(-2.5,180)(2.5,180) \rput(-9,20){20} \rput(-9,40){40} \rput(-9,60){60} \rput(-9,80){80} \rput(-12,100){100} \rput(-12,120){120} \rput(-12,140){140} \rput(-12,160){160} \rput(-12,180){180}
\psline(0,152) (1,111) (2,102) (3,96) (4,85) (5,74) (6,70) (7,65) (8,61) (9,55) (10,51) (11,47) (12,47) (13,45) (14,42) (15,38) (16,36) (17,36) (18,32) (19,26) (20,24) (21,24) (22,24) (23,24) (24,20) (25,20) (26,20) (27,20) (28,20) (29,17) (30,16) (31,15) (32,15) (33,15) (34,14) (35,14) (36,14) (37,14) (38,14) (39,14) (40,14) (41,13) (42,13) (43,13) (44,12) (45,12) (46,12) (47,12) (48,12) (49,11) (50,11) (51,11) (52,11) (53,10) (54,10) (55,10) (56,10) (57,10) (58,10) (59,10) (60,10) (61,10) (62,10) (63,10) (64,10) (65,10) (66,10) (67,10) (68,10) (69,10) (70,10) (71,10) (72,10) (73,10) (74,10) (75,10) (76,10) (77,10) (78,10) (79,9) (80,9) (81,9) (82,9) (83,9) (84,9) (85,9) (86,9) (87,9) (88,9) (89,9) (90,9) (91,9) (92,9) (93,9) (94,9) (95,9) (96,9) (97,9) (98,9) (99,9) (100,9) (101,9) (102,9) (103,9) (104,9) (105,9) (106,9) (107,9) (108,9) (109,9) (110,9) (111,9) (112,9) (113,9) (114,9) (115,9) (116,9) (117,9) (118,9) (119,9) (120,9) (121,9) (122,8) (123,8) (124,8) (125,8) (126,8) (127,8) (128,8) (129,8) (130,8) (131,8) (132,8) (133,8) (134,8) (135,8) (136,8) (137,8) (138,6) (139,6) (140,6) (141,6) (142,6) (143,6) (144,6) (145,6) (146,6) (147,6) (148,6) (149,6) (150,6) (151,4) (152,4) (153,4) (154,3) (155,3) (156,3) (157,3) (158,3) (159,3) (160,3) (161,3) (162,3) (163,3) (164,3) (165,3) (166,3) (167,3) (168,3) (169,3) (170,3) (171,3) (172,3) (173,3) (174,3) (175,3) (176,3) (177,3) (178,3) (179,3) (180,3) (181,3) (182,3) (183,3) (184,3) (185,3) (186,3) (187,3) (188,3) (189,3) (190,3)
\psline[linestyle=dashed](0,152) (1,97.139) (2,80.045) (3,68.244) (4,59.364) (5,52.309) (6,46.333) (7,41.153) (8,36.852) (9,33.052) (10,29.981) (11,27.133) (12,24.757) (13,22.543) (14,20.664) (15,18.801) (16,17.261) (17,15.762) (18,14.406) (19,13.217) (20,12.121) (21,11.124) (22,10.209) (23,9.368) (24,8.573) (25,7.914) (26,7.281) (27,6.681) (28,6.15) (29,5.673) (30,5.207) (31,4.775) (32,4.387) (33,4.033) (34,3.708) (35,3.377) (36,3.118) (37,2.849) (38,2.606) (39,2.388) (40,2.193) (41,2.001) (42,1.818) (43,1.668) (44,1.516) (45,1.39) (46,1.277) (47,1.158) (48,1.063) (49,0.966) (50,0.883) (51,0.794) (52,0.728) (53,0.644) (54,0.569) (55,0.525) (56,0.471) (57,0.413) (58,0.358) (59,0.317) (60,0.283) (61,0.253) (62,0.229) (63,0.202) (64,0.185) (65,0.173) (66,0.159) (67,0.142) (68,0.128) (69,0.116) (70,0.105) (71,0.099) (72,0.088) (73,0.08) (74,0.072) (75,0.065) (76,0.061) (77,0.053) (78,0.048) (79,0.044) (80,0.041) (81,0.034) (82,0.028) (83,0.024) (84,0.021) (85,0.018) (86,0.017) (87,0.015) (88,0.015) (89,0.014) (90,0.012) (91,0.012) (92,0.012) (93,0.012) (94,0.01) (95,0.01) (96,0.009) (97,0.007) (98,0.007) (99,0.007) (100,0.007) (101,0.007) (102,0.006) (103,0.005) (104,0.005) (105,0.005) (106,0.004) (107,0.004) (108,0.004) (109,0.003) (110,0.003) (111,0.003) (112,0.003) (113,0.003) (114,0.003) (115,0.003) (116,0.003) (117,0.002) (118,0.002) (119,0.002) (120,0.002) (121,0.001) (122,0.001) (123,0.001) (124,0.001) (125,0.001) (126,0.001) (127,0.001) (128,0.001) (129,0.001) (130,0.001) (131,0.001) (132,0.001) (133,0) (190,0)
\psline[linestyle=dotted](0,152.000) (1,88.940) (2,72.083) (3,60.361) (4,51.792) (5,44.969) (6,39.424) (7,34.604) (8,30.625) (9,27.150) (10,24.328) (11,21.885) (12,19.779) (13,17.756) (14,16.091) (15,14.451) (16,13.063) (17,11.738) (18,10.597) (19,9.586) (20,8.624) (21,7.784) (22,7.025) (23,6.366) (24,5.739) (25,5.131) (26,4.612) (27,4.121) (28,3.682) (29,3.295) (30,2.934) (31,2.623) (32,2.324) (33,2.037) (34,1.791) (35,1.539) (36,1.345) (37,1.181) (38,0.994) (39,0.852) (40,0.692) (41,0.549) (42,0.450) (43,0.371) (44,0.272) (45,0.194) (46,0.143) (47,0.088) (48,0.042) (49,-0.011) (50,-0.058) (51,-0.102) (52,-0.141) (53,-0.165) (54,-0.199) (55,-0.218) (56,-0.234) (57,-0.237) (58,-0.239) (59,-0.235) (60,-0.238) (61,-0.242) (62,-0.243) (63,-0.245) (64,-0.245) (65,-0.243) (66,-0.243) (67,-0.240) (68,-0.229) (69,-0.226) (70,-0.221) (71,-0.219) (72,-0.216) (73,-0.209) (74,-0.205) (75,-0.197) (76,-0.195) (77,-0.188) (78,-0.179) (79,-0.171) (80,-0.162) (81,-0.147) (82,-0.137) (83,-0.129) (84,-0.122) (85,-0.115) (86,-0.112) (87,-0.107) (88,-0.107) (89,-0.104) (90,-0.097) (91,-0.097) (92,-0.097) (93,-0.097) (94,-0.090) (95,-0.090) (96,-0.085) (97,-0.076) (98,-0.076) (99,-0.076) (100,-0.076) (101,-0.076) (102,-0.071) (103,-0.066) (104,-0.066) (105,-0.066) (106,-0.059) (107,-0.059) (108,-0.059) (109,-0.052) (110,-0.052) (111,-0.052) (112,-0.052) (113,-0.052) (114,-0.052) (115,-0.052) (116,-0.052) (117,-0.043) (118,-0.043) (119,-0.043) (120,-0.043) (121,-0.031) (122,-0.031) (123,-0.031) (124,-0.031) (125,-0.031) (126,-0.031) (127,-0.031) (128,-0.031) (129,-0.031) (130,-0.031) (131,-0.031) (132,-0.031) (133,0.000) (190,0.000)
\psline[linestyle=dotted](0,152.000) (1,105.338) (2,88.007) (3,76.127) (4,66.936) (5,59.649) (6,53.242) (7,47.702) (8,43.079) (9,38.954) (10,35.634) (11,32.381) (12,29.735) (13,27.330) (14,25.237) (15,23.151) (16,21.459) (17,19.786) (18,18.215) (19,16.848) (20,15.618) (21,14.464) (22,13.393) (23,12.370) (24,11.407) (25,10.697) (26,9.950) (27,9.241) (28,8.618) (29,8.051) (30,7.480) (31,6.927) (32,6.450) (33,6.029) (34,5.625) (35,5.215) (36,4.891) (37,4.517) (38,4.218) (39,3.924) (40,3.694) (41,3.453) (42,3.186) (43,2.965) (44,2.760) (45,2.586) (46,2.411) (47,2.228) (48,2.084) (49,1.943) (50,1.824) (51,1.690) (52,1.597) (53,1.453) (54,1.337) (55,1.268) (56,1.176) (57,1.063) (58,0.955) (59,0.869) (60,0.804) (61,0.748) (62,0.701) (63,0.649) (64,0.615) (65,0.589) (66,0.561) (67,0.524) (68,0.485) (69,0.458) (70,0.431) (71,0.417) (72,0.392) (73,0.369) (74,0.349) (75,0.327) (76,0.317) (77,0.294) (78,0.275) (79,0.259) (80,0.244) (81,0.215) (82,0.193) (83,0.177) (84,0.164) (85,0.151) (86,0.146) (87,0.137) (88,0.137) (89,0.132) (90,0.121) (91,0.121) (92,0.121) (93,0.121) (94,0.110) (95,0.110) (96,0.103) (97,0.090) (98,0.090) (99,0.090) (100,0.090) (101,0.090) (102,0.083) (103,0.076) (104,0.076) (105,0.076) (106,0.067) (107,0.067) (108,0.067) (109,0.058) (110,0.058) (111,0.058) (112,0.058) (113,0.058) (114,0.058) (115,0.058) (116,0.058) (117,0.047) (118,0.047) (119,0.047) (120,0.047) (121,0.033) (122,0.033) (123,0.033) (124,0.033) (125,0.033) (126,0.033) (127,0.033) (128,0.033) (129,0.033) (130,0.033) (131,0.033) (132,0.033) (133,0.000) (190,0.000)
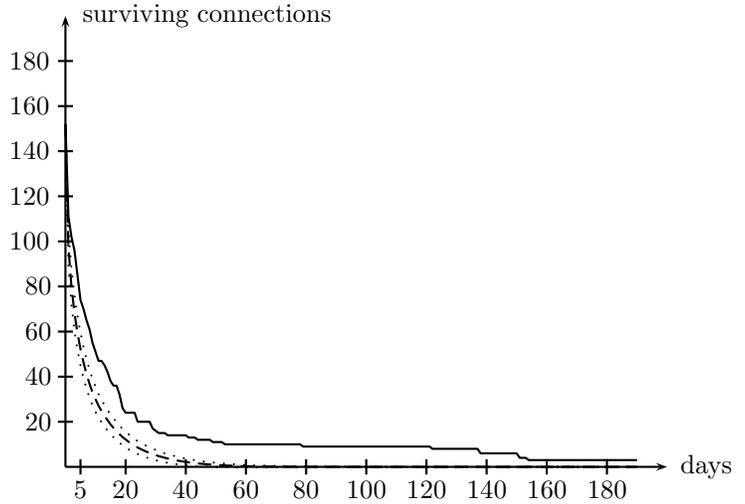
\rput(100,-34){\normalsize Figure 21: surviving connections as a function}
\rput(43,-49){\normalsize of time for $A+A^2$.}
\end{pspicture}
\end{minipage}

\vskip 0.2 cm

So, now we have a measure of how many connections (first and second order) survive through time. The problem is, we are measuring the survival of connections just for the first days of the year, so every connection is being judged by its behavior in a specific period of time. In order to fix that, one may measure the survivability of connections starting from different times during the year, or during a semester, since we are calculating the minimum spanning trees based on a half-yearly data. Then, one may sum over all the resulting measures of survavibility and obtain a number that states the percentage of times that a certain connection exists between two indices.

As an example, for the first semester of 2008, using a moving window of 60 days, the connections (first and second order) between S\&P and Nasdaq survive 100\% of the time. The connections (first and second order) between S\&P and Canada survive 49\% of the time, and those between the S\&P and Panama do not survive at all for five days in any period of time. We may hypothesise that there is a direct relation between survivability and the strengh of the connection (first order), given by one minus the distance calculated for the minimum spanning tree. The graph in figure 22 shows that this hypothesis is wrong: there are several connections that endure in time yet are not considered in the minimum spanning tree and many connections in the minimum spanning tree do not endure in time. The same type of graphic is obtained if one includes all correlations between the indices, and not just the ones relative to the connections that appear in the minimum spanning tree graph.

\begin{minipage}{8 cm }
\begin{pspicture}(-0.5,-0.25)(6,5.4)
\psset{xunit=4,yunit=4} \small \psline{->}(0,0)(1.2,0) \psline{->}(0,0)(0,1.2) \rput(1.45,0){correlation} \rput(0.3,1.2){survivability}
\psline(0.2,-0.025)(0.2,0.025) \psline(0.4,-0.025)(0.4,0.025) \psline(0.6,-0.025)(0.6,0.025) \psline(0.8,-0.025)(0.8,0.025) \psline(1,-0.025)(1,0.025) \rput(0.2,-0.075){0.2} \rput(0.4,-0.075){0.4} \rput(0.6,-0.075){0.6} \rput(0.8,-0.075){0.8} \rput(1,-0.075){1} \psline(-0.025,0.2)(0.025,0.2) \psline(-0.025,0.4)(0.025,0.4) \psline(-0.025,0.6)(0.025,0.6) \psline(-0.025,0.8)(0.025,0.8) \psline(-0.025,1)(0.025,1) \rput(-0.085,0.2){0.2} \rput(-0.085,0.4){0.4} \rput(-0.085,0.6){0.6} \rput(-0.085,0.8){0.8} \rput(-0.085,1){1}
\psdots(0.000,0.000) (0.909,1.000) (0.683,0.492) (0.635,0.410) (0.000,0.033) (0.000,0.311) (0.000,0.049) (0.000,0.016) (0.000,0.230) (0.000,0.033) (0.000,0.180) (0.000,0.180) (0.000,0.082) (0.677,0.721) (0.000,0.148) (0.000,0.115) (0.310,0.000) (0.000,0.262) (0.000,0.525) (0.000,0.262) (0.000,0.098) (0.280,0.000) (0.212,0.623) (0.000,0.131) (0.000,0.197) (0.241,0.000) (0.000,0.049) (0.000,0.049) (0.000,0.049) (0.300,0.148) (0.000,0.377) (0.000,0.033) (0.187,0.246) (0.230,0.557) (0.150,0.000) (0.650,0.689) (0.602,0.000) (0.000,0.016) (0.000,0.197) (0.000,0.016) (0.000,0.295) (0.000,0.311) (0.544,0.000) (0.000,0.213) (0.000,0.164) (0.000,0.197) (0.000,0.115) (0.000,0.131) (0.486,0.410) (0.202,0.311) (0.000,0.033) (0.000,0.066) (0.000,0.016) (0.000,0.066) (0.000,0.098) (0.228,0.197) (0.000,0.393) (0.587,0.279) (0.000,0.197) (0.000,0.016) (0.746,0.164) (0.891,0.393) (0.000,0.016) (0.000,0.082) (0.000,0.213) (0.000,0.016) (0.000,0.180) (0.000,0.131) (0.000,0.131) (0.000,0.246) (0.000,0.115) (0.000,0.016) (0.000,0.148) (0.000,0.344) (0.000,0.016) (0.934,0.885) (0.835,0.066) (0.816,0.000) (0.874,0.033) (0.865,0.475) (0.928,1.000) (0.883,0.164) (0.000,0.016) (0.000,0.066) (0.000,0.115) (0.534,0.426) (0.000,0.213) (0.000,0.475) (0.000,0.770) (0.000,0.164) (0.000,0.016) (0.000,0.115) (0.903,0.656) (0.673,0.000) (0.000,0.131) (0.000,0.049) (0.000,0.066) (0.664,0.000) (0.796,1.000) (0.000,0.016) (0.690,0.131) (0.000,0.016) (0.000,0.262) (0.000,0.033) (0.000,0.262) (0.000,0.066) (0.000,0.082) (0.000,0.098) (0.156,0.000) (0.000,0.197) (0.771,1.000) (0.000,0.246) (0.000,0.148) (0.000,0.148) (0.423,0.066) (0.000,0.311) (0.000,0.131) (0.000,0.082) (0.000,0.049) (0.000,0.016) (0.761,0.033) (0.838,0.262) (0.000,0.279) (0.763,0.361) (0.000,0.033) (0.000,0.115) (0.740,0.230) (0.000,0.082) (0.000,0.016) (0.000,0.279) (0.289,0.279) (0.582,0.000) (0.000,0.033) (0.000,0.016) (0.000,0.475) (0.614,0.836) (0.831,1.000) (0.000,0.016) (0.523,0.918) (0.000,0.098) (0.589,0.148) (0.685,0.230) (0.000,0.361) (0.254,0.082) (0.000,0.311) (0.000,0.230) (0.177,0.000) (0.000,0.033) (0.000,0.016) (0.232,0.443) (0.241,0.082) (0.000,0.131) (0.000,0.180) (0.000,0.066) (0.449,0.803) (0.000,0.049) (0.277,0.000) (0.287,0.656) (0.000,0.016) (0.155,0.000) (0.000,0.082) (0.000,0.230) (0.000,0.164) (0.000,0.033) (0.000,0.098) (0.000,0.115) (0.000,0.148) (0.272,0.000) (0.244,0.180) (0.000,0.377) (0.286,0.016) (0.000,0.393) (0.304,0.098) (0.000,0.492) (0.542,0.393) (0.000,0.180) (0.321,0.492) (0.000,0.033) (0.000,0.033) (0.000,0.180) (0.378,0.082) (0.000,0.557) (0.000,0.311) (0.000,0.148) (0.000,0.148) (0.114,0.000) (0.000,0.016) (0.000,0.344) (0.476,0.738) (0.000,0.066) (0.276,0.000) (0.000,0.262) (0.417,0.230) (0.000,0.180) (0.000,0.066) (0.000,0.197) (0.000,0.033) (0.000,0.148) (0.000,0.279) (0.000,0.197) (0.364,0.508) (0.000,0.098) (0.253,0.000) (0.000,0.230) (0.000,0.115) (0.000,0.131) (0.000,0.295) (0.000,0.033) (0.000,0.049) (0.343,0.295) (0.000,0.393) (0.000,0.082) (0.000,0.033) (0.000,0.033) (0.394,0.492) (0.000,0.115) (0.000,0.361) (0.000,0.082) (0.405,0.016) (0.393,0.000) (0.000,0.295) (0.345,0.000) (0.000,0.098) (0.000,0.115) (0.000,0.033) (0.000,0.016) (0.298,0.000) (0.000,0.426) (0.531,0.262) (0.000,0.279) (0.202,0.000) (0.000,0.049) (0.756,1.000) (0.448,0.148) (0.000,0.361) (0.000,0.213) (0.000,0.016) (0.000,0.311) (0.348,0.000) (0.000,0.574) (0.441,0.803) (0.702,1.000) (0.316,0.000) (0.282,0.033) (0.617,1.000) (0.661,0.033) (0.000,0.049) (0.615,0.066) (0.489,0.066) (0.000,0.016) (0.000,0.377) (0.446,0.066) (0.470,0.541) (0.000,0.016) (0.000,0.492) (0.851,1.000)
\end{pspicture}

\vskip 0.2 cm

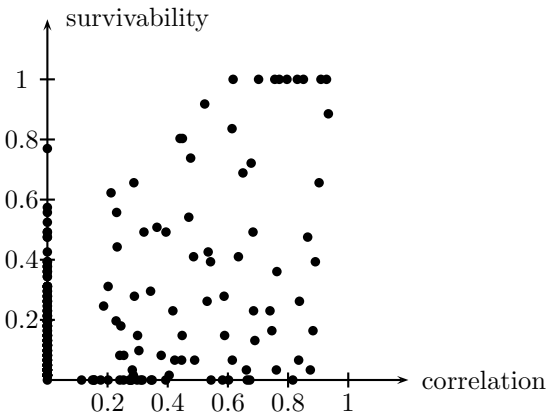
Figure 22: survivability of connections as a function of their Spearman rank correlation.
\end{minipage} \hskip 0.4 cm
\begin{minipage}{9 cm }
\begin{pspicture}(-1.5,2.5)(5.5,6)
\psset{xunit=5,yunit=0.0125} \small \psline{->}(0,0)(1.2,0) \psline{->}(0,0)(0,450) \rput(0.6,-57.6){survivability (threshold)} \rput(0.37,450){surviving connections}
\psline(0.2,-8)(0.2,8) \psline(0.4,-8)(0.4,8) \psline(0.6,-8)(0.6,8) \psline(0.8,-8)(0.8,8) \psline(1,-8)(1,8) \rput(0.2,-24){0.2} \rput(0.4,-24){0.4} \rput(0.6,-24){0.6} \rput(0.8,-24){0.8} \rput(1,-24){1} \psline(-0.02,100)(0.02,100) \psline(-0.02,200)(0.02,200) \psline(-0.02,300)(0.02,300) \psline(-0.02,400)(0.02,400) \rput(-0.09,100){100} \rput(-0.09,200){200} \rput(-0.09,300){300} \rput(-0.09,400){400}
\psline(0,387)(0.1,225)(0.2,148)(0.3,100)(0.4,68)(0.5,45)(0.6,31)(0.7,24)(0.8,20)(0.9,13)(1,10)
\psdots[linecolor=red](0,387)(0.1,225)(0.2,148)(0.3,100)(0.4,68)(0.5,45)(0.6,31)(0.7,24)(0.8,20)(0.9,13)(1,10)
\psline[linestyle=dashed](0,386.62)(0.1,172.05)(0.2,91.37)(0.3,50.94)(0.4,28.77)(0.5,16.02)(0.6,8.23)(0.7,4.14)
(0.8,1.86)(0.9,0.62)(1,0.17)
\psdots(0,386.62)(0.1,172.05)(0.2,91.37)(0.3,50.94)(0.4,28.77)(0.5,16.02)(0.6,8.23)(0.7,4.14)
(0.8,1.86)(0.9,0.62)(1,0.17)
\psline(0,371.9225255)(0,401.3174745) \psline(0.1,162.3771963)(0.1,181.7228037) \psline(0.2,84.22411046)(0.2,98.51588954) \psline(0.3,44.89501711)(0.3,56.98498289) \psline(0.4,23.6732492)(0.4,33.8667508) \psline(0.5,11.73240926)(0.5,20.30759074) \psline(0.6,5.489156659)(0.6,10.97084334) \psline(0.7,2.196245433)(0.7,6.083754567) \psline(0.8,0.687762219)(0.8,3.032237781) \psline(0.9,-0.115533888)(0.9,1.355533888) \psline(1,-0.257702342)(1,0.597702342)
\end{pspicture}

\vskip 3.4 cm

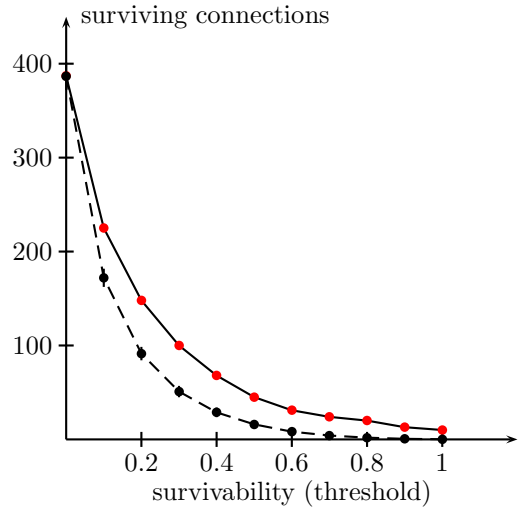
Figure 23: surviving connections as a function of minimum survivability.
\end{minipage}

\vskip 0.3 cm

Two examples help shed some light into this relationship (or lack of relationship) between survivability and correlation. One of the dots of the graph in figure 22 corresponds to the connection Germany-Italy, which has survivability 77\% but does not appear in the minimum spanning tree. The connection France-Italy, which has survivability 3\%, appears in the minimum spanning tree. The reason for that is that the correlation between France and Italy is slightly larger, 0.874 against 0.868, than the correlation between Germany and Italy. So, when the minimum spanning tree is being built, the link France-Italy is preferable to the link Germany-Italy. Another interesting example is the link Costa Rica - Montenegro, which has survivability 62\% but correlation 0.21.

So, survivability and correlation do not go hand in hand: there are strong connections that do not survive long and weak connections that survive for longer periods of time, although connections that are stronger tend to survive longer. We must now decide if survivability may be used in order to reduce the size of a minimum spanning tree without damaging links that are strong or enduring in time, and at the same time maintaining the bi-dimensional, non-intersecting characteristics that make the minimum spanning tree such a valuable representation of the information in a network.

Figure 23 shows the number of surviving connections of the minimum spanning tree built for the first semester of 2008 as a function of a survivability threshold (solid line). The first dot corresponds to the number of all connections of first and second orders in the minimum spanning tree (387 connections), the second dot corresponds to the number of connections that survive at least in 10\% of the windows considered for at least five days, and so on.

The only connections with 100\% survivability and which exist in the minimum spanning tree are S\&P - Nasdaq, France - Netherlands, Austria - Czech Republic, Belgium - Luxembourg, Greece - Cyprus, Japan - South Korea, Hong Kong - Singapore, Taiwan - South Korea, and Namibia - South Africa, all of them strong connections. There is no connection that is weak and that survives 100\% of the time.

On the same graph of figure 23, we plot (dashed line) the survivability of connections of the average of 100 simulations with randomized data (the error bars are displayed in the figure, but are too small to be seen). One can see that the survivability of connections obtained from randomized data decreases more rapidly than the survivability obtained from the real data. One may use this graph in order to decide which is the minimum survivability requirement for a connection to be considered real. The proximity of the two curves does not make it an easy task.

Some light may be shed if one studies the survivability of the connections that are above the distance threshold obtained by considering randomized data. For the first semester of 2008, that threshold was $0.68$, indicating that distances above this value in the minimum spanning tree have a stronger possibility of being due to noise. Above this threshold, the connections that survive the most are those between Slovenia and Palestine, and between Costa Rica and Montenegro, which survive 60\% of the time, both of them being unlikely connections if one considers other factors others than the correlation between the indices. Only strong connections (distances below $0.68$) survive more than 70\% of the time. For the data concerning other years, weak correlations may survive more than 80\% of the time.

We shall then consider the following rule: if a connection is below the distance threshold established by using randomized data, then it shall be considered strong and real; if the connection is above the distance threshold but survives more than 80\% of the time being considered in the computation of the minimum spanning tree, then it shall also be considered real, for it is unlikely that a random connection would survive to such extent; otherwise, the connection is considered as due to random noise and removed from the graph. This is obviously a somewhat biased procedure, although based on the comparison of results between real and randomized data, and by careful analysis of the types of connections that are made below that threshold, and it is the main result we are extracting from our calculations so far.

Similar behavior occurs for other data, sampled from different time intervals. By establishing the same rules for defining which connections are strong and which are not, we then obtain the following pruned minimum spanning trees (figures 24 to 39). The only weak connections that survive the pruning procedure are those of Hong Kong - Malaysia for the second semester of 1986, Nasdaq-Netherlands and Netherlands-Japan for the first semester of 1987, Chile-Morocco and Malta-Tunisia for the first semester of 2000, and China-Tunisia for the second semester of 2000.



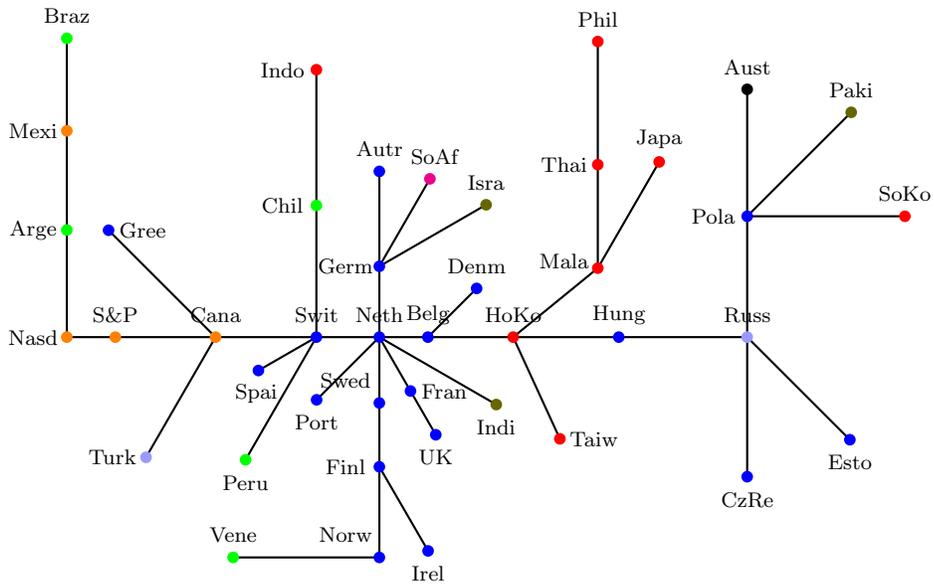
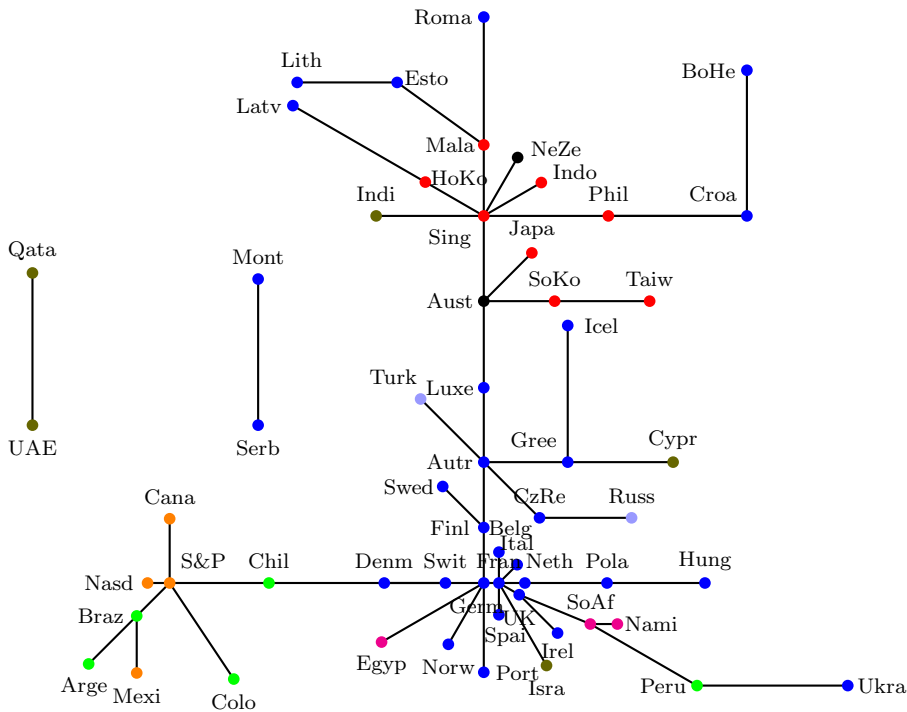
\begin{center}
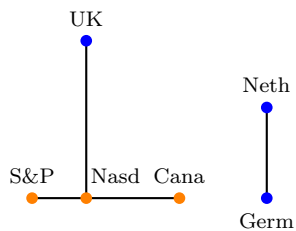
Figure 24: pruned minimum spanning tree for the first semester of 1986.
\end{center}

%

\begin{center}
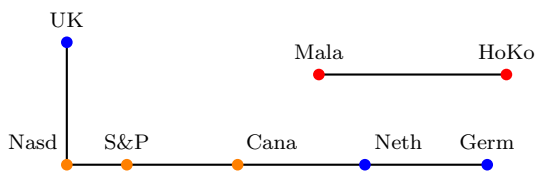
Figure 25: pruned minimum spanning tree for the second semester of 1986.
\end{center}

%

\begin{center}
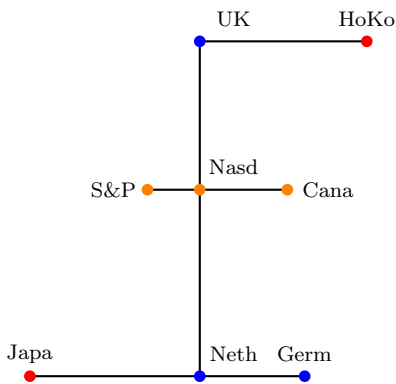
Figure 26: pruned minimum spanning tree for the first semester of 1987.
\end{center}

%

\begin{center}
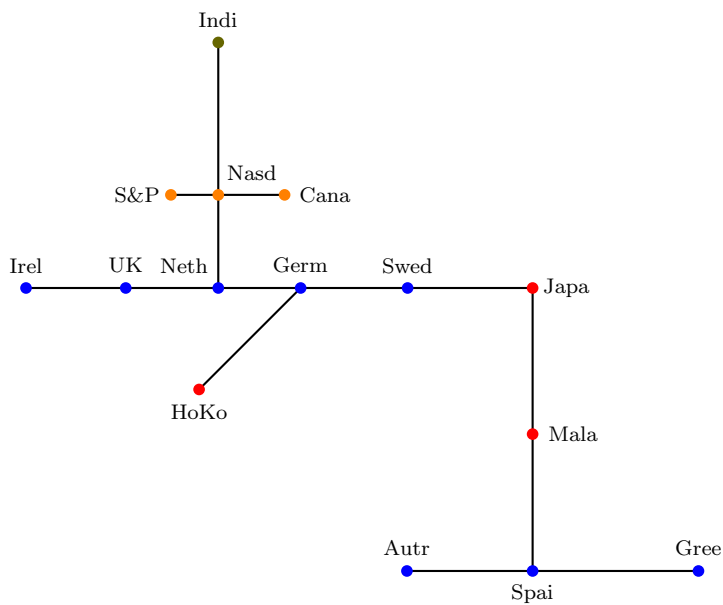
Figure 27: pruned minimum spanning tree for the second semester of 1987.
\end{center}

%

\begin{center}
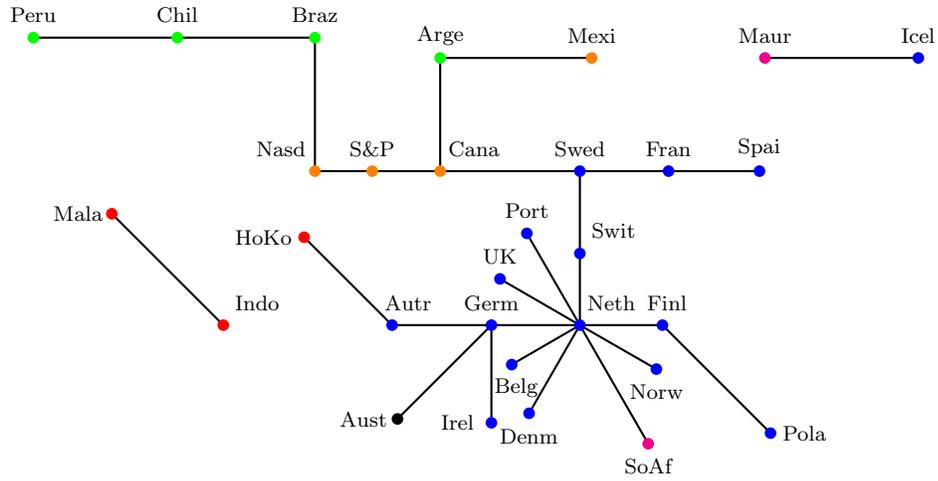
Figure 28: pruned minimum spanning tree for the first semester of 1997.
\end{center}

%

\begin{center}
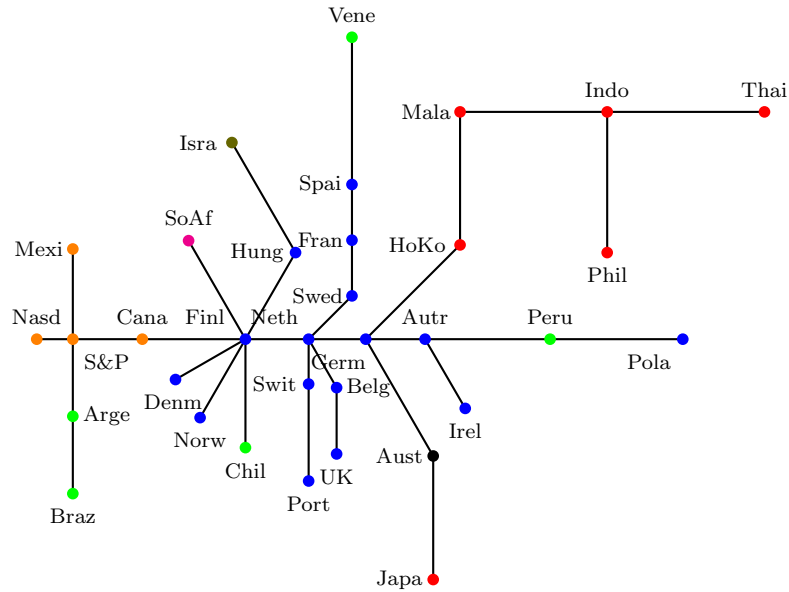
Figure 29: pruned minimum spanning tree for the second semester of 1997.
\end{center}

%

\vskip 0.3 cm

\noindent Figure 36: pruned minimum spanning tree for the first semester of 2007. The inset shows an amplification of the connections around France and Germany.

\vskip -8 cm
\hskip 11 cm
\begin{minipage}{4 cm }
%

\vskip 0.4 cm

\noindent Figure 37: pruned minimum spanning tree for the second semester of 2007. The inset shows an amplification of the connections closest to France.

\vskip -6 cm
\hskip 10.5 cm
\begin{minipage}{4 cm }
%

\noindent Figure 38: pruned minimum spanning tree for the first semester of 2008. The inset shows an amplification of the connections closest to France.

\vskip -9 cm
\hskip 12 cm
\begin{minipage}{4 cm }
%

\vskip 0.3 cm

\noindent Figure 39: pruned minimum spanning tree for the second semester of 2008. The inset shows an amplification of the connections closest to France.

\vskip -10 cm
\hskip 13.3 cm
\begin{minipage}{4 cm }
%
\end{minipage}

\vskip 8.4 cm

Note that the networks (many are not trees anymore) are much more compact, and most connections make sense if one considers other factors like geographic proximity, culture affinity, and economic ties between countries. Once again, one may notice that the networks shrink substantially during periods of international financial crises.

A last discussion shall be made in this section. The approach of choosing only strongly connected nodes, and also those weakly connected ones that endure in time, has produced graphs that are not substantially different from the ones we could have obtained by just pruning away every weakly interacting nodes. As checking for survivability is a time consuming task, most of the times there will be no real damage in pruning a minimum spanning tree using only the first criterium. Nevertheless, other types of networks, based solely on survivability, may also be built, with their own merits and maladies.

\section{Centrality measures}

A substantial part of network theory is devoted to the idea of the centrality of a node \cite{Newman}, or how influential the vertex is in the network. It is an important measurement which is handled in a number of diferent ways. In what follows, I perform an analysis of the centrality of vertices in the networks depicted by the minimum spanning trees according to three different definitions of centrality. I also do some analysis of the frequency distribution of each centrality measure in the network and which are the stocks that are more central according to each definition.

\subsection{Node degree}

The node degree of a node, or stock exchange index in this article, is the number of connections it has in the network. Most of the indices have low node degree, and some of them have a large node degree associated with them. The latter are called hubs, and are generally nodes that are more important in events that can change the network. Figures 40 to 43 show the evolution in the four different crisis periods whe are studying of the frequency distribution of the node degrees of the indices in the original minimum spanning trees (not the pruned ones). The graphs for the pruned networks are nearly identical to these ones. Note that the frequencies all drop exponentialy, as it would be expected from scale free networks.

During the first crisis, 1986/1987, the indices with the largest node degree are Nasdaq and Japan (both with node degree 5), S\&P, Canada, Netherlands, and Ireland (all with node degree 4); for 1997/1998, the indices with largest node degree are the Netherlands and Finland; for 200/2001, we have France, Netherlands, and Hong Kong occupying the first positions; for 2007/2008, this role is played by France, Germany, Singapore, and Australia.

\vskip 0.3 cm

\begin{minipage}{8 cm}
\begin{pspicture}(-4,-2.1)(1,1.7)
\psset{xunit=1,yunit=1} \scriptsize
\pstThreeDLine[linecolor=black](0,0,0)(0,0,1.875)(0,0.3,1.875)(0,0.3,0)(0,0,0)
\pstThreeDLine[linecolor=black](0,0.3,0)(0,0.3,0.375)(0,0.6,0.375)(0,0.6,0)(0,0.3,0)
\pstThreeDLine[linecolor=black](0,0.6,0)(0,0.6,0.375)(0,0.9,0.375)(0,0.9,0)(0,0.6,0)
\pstThreeDLine[linecolor=black](0,1.2,0)(0,1.2,0.375)(0,1.5,0.375)(0,1.5,0)(0,1.2,0)
\pstThreeDPut(0,2.1,0){01/1986}
\pstThreeDLine[linecolor=black](1,0,0)(1,0,1.3125)(1,0.3,1.3125)(1,0.3,0)(1,0,0)
\pstThreeDLine[linecolor=black](1,0.3,0)(1,0.3,1.125)(1,0.6,1.125)(1,0.6,0)(1,0.3,0)
\pstThreeDLine[linecolor=black](1,0.6,0)(1,0.6,0.1875)(1,0.9,0.1875)(1,0.9,0)(1,0.6,0)
\pstThreeDLine[linecolor=black](1,0.9,0)(1,0.9,0.375)(1,1.2,0.375)(1,1.2,0)(1,0.9,0)
\pstThreeDPut(1,2.1,0){02/1986}
\pstThreeDLine[linecolor=black](2,0,0)(2,0,1.435)(2,0.3,1.435)(2,0.3,0)(2,0,0)
\pstThreeDLine[linecolor=black](2,0.3,0)(2,0.3,0.783)(2,0.6,0.783)(2,0.6,0)(2,0.3,0)
\pstThreeDLine[linecolor=black](2,0.6,0)(2,0.6,0.391)(2,0.9,0.391)(2,0.9,0)(2,0.6,0)
\pstThreeDLine[linecolor=black](2,0.9,0)(2,0.9,0.391)(2,1.2,0.391)(2,1.2,0)(2,0.9,0)
\pstThreeDPut(2,2.1,0){01/1987}
\pstThreeDLine[linecolor=black](3,0,0)(3,0,1.174)(3,0.3,1.174)(3,0.3,0)(3,0,0)
\pstThreeDLine[linecolor=black](3,0.3,0)(3,0.3,1.174)(3,0.6,1.174)(3,0.6,0)(3,0.3,0)
\pstThreeDLine[linecolor=black](3,0.6,0)(3,0.6,0.391)(3,0.9,0.391)(3,0.9,0)(3,0.6,0)
\pstThreeDLine[linecolor=black](3,0.9,0)(3,0.9,0.261)(3,1.2,0.261)(3,1.2,0)(3,0.9,0)
\pstThreeDPut(3,2.1,0){02/1987}
\end{pspicture}

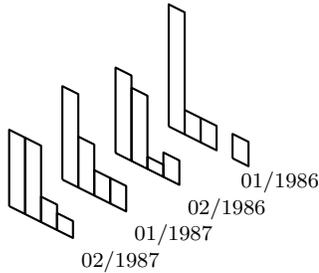
{\noindent \small Figure 40: probability distributions of the node degree for the first and second semesters of 1986 and 1987.}
\end{minipage}
\hskip 1.3 cm \begin{minipage}{8 cm}
\begin{pspicture}(-4,-2.7)(1,1.7)
\psset{xunit=1,yunit=1} \scriptsize
\pstThreeDLine[linecolor=black](0,0,0)(0,0,1.895)(0,0.3,1.895)(0,0.3,0)(0,0,0)
\pstThreeDLine[linecolor=black](0,0.3,0)(0,0.3,1.123)(0,0.6,1.123)(0,0.6,0)(0,0.3,0)
\pstThreeDLine[linecolor=black](0,0.6,0)(0,0.6,0.702)(0,0.9,0.702)(0,0.9,0)(0,0.6,0)
\pstThreeDLine[linecolor=black](0,0.9,0)(0,0.9,0.140)(0,1.2,0.140)(0,1.2,0)(0,0.9,0)
\pstThreeDLine[linecolor=black](0,1.2,0)(0,1.2,0.070)(0,1.5,0.070)(0,1.5,0)(0,1.2,0)
\pstThreeDLine[linecolor=black](0,2.7,0)(0,2.7,0.070)(0,3,0.070)(0,3,0)(0,2.7,0)
\pstThreeDPut(0,4,0){01/1997}
\pstThreeDLine[linecolor=black](1,0,0)(1,0,1.895)(1,0.3,1.895)(1,0.3,0)(1,0,0)
\pstThreeDLine[linecolor=black](1,0.3,0)(1,0.3,1.263)(1,0.6,1.263)(1,0.6,0)(1,0.3,0)
\pstThreeDLine[linecolor=black](1,0.6,0)(1,0.6,0.421)(1,0.9,0.421)(1,0.9,0)(1,0.6,0)
\pstThreeDLine[linecolor=black](1,0.9,0)(1,0.9,0.140)(1,1.2,0.140)(1,1.2,0)(1,0.9,0)
\pstThreeDLine[linecolor=black](1,1.2,0)(1,1.2,0.140)(1,1.5,0.140)(1,1.5,0)(1,1.2,0)
\pstThreeDLine[linecolor=black](1,1.5,0)(1,1.5,0.070)(1,1.8,0.070)(1,1.8,0)(1,1.5,0)
\pstThreeDLine[linecolor=black](1,1.8,0)(1,1.8,0.070)(1,2.1,0.070)(1,2.1,0)(1,1.8,0)
\pstThreeDPut(1,4,0){02/1997}
\pstThreeDLine[linecolor=black](2,0,0)(2,0,2.069)(2,0.3,2.069)(2,0.3,0)(2,0,0)
\pstThreeDLine[linecolor=black](2,0.3,0)(2,0.3,1.034)(2,0.6,1.034)(2,0.6,0)(2,0.3,0)
\pstThreeDLine[linecolor=black](2,0.6,0)(2,0.6,0.414)(2,0.9,0.414)(2,0.9,0)(2,0.6,0)
\pstThreeDLine[linecolor=black](2,0.9,0)(2,0.9,0.138)(2,1.2,0.138)(2,1.2,0)(2,0.9,0)
\pstThreeDLine[linecolor=black](2,1.2,0)(2,1.2,0.207)(2,1.5,0.207)(2,1.5,0)(2,1.2,0)
\pstThreeDLine[linecolor=black](2,1.5,0)(2,1.5,0.069)(2,1.8,0.069)(2,1.8,0)(2,1.5,0)
\pstThreeDLine[linecolor=black](2,1.8,0)(2,1.8,0.069)(2,2.1,0.069)(2,2.1,0)(2,1.8,0)
\pstThreeDPut(2,4,0){01/1998}
\pstThreeDLine[linecolor=black](3,0,0)(3,0,1.895)(3,0.3,1.895)(3,0.3,0)(3,0,0)
\pstThreeDLine[linecolor=black](3,0.3,0)(3,0.3,1.123)(3,0.6,1.123)(3,0.6,0)(3,0.3,0)
\pstThreeDLine[linecolor=black](3,0.6,0)(3,0.6,0.845)(3,0.9,0.845)(3,0.9,0)(3,0.6,0)
\pstThreeDLine[linecolor=black](3,0.9,0)(3,0.9,0.140)(3,1.2,0.140)(3,1.2,0)(3,0.9,0)
\pstThreeDLine[linecolor=black](3,3,0)(3,3,0.070)(3,3.3,0.070)(3,3.3,0)(3,3,0)
\pstThreeDPut(3,4,0){02/1998}
\end{pspicture}

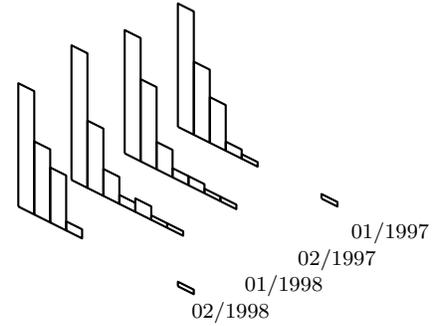
{\noindent \small Figure 41: probability distributions of the node degree for the first and second semesters of 1997 and 1998.}
\end{minipage}

\vskip 0.6 cm

\begin{minipage}{8 cm}
\begin{pspicture}(-4,-2.7)(1,1.7)
\psset{xunit=1,yunit=1} \scriptsize
\pstThreeDLine[linecolor=black](0,0,0)(0,0,1.784)(0,0.3,1.784)(0,0.3,0)(0,0,0)
\pstThreeDLine[linecolor=black](0,0.3,0)(0,0.3,1.243)(0,0.6,1.243)(0,0.6,0)(0,0.3,0)
\pstThreeDLine[linecolor=black](0,0.6,0)(0,0.6,0.649)(0,0.9,0.649)(0,0.9,0)(0,0.6,0)
\pstThreeDLine[linecolor=black](0,0.9,0)(0,0.9,0.162)(0,1.2,0.162)(0,1.2,0)(0,0.9,0)
\pstThreeDLine[linecolor=black](0,1.2,0)(0,1.2,0.054)(0,1.5,0.054)(0,1.5,0)(0,1.2,0)
\pstThreeDLine[linecolor=black](0,1.8,0)(0,1.8,0.108)(0,2.1,0.108)(0,2.1,0)(0,1.8,0)
\pstThreeDPut(0,4,0){01/2000}
\pstThreeDLine[linecolor=black](1,0,0)(1,0,1.784)(1,0.3,1.784)(1,0.3,0)(1,0,0)
\pstThreeDLine[linecolor=black](1,0.3,0)(1,0.3,1.351)(1,0.6,1.351)(1,0.6,0)(1,0.3,0)
\pstThreeDLine[linecolor=black](1,0.6,0)(1,0.6,0.541)(1,0.9,0.541)(1,0.9,0)(1,0.6,0)
\pstThreeDLine[linecolor=black](1,0.9,0)(1,0.9,0.162)(1,1.2,0.162)(1,1.2,0)(1,0.9,0)
\pstThreeDLine[linecolor=black](1,1.2,0)(1,1.2,0.054)(1,1.5,0.054)(1,1.5,0)(1,1.2,0)
\pstThreeDLine[linecolor=black](1,1.8,0)(1,1.8,0.054)(1,2.1,0.054)(1,2.1,0)(1,1.8,0)
\pstThreeDLine[linecolor=black](1,2.4,0)(1,2.4,0.054)(1,2.7,0.054)(1,2.7,0)(1,2.4,0)
\pstThreeDPut(1,4,0){02/2000}
\pstThreeDLine[linecolor=black](2,0,0)(2,0,2.069)(2,0.3,2.069)(2,0.3,0)(2,0,0)
\pstThreeDLine[linecolor=black](2,0.3,0)(2,0.3,1.034)(2,0.6,1.034)(2,0.6,0)(2,0.3,0)
\pstThreeDLine[linecolor=black](2,0.6,0)(2,0.6,0.414)(2,0.9,0.414)(2,0.9,0)(2,0.6,0)
\pstThreeDLine[linecolor=black](2,0.9,0)(2,0.9,0.138)(2,1.2,0.138)(2,1.2,0)(2,0.9,0)
\pstThreeDLine[linecolor=black](2,1.2,0)(2,1.2,0.207)(2,1.5,0.207)(2,1.5,0)(2,1.2,0)
\pstThreeDLine[linecolor=black](2,1.5,0)(2,1.5,0.069)(2,1.8,0.069)(2,1.8,0)(2,1.5,0)
\pstThreeDLine[linecolor=black](2,1.8,0)(2,1.8,0.069)(2,2.1,0.069)(2,2.1,0)(2,1.8,0)
\pstThreeDPut(2,4,0){01/2001}
\pstThreeDLine[linecolor=black](3,0,0)(3,0,2)(3,0.3,2)(3,0.3,0)(3,0,0)
\pstThreeDLine[linecolor=black](3,0.3,0)(3,0.3,0.919)(3,0.6,0.919)(3,0.6,0)(3,0.3,0)
\pstThreeDLine[linecolor=black](3,0.6,0)(3,0.6,0.646)(3,0.9,0.646)(3,0.9,0)(3,0.6,0)
\pstThreeDLine[linecolor=black](3,0.9,0)(3,0.9,0.216)(3,1.2,0.216)(3,1.2,0)(3,0.9,0)
\pstThreeDLine[linecolor=black](3,1.2,0)(3,1.2,0.162)(3,1.5,0.162)(3,1.5,0)(3,1.2,0)
\pstThreeDLine[linecolor=black](3,3,0)(3,3,0.054)(3,3.3,0.054)(3,3.3,0)(3,3,0)
\pstThreeDPut(3,4,0){02/2001}
\end{pspicture}

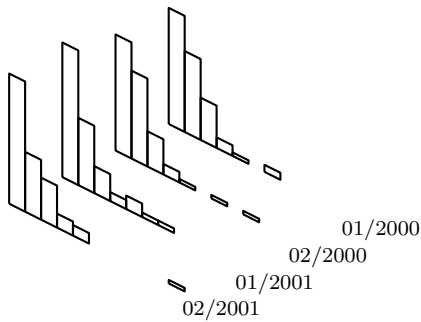
{\noindent \small Figure 42: probability distributions of the node degree for the first and second semesters of 2000 and 2001.}
\end{minipage}
\hskip 1.3 cm \begin{minipage}{8 cm}
\begin{pspicture}(-4,-2.7)(1,1.7)
\psset{xunit=1,yunit=1} \scriptsize
\pstThreeDLine[linecolor=black](0,0,0)(0,0,1.870)(0,0.3,1.870)(0,0.3,0)(0,0,0)
\pstThreeDLine[linecolor=black](0,0.3,0)(0,0.3,1.174)(0,0.6,1.174)(0,0.6,0)(0,0.3,0)
\pstThreeDLine[linecolor=black](0,0.6,0)(0,0.6,0.609)(0,0.9,0.609)(0,0.9,0)(0,0.6,0)
\pstThreeDLine[linecolor=black](0,0.9,0)(0,0.9,0.130)(0,1.2,0.130)(0,1.2,0)(0,0.9,0)
\pstThreeDLine[linecolor=black](0,1.2,0)(0,1.2,0.087)(0,1.5,0.087)(0,1.5,0)(0,1.2,0)
\pstThreeDLine[linecolor=black](0,1.8,0)(0,1.8,0.130)(0,2.1,0.130)(0,2.1,0)(0,1.8,0)
\pstThreeDPut(0,4.3,0){01/2007}
\pstThreeDLine[linecolor=black](1,0,0)(1,0,2.087)(1,0.3,2.087)(1,0.3,0)(1,0,0)
\pstThreeDLine[linecolor=black](1,0.3,0)(1,0.3,0.870)(1,0.6,0.870)(1,0.6,0)(1,0.3,0)
\pstThreeDLine[linecolor=black](1,0.6,0)(1,0.6,0.652)(1,0.9,0.652)(1,0.9,0)(1,0.6,0)
\pstThreeDLine[linecolor=black](1,0.9,0)(1,0.9,0.174)(1,1.2,0.174)(1,1.2,0)(1,0.9,0)
\pstThreeDLine[linecolor=black](1,1.2,0)(1,1.2,0.130)(1,1.5,0.130)(1,1.5,0)(1,1.2,0)
\pstThreeDLine[linecolor=black](1,1.5,0)(1,1.5,0.043)(1,1.8,0.043)(1,1.8,0)(1,1.5,0)
\pstThreeDLine[linecolor=black](1,3.3,0)(1,3.3,0.043)(1,3.6,0.043)(1,3.6,0)(1,3.3,0)
\pstThreeDPut(1,4.3,0){02/2007}
\pstThreeDLine[linecolor=black](2,0,0)(2,0,1.870)(2,0.3,1.870)(2,0.3,0)(2,0,0)
\pstThreeDLine[linecolor=black](2,0.3,0)(2,0.3,1.217)(2,0.6,1.217)(2,0.6,0)(2,0.3,0)
\pstThreeDLine[linecolor=black](2,0.6,0)(2,0.6,0.435)(2,0.9,0.435)(2,0.9,0)(2,0.6,0)
\pstThreeDLine[linecolor=black](2,0.9,0)(2,0.9,0.304)(2,1.2,0.304)(2,1.2,0)(2,0.9,0)
\pstThreeDLine[linecolor=black](2,1.2,0)(2,1.2,0.130)(2,1.5,0.130)(2,1.5,0)(2,1.2,0)
\pstThreeDLine[linecolor=black](2,2.7,0)(2,2.7,0.043)(2,3,0.043)(2,3,0)(2,2.7,0)
\pstThreeDLine[linecolor=black](2,3.3,0)(2,3.3,0.043)(2,3.6,0.043)(2,3.6,0)(2,3.3,0)
\pstThreeDPut(2,4.3,0){01/2008}
\pstThreeDLine[linecolor=black](3,0,0)(3,0,1.957)(3,0.3,1.957)(3,0.3,0)(3,0,0)
\pstThreeDLine[linecolor=black](3,0.3,0)(3,0.3,1.043)(3,0.6,1.043)(3,0.6,0)(3,0.3,0)
\pstThreeDLine[linecolor=black](3,0.6,0)(3,0.6,0.696)(3,0.9,0.696)(3,0.9,0)(3,0.6,0)
\pstThreeDLine[linecolor=black](3,0.9,0)(3,0.9,0.087)(3,1.2,0.087)(3,1.2,0)(3,0.9,0)
\pstThreeDLine[linecolor=black](3,1.2,0)(3,1.2,0.043)(3,1.5,0.043)(3,1.5,0)(3,1.2,0)
\pstThreeDLine[linecolor=black](3,1.5,0)(3,1.5,0.087)(3,1.8,0.087)(3,1.8,0)(3,1.5,0)
\pstThreeDLine[linecolor=black](3,1.8,0)(3,1.8,0.043)(3,2.1,0.043)(3,2.1,0)(3,1.8,0)
\pstThreeDPut(3,4.3,0){02/2008}
\end{pspicture}

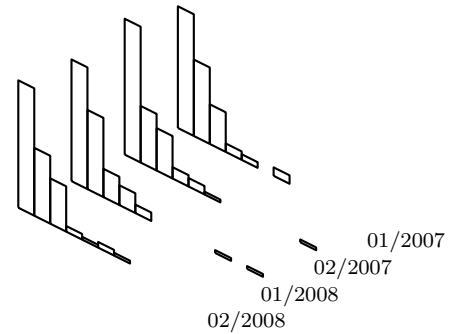
{\noindent \small Figure 43: probability distributions of the node degree for the first and second semesters of 2007 and 2008.}
\end{minipage}

\vskip 1 cm

\subsection{Node strength}

The strength of a node is the sum of the correlations of the node with all other nodes to which it is connected. If $C$ is the matrix that stores the correlations between nodes that are linked in the minimum spanning tree, then the node strength is given by
\begin{equation}N_s^k=\sum_{i=1}^nC_{ik}+\sum_{j=1}^nC_{kj}\ ,\end{equation}where $C_{ij}$ is an element of matrix $C$.

Figures 44 to 47 show the evolution in the four different crisis periods of the frequency distribution of the node strengths of the indices in the original minimum spanning trees. The exponential drop seen for the node degrees distributions is now substituted by a strongly tilted distribution which is not exponentially decreasing.

\vskip 0.3 cm

\begin{minipage}{8 cm}
\begin{pspicture}(-3,-2.4)(1,2)
\psset{xunit=1,yunit=1} \scriptsize
\pstThreeDLine[linecolor=black](0,0.27,0)(0,0.27,0.313)(0,0.54,0.313)(0,0.54,0)(0,0.27,0)
\pstThreeDLine[linecolor=black](0,0.54,0)(0,0.54,2.500)(0,0.81,2.500)(0,0.81,0)(0,0.54,0)
\pstThreeDLine[linecolor=black](0,0.81,0)(0,0.81,0.313)(0,1.08,0.313)(0,1.08,0)(0,0.81,0)
\pstThreeDLine[linecolor=black](0,1.08,0)(0,1.08,0.313)(0,1.35,0.313)(0,1.35,0)(0,1.08,0)
\pstThreeDLine[linecolor=black](0,1.35,0)(0,1.35,0.313)(0,1.62,0.313)(0,1.62,0)(0,1.35,0)
\pstThreeDLine[linecolor=black](0,1.62,0)(0,1.62,0.625)(0,1.89,0.625)(0,1.89,0)(0,1.62,0)
\pstThreeDLine[linecolor=black](0,3.24,0)(0,3.24,0.313)(0,3.51,0.313)(0,3.51,0)(0,3.24,0)
\pstThreeDLine[linecolor=black](0,3.51,0)(0,3.51,0.313)(0,3.78,0.313)(0,3.78,0)(0,3.51,0)
\pstThreeDPut(0,5,0){01/1986}
\pstThreeDLine[linecolor=black](1,0.27,0)(1,0.27,0.313)(1,0.54,0.313)(1,0.54,0)(1,0.27,0)
\pstThreeDLine[linecolor=black](1,0.54,0)(1,0.54,1.875)(1,0.81,1.875)(1,0.81,0)(1,0.54,0)
\pstThreeDLine[linecolor=black](1,0.81,0)(1,0.81,0.313)(1,1.08,0.313)(1,1.08,0)(1,0.81,0)
\pstThreeDLine[linecolor=black](1,1.08,0)(1,1.08,0.313)(1,1.35,0.313)(1,1.35,0)(1,1.08,0)
\pstThreeDLine[linecolor=black](1,1.35,0)(1,1.35,0.938)(1,1.62,0.938)(1,1.62,0)(1,1.35,0)
\pstThreeDLine[linecolor=black](1,1.62,0)(1,1.62,0.625)(1,1.89,0.625)(1,1.89,0)(1,1.62,0)
\pstThreeDLine[linecolor=black](1,1.89,0)(1,1.89,0.313)(1,2.16,0.313)(1,2.16,0)(1,1.89,0)
\pstThreeDLine[linecolor=black](1,2.43,0)(1,2.43,0.313)(1,2.70,0.313)(1,2.70,0)(1,2.43,0)
\pstThreeDPut(1,3.4,0){02/1986}
\pstThreeDLine[linecolor=black](2,0,0)(2,0,0.217)(2,0.27,0.217)(2,0.27,0)(2,0,0)
\pstThreeDLine[linecolor=black](2,0.27,0)(2,0.27,0.217)(2,0.54,0.217)(2,0.54,0)(2,0.27,0)
\pstThreeDLine[linecolor=black](2,0.54,0)(2,0.54,1.957)(2,0.81,1.957)(2,0.81,0)(2,0.54,0)
\pstThreeDLine[linecolor=black](2,1.08,0)(2,1.08,0.217)(2,1.35,0.217)(2,1.35,0)(2,1.08,0)
\pstThreeDLine[linecolor=black](2,1.35,0)(2,1.35,1.087)(2,1.62,1.087)(2,1.62,0)(2,1.35,0)
\pstThreeDLine[linecolor=black](2,1.89,0)(2,1.89,0.435)(2,2.16,0.435)(2,2.16,0)(2,1.89,0)
\pstThreeDLine[linecolor=black](2,2.16,0)(2,2.16,0.435)(2,2.43,0.435)(2,2.43,0)(2,2.16,0)
\pstThreeDLine[linecolor=black](2,2.43,0)(2,2.43,0.217)(2,2.70,0.217)(2,2.70,0)(2,2.42,0)
\pstThreeDLine[linecolor=black](2,2.70,0)(2,2.70,0.217)(2,2.97,0.217)(2,2.97,0)(2,2.70,0)
\pstThreeDPut(2,3.7,0){01/1987}
\pstThreeDLine[linecolor=black](3,0,0)(3,0,0.217)(3,0.27,0.217)(3,0.27,0)(3,0,0)
\pstThreeDLine[linecolor=black](3,0.27,0)(3,0.27,0.217)(3,0.54,0.217)(3,0.54,0)(3,0.27,0)
\pstThreeDLine[linecolor=black](3,0.54,0)(3,0.54,1.739)(3,0.81,1.739)(3,0.81,0)(3,0.54,0)
\pstThreeDLine[linecolor=black](3,0.81,0)(3,0.81,0.652)(3,1.08,0.652)(3,1.08,0)(3,0.81,0)
\pstThreeDLine[linecolor=black](3,1.08,0)(3,1.08,0.652)(3,1.35,0.652)(3,1.35,0)(3,1.08,0)
\pstThreeDLine[linecolor=black](3,1.35,0)(3,1.35,0.870)(3,1.62,0.870)(3,1.62,0)(3,1.35,0)
\pstThreeDLine[linecolor=black](3,1.62,0)(3,1.62,0.217)(3,1.89,0.217)(3,1.89,0)(3,1.62,0)
\pstThreeDLine[linecolor=black](3,1.89,0)(3,1.89,0.435)(3,2.16,0.435)(3,2.16,0)(3,1.89,0)
\pstThreeDPut(3,2.8,0){02/1987}
\end{pspicture}

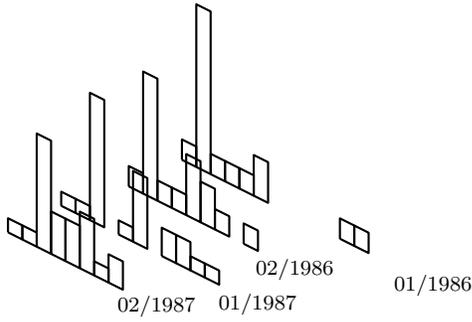
{\noindent \small Figure 44: probability distributions of the node strength for the first and second semesters of 1986 and 1987.}
\end{minipage}
\hskip 1.3 cm \begin{minipage}{8 cm}
\begin{pspicture}(-3,-2.7)(1,1.7)
\psset{xunit=1,yunit=1} \scriptsize
\pstThreeDLine[linecolor=black](0,0.27,0)(0,0.27,0.614)(0,0.54,0.614)(0,0.54,0)(0,0.27,0)
\pstThreeDLine[linecolor=black](0,0.54,0)(0,0.54,1.930)(0,0.81,1.930)(0,0.81,0)(0,0.54,0)
\pstThreeDLine[linecolor=black](0,0.81,0)(0,0.81,0.175)(0,1.08,0.175)(0,1.08,0)(0,0.81,0)
\pstThreeDLine[linecolor=black](0,1.08,0)(0,1.08,0.877)(0,1.35,0.877)(0,1.35,0)(0,1.08,0)
\pstThreeDLine[linecolor=black](0,1.35,0)(0,1.35,0.526)(0,1.62,0.526)(0,1.62,0)(0,1.35,0)
\pstThreeDLine[linecolor=black](0,1.62,0)(0,1.62,0.263)(0,1.89,0.263)(0,1.89,0)(0,1.62,0)
\pstThreeDLine[linecolor=black](0,1.89,0)(0,1.89,0.263)(0,2.16,0.263)(0,2.16,0)(0,1.89,0)
\pstThreeDLine[linecolor=black](0,2.16,0)(0,2.16,0.175)(0,2.43,0.175)(0,2.43,0)(0,2.16,0)
\pstThreeDLine[linecolor=black](0,3.24,0)(0,3.24,0.088)(0,3.51,0.088)(0,3.51,0)(0,3.24,0)
\pstThreeDLine[linecolor=black](0,4.05,0)(0,4.05,0.088)(0,4.32,0.088)(0,4.32,0)(0,4.05,0)
\pstThreeDPut(0,5,0){01/1997}
\pstThreeDLine[linecolor=black](1,0,0)(1,0,0.175)(1,0.27,0.175)(1,0.27,0)(1,0,0)
\pstThreeDLine[linecolor=black](1,0.27,0)(1,0.27,0.877)(1,0.54,0.877)(1,0.54,0)(1,0.27,0)
\pstThreeDLine[linecolor=black](1,0.54,0)(1,0.54,1.842)(1,0.81,1.842)(1,0.81,0)(1,0.54,0)
\pstThreeDLine[linecolor=black](1,0.81,0)(1,0.81,0.351)(1,1.08,0.351)(1,1.08,0)(1,0.81,0)
\pstThreeDLine[linecolor=black](1,1.08,0)(1,1.08,0.614)(1,1.35,0.614)(1,1.35,0)(1,1.08,0)
\pstThreeDLine[linecolor=black](1,1.35,0)(1,1.35,0.614)(1,1.62,0.614)(1,1.62,0)(1,1.35,0)
\pstThreeDLine[linecolor=black](1,1.62,0)(1,1.62,0.088)(1,1.89,0.088)(1,1.89,0)(1,1.62,0)
\pstThreeDLine[linecolor=black](1,1.89,0)(1,1.89,0.088)(1,2.16,0.088)(1,2.16,0)(1,1.89,0)
\pstThreeDLine[linecolor=black](1,2.16,0)(1,2.16,0.088)(1,2.43,0.088)(1,2.43,0)(1,2.16,0)
\pstThreeDLine[linecolor=black](1,2.43,0)(1,2.43,0.088)(1,2.70,0.088)(1,2.70,0)(1,2.43,0)
\pstThreeDLine[linecolor=black](1,2.97,0)(1,2.97,0.088)(1,3.24,0.088)(1,3.24,0)(1,2.970,0)
\pstThreeDLine[linecolor=black](1,3.78,0)(1,3.78,0.088)(1,4.05,0.088)(1,4.05,0)(1,3.78,0)
\pstThreeDPut(1,5,0){02/1997}
\pstThreeDLine[linecolor=black](2,0,0)(2,0,0.172)(2,0.27,0.172)(2,0.27,0)(2,0,0)
\pstThreeDLine[linecolor=black](2,0.27,0)(2,0.27,1.121)(2,0.54,1.121)(2,0.54,0)(2,0.27,0)
\pstThreeDLine[linecolor=black](2,0.54,0)(2,0.54,1.724)(2,0.81,1.724)(2,0.81,0)(2,0.54,0)
\pstThreeDLine[linecolor=black](2,0.81,0)(2,0.81,0.431)(2,1.08,0.431)(2,1.08,0)(2,0.81,0)
\pstThreeDLine[linecolor=black](2,1.08,0)(2,1.08,0.431)(2,1.35,0.431)(2,1.35,0)(2,1.08,0)
\pstThreeDLine[linecolor=black](2,1.35,0)(2,1.35,0.431)(2,1.62,0.431)(2,1.62,0)(2,1.35,0)
\pstThreeDLine[linecolor=black](2,1.62,0)(2,1.62,0.086)(2,1.89,0.086)(2,1.89,0)(2,1.62,0)
\pstThreeDLine[linecolor=black](2,1.89,0)(2,1.89,0.172)(2,2.16,0.172)(2,2.16,0)(2,1.89,0)
\pstThreeDLine[linecolor=black](2,2.16,0)(2,2.16,0.172)(2,2.43,0.172)(2,2.43,0)(2,2.16,0)
\pstThreeDLine[linecolor=black](2,2.43,0)(2,2.43,0.086)(2,2.70,0.086)(2,2.70,0)(2,2.43,0)
\pstThreeDLine[linecolor=black](2,2.70,0)(2,2.70,0.172)(2,2.97,0.172)(2,2.97,0)(2,2.70,0)
\pstThreeDPut(2,3.7,0){01/1998}
\pstThreeDLine[linecolor=black](3,0,0)(3,0,0.603)(3,0.27,0.603)(3,0.27,0)(3,0,0)
\pstThreeDLine[linecolor=black](3,0.27,0)(3,0.27,0.776)(3,0.54,0.776)(3,0.54,0)(3,0.27,0)
\pstThreeDLine[linecolor=black](3,0.54,0)(3,0.54,1.638)(3,0.81,1.638)(3,0.81,0)(3,0.54,0)
\pstThreeDLine[linecolor=black](3,0.81,0)(3,0.81,0.172)(3,1.08,0.172)(3,1.08,0)(3,0.81,0)
\pstThreeDLine[linecolor=black](3,1.08,0)(3,1.08,0.603)(3,1.35,0.603)(3,1.35,0)(3,1.08,0)
\pstThreeDLine[linecolor=black](3,1.35,0)(3,1.35,0.690)(3,1.62,0.690)(3,1.62,0)(3,1.35,0)
\pstThreeDLine[linecolor=black](3,1.62,0)(3,1.62,0.172)(3,1.89,0.172)(3,1.89,0)(3,1.62,0)
\pstThreeDLine[linecolor=black](3,1.89,0)(3,1.89,0.259)(3,2.16,0.259)(3,2.16,0)(3,1.89,0)
\pstThreeDLine[linecolor=black](3,2.16,0)(3,2.16,0.086)(3,2.43,0.086)(3,2.43,0)(3,2.16,0)
\pstThreeDPut(3,3.1,0){02/1998}
\end{pspicture}

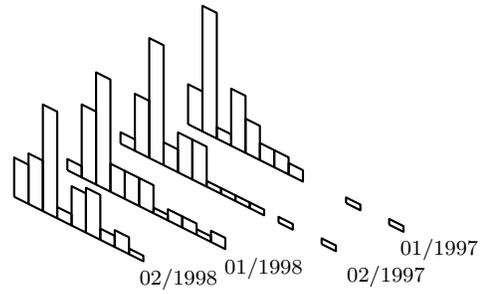
{\noindent \small Figure 45: probability distributions of the node strength for the first and second semesters of 1997 and 1998.}
\end{minipage}

\vskip 0.6 cm

\begin{minipage}{8 cm}
\begin{pspicture}(-3,-2.7)(1,1.7)
\psset{xunit=1,yunit=1} \scriptsize
\pstThreeDLine[linecolor=black](0,0,0)(0,0,0.068)(0,0.27,0.068)(0,0.27,0)(0,0,0)
\pstThreeDLine[linecolor=black](0,0.27,0)(0,0.27,0.405)(0,0.54,0.405)(0,0.54,0)(0,0.27,0)
\pstThreeDLine[linecolor=black](0,0.54,0)(0,0.54,2.027)(0,0.81,2.027)(0,0.81,0)(0,0.54,0)
\pstThreeDLine[linecolor=black](0,0.81,0)(0,0.81,0.338)(0,1.08,0.338)(0,1.08,0)(0,0.81,0)
\pstThreeDLine[linecolor=black](0,1.08,0)(0,1.08,0.405)(0,1.35,0.405)(0,1.35,0)(0,1.08,0)
\pstThreeDLine[linecolor=black](0,1.35,0)(0,1.35,0.878)(0,1.62,0.878)(0,1.62,0)(0,1.35,0)
\pstThreeDLine[linecolor=black](0,1.62,0)(0,1.62,0.203)(0,1.89,0.203)(0,1.89,0)(0,1.62,0)
\pstThreeDLine[linecolor=black](0,1.89,0)(0,1.89,0.203)(0,2.16,0.203)(0,2.16,0)(0,1.89,0)
\pstThreeDLine[linecolor=black](0,2.16,0)(0,2.16,0.270)(0,2.43,0.270)(0,2.43,0)(0,2.16,0)
\pstThreeDLine[linecolor=black](0,2.43,0)(0,2.43,0.135)(0,2.70,0.135)(0,2.70,0)(0,2.43,0)
\pstThreeDLine[linecolor=black](0,2.97,0)(0,2.97,0.068)(0,3.24,0.068)(0,3.24,0)(0,2.970,0)
\pstThreeDPut(0,4,0){01/2000}
\pstThreeDLine[linecolor=black](1,0,0)(1,0,0.068)(1,0.27,0.068)(1,0.27,0)(1,0,0)
\pstThreeDLine[linecolor=black](1,0.27,0)(1,0.27,0.616)(1,0.54,0.616)(1,0.54,0)(1,0.27,0)
\pstThreeDLine[linecolor=black](1,0.54,0)(1,0.54,1.849)(1,0.81,1.849)(1,0.81,0)(1,0.54,0)
\pstThreeDLine[linecolor=black](1,0.81,0)(1,0.81,0.205)(1,1.08,0.205)(1,1.08,0)(1,0.81,0)
\pstThreeDLine[linecolor=black](1,1.08,0)(1,1.08,0.753)(1,1.35,0.753)(1,1.35,0)(1,1.08,0)
\pstThreeDLine[linecolor=black](1,1.35,0)(1,1.35,0.822)(1,1.62,0.822)(1,1.62,0)(1,1.35,0)
\pstThreeDLine[linecolor=black](1,1.62,0)(1,1.62,0.342)(1,1.89,0.342)(1,1.89,0)(1,1.62,0)
\pstThreeDLine[linecolor=black](1,1.89,0)(1,1.89,0.274)(1,2.16,0.274)(1,2.16,0)(1,1.89,0)
\pstThreeDLine[linecolor=black](1,2.97,0)(1,2.97,0.068)(1,3.24,0.068)(1,3.24,0)(1,2.97,0)
\pstThreeDLine[linecolor=black](1,4.86,0)(1,4.86,0.068)(1,5.13,0.068)(1,5.13,0)(1,4.86,0)
\pstThreeDPut(1,6,0){02/2000}
\pstThreeDLine[linecolor=black](2,0,0)(2,0,0.203)(2,0.27,0.203)(2,0.27,0)(2,0,0)
\pstThreeDLine[linecolor=black](2,0.27,0)(2,0.27,0.541)(2,0.54,0.541)(2,0.54,0)(2,0.27,0)
\pstThreeDLine[linecolor=black](2,0.54,0)(2,0.54,2.027)(2,0.81,2.027)(2,0.81,0)(2,0.54,0)
\pstThreeDLine[linecolor=black](2,0.81,0)(2,0.81,0.338)(2,1.08,0.338)(2,1.08,0)(2,0.81,0)
\pstThreeDLine[linecolor=black](2,1.08,0)(2,1.08,0.338)(2,1.35,0.338)(2,1.35,0)(2,1.08,0)
\pstThreeDLine[linecolor=black](2,1.35,0)(2,1.35,0.541)(2,1.62,0.541)(2,1.62,0)(2,1.35,0)
\pstThreeDLine[linecolor=black](2,1.62,0)(2,1.62,0.203)(2,1.89,0.203)(2,1.89,0)(2,1.62,0)
\pstThreeDLine[linecolor=black](2,1.89,0)(2,1.89,0.338)(2,2.16,0.338)(2,2.16,0)(2,1.89,0)
\pstThreeDLine[linecolor=black](2,2.16,0)(2,2.16,0.203)(2,2.43,0.203)(2,2.43,0)(2,2.16,0)
\pstThreeDLine[linecolor=black](2,2.43,0)(2,2.43,0.068)(2,2.70,0.068)(2,2.70,0)(2,2.43,0)
\pstThreeDLine[linecolor=black](2,2.70,0)(2,2.70,0.068)(2,2.97,0.068)(2,2.97,0)(2,2.70,0)
\pstThreeDLine[linecolor=black](2,2.97,0)(2,2.97,0.135)(2,3.24,0.135)(2,3.24,0)(2,2.970,0)
\pstThreeDPut(2,4,0){01/2001}
\pstThreeDLine[linecolor=black](3,0,0)(3,0,0.137)(3,0.27,0.137)(3,0.27,0)(3,0,0)
\pstThreeDLine[linecolor=black](3,0.27,0)(3,0.27,0.753)(3,0.54,0.753)(3,0.54,0)(3,0.27,0)
\pstThreeDLine[linecolor=black](3,0.54,0)(3,0.54,2.260)(3,0.81,2.260)(3,0.81,0)(3,0.54,0)
\pstThreeDLine[linecolor=black](3,0.81,0)(3,0.81,0.342)(3,1.08,0.342)(3,1.08,0)(3,0.81,0)
\pstThreeDLine[linecolor=black](3,1.08,0)(3,1.08,0.274)(3,1.35,0.274)(3,1.35,0)(3,1.08,0)
\pstThreeDLine[linecolor=black](3,1.35,0)(3,1.35,0.205)(3,1.62,0.205)(3,1.62,0)(3,1.35,0)
\pstThreeDLine[linecolor=black](3,1.62,0)(3,1.62,0.274)(3,1.89,0.274)(3,1.89,0)(3,1.62,0)
\pstThreeDLine[linecolor=black](3,1.89,0)(3,1.89,0.342)(3,2.16,0.342)(3,2.16,0)(3,1.89,0)
\pstThreeDLine[linecolor=black](3,2.16,0)(3,2.16,0.274)(3,2.43,0.274)(3,2.43,0)(3,2.16,0)
\pstThreeDLine[linecolor=black](3,2.43,0)(3,2.43,0.068)(3,2.70,0.068)(3,2.70,0)(3,2.43,0)
\pstThreeDLine[linecolor=black](3,2.70,0)(3,2.70,0.068)(3,2.97,0.068)(3,2.97,0)(3,2.70,0)
\pstThreeDPut(3,4,0){02/2001}
\end{pspicture}

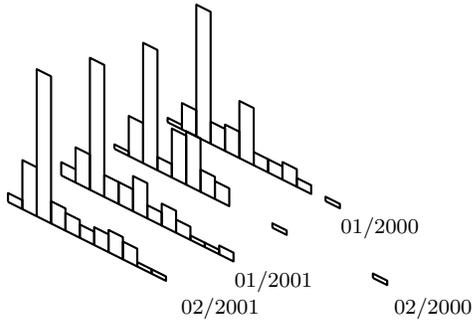
{\noindent \small Figure 46: probability distributions of the node strength for the first and second semesters of 2000 and 2001.}
\end{minipage}
\hskip 1.3 cm \begin{minipage}{8 cm}
\begin{pspicture}(-4,-2.7)(1,1.7)
\psset{xunit=1,yunit=1} \scriptsize
\pstThreeDLine[linecolor=black](0,0,0)(0,0,0.587)(0,0.27,0.587)(0,0.27,0)(0,0,0)
\pstThreeDLine[linecolor=black](0,0.27,0)(0,0.27,0.717)(0,0.54,0.717)(0,0.54,0)(0,0.27,0)
\pstThreeDLine[linecolor=black](0,0.54,0)(0,0.54,2.087)(0,0.81,2.087)(0,0.81,0)(0,0.54,0)
\pstThreeDLine[linecolor=black](0,0.81,0)(0,0.81,0.457)(0,1.08,0.457)(0,1.08,0)(0,0.81,0)
\pstThreeDLine[linecolor=black](0,1.08,0)(0,1.08,0.587)(0,1.35,0.587)(0,1.35,0)(0,1.08,0)
\pstThreeDLine[linecolor=black](0,1.35,0)(0,1.35,0.978)(0,1.62,0.978)(0,1.62,0)(0,1.35,0)
\pstThreeDLine[linecolor=black](0,1.62,0)(0,1.62,0.130)(0,1.89,0.130)(0,1.89,0)(0,1.62,0)
\pstThreeDLine[linecolor=black](0,1.89,0)(0,1.89,0.261)(0,2.16,0.261)(0,2.16,0)(0,1.89,0)
\pstThreeDLine[linecolor=black](0,2.16,0)(0,2.16,0.065)(0,2.43,0.065)(0,2.43,0)(0,2.16,0)
\pstThreeDLine[linecolor=black](0,2.43,0)(0,2.43,0.130)(0,2.70,0.130)(0,2.70,0)(0,2.42,0)
\pstThreeDPut(0,3.7,0){01/2007}
\pstThreeDLine[linecolor=black](1,0,0)(1,0,0.913)(1,0.27,0.913)(1,0.27,0)(1,0,0)
\pstThreeDLine[linecolor=black](1,0.27,0)(1,0.27,1.109)(1,0.54,1.109)(1,0.54,0)(1,0.27,0)
\pstThreeDLine[linecolor=black](1,0.54,0)(1,0.54,2.087)(1,0.81,2.087)(1,0.81,0)(1,0.54,0)
\pstThreeDLine[linecolor=black](1,0.81,0)(1,0.81,0.261)(1,1.08,0.261)(1,1.08,0)(1,0.81,0)
\pstThreeDLine[linecolor=black](1,1.08,0)(1,1.08,0.326)(1,1.35,0.326)(1,1.35,0)(1,1.08,0)
\pstThreeDLine[linecolor=black](1,1.35,0)(1,1.35,0.457)(1,1.62,0.457)(1,1.62,0)(1,1.35,0)
\pstThreeDLine[linecolor=black](1,1.62,0)(1,1.62,0.130)(1,1.89,0.130)(1,1.89,0)(1,1.62,0)
\pstThreeDLine[linecolor=black](1,1.89,0)(1,1.89,0.522)(1,2.16,0.522)(1,2.16,0)(1,1.89,0)
\pstThreeDLine[linecolor=black](1,2.16,0)(1,2.16,0.130)(1,2.43,0.130)(1,2.43,0)(1,2.16,0)
\pstThreeDLine[linecolor=black](1,2.43,0)(1,2.43,0.065)(1,2.70,0.065)(1,2.70,0)(1,2.42,0)
\pstThreeDPut(1,3.7,0){02/2007}
\pstThreeDLine[linecolor=black](2,0,0)(2,0,0.522)(2,0.27,0.522)(2,0.27,0)(2,0,0)
\pstThreeDLine[linecolor=black](2,0.27,0)(2,0.27,1.174)(2,0.54,1.174)(2,0.54,0)(2,0.27,0)
\pstThreeDLine[linecolor=black](2,0.54,0)(2,0.54,1.826)(2,0.81,1.826)(2,0.81,0)(2,0.54,0)
\pstThreeDLine[linecolor=black](2,0.81,0)(2,0.81,0.522)(2,1.08,0.522)(2,1.08,0)(2,0.81,0)
\pstThreeDLine[linecolor=black](2,1.08,0)(2,1.08,0.848)(2,1.35,0.848)(2,1.35,0)(2,1.08,0)
\pstThreeDLine[linecolor=black](2,1.35,0)(2,1.35,0.326)(2,1.62,0.326)(2,1.62,0)(2,1.35,0)
\pstThreeDLine[linecolor=black](2,1.62,0)(2,1.62,0.261)(2,1.89,0.261)(2,1.89,0)(2,1.62,0)
\pstThreeDLine[linecolor=black](2,1.89,0)(2,1.89,0.196)(2,2.16,0.196)(2,2.16,0)(2,1.89,0)
\pstThreeDLine[linecolor=black](2,2.16,0)(2,2.16,0.261)(2,2.43,0.261)(2,2.43,0)(2,2.16,0)
\pstThreeDLine[linecolor=black](2,2.43,0)(2,2.43,0.065)(2,2.70,0.065)(2,2.70,0)(2,2.42,0)
\pstThreeDPut(2,3.7,0){01/2008}
\pstThreeDLine[linecolor=black](3,0,0)(3,0,0.522)(3,0.27,0.522)(3,0.27,0)(3,0,0)
\pstThreeDLine[linecolor=black](3,0.27,0)(3,0.27,2.087)(3,0.54,2.087)(3,0.54,0)(3,0.27,0)
\pstThreeDLine[linecolor=black](3,0.54,0)(3,0.54,1.500)(3,0.81,1.500)(3,0.81,0)(3,0.54,0)
\pstThreeDLine[linecolor=black](3,0.81,0)(3,0.81,0.717)(3,1.08,0.717)(3,1.08,0)(3,0.81,0)
\pstThreeDLine[linecolor=black](3,1.08,0)(3,1.08,0.587)(3,1.35,0.587)(3,1.35,0)(3,1.08,0)
\pstThreeDLine[linecolor=black](3,1.35,0)(3,1.35,0.196)(3,1.62,0.196)(3,1.62,0)(3,1.35,0)
\pstThreeDLine[linecolor=black](3,1.62,0)(3,1.62,0.065)(3,1.89,0.065)(3,1.89,0)(3,1.62,0)
\pstThreeDLine[linecolor=black](3,1.89,0)(3,1.89,0.196)(3,2.16,0.196)(3,2.16,0)(3,1.89,0)
\pstThreeDLine[linecolor=black](3,2.43,0)(3,2.43,0.065)(3,2.70,0.065)(3,2.70,0)(3,2.42,0)
\pstThreeDLine[linecolor=black](3,2.70,0)(3,2.70,0.065)(3,2.97,0.065)(3,2.97,0)(3,2.70,0)
\pstThreeDPut(3,3.7,0){02/2008}
\end{pspicture}

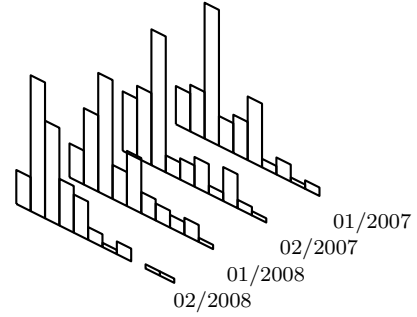
{\noindent \small Figure 47: probability distributions of the node strength for the first and second semesters of 2007 and 2008.}
\end{minipage}

\vskip 1 cm

\subsection{Betweenness}

The betweennes centrality measures how much a node lies on the shortest paths between other vertices. It is an important measure of how much a node is important as an intermediate between other nodes. It may be defined as
\begin{equation}
B_c^k=\sum_{i,j=1}^n\frac{n_{ij}^k}{m_{ij}}\ ,
\end{equation}
where $n_{ij}$ is the number of shortest paths (geodesic paths) between nodes $i$ and $j$ that pass through node $k$ and $m_{ij}$ is the total number of shortest paths between nodes $i$ and $j$. Our network is fully connected, so we need not worry about $m_{ij}$ being zero. The frequency distribution of the betweenness of the indices in the original minimum spanning trees is shown in figyres 48 to 51. The distributions drop substantially after the first interval.

\vskip 0.3 cm

\begin{minipage}{8 cm}
\begin{pspicture}(-3.5,-2.7)(1,2)
\psset{xunit=1,yunit=1} \scriptsize
\pstThreeDLine[linecolor=black](0,0,0)(0,0,1.875)(0,0.3,1.875)(0,0.3,0)(0,0,0)
\pstThreeDLine[linecolor=black](0,0.3,0)(0,0.3,0.375)(0,0.6,0.375)(0,0.6,0)(0,0.3,0)
\pstThreeDLine[linecolor=black](0,0.9,0)(0,0.9,0.375)(0,1.2,0.375)(0,1.2,0)(0,0.9,0)
\pstThreeDLine[linecolor=black](0,2.1,0)(0,2.1,0.188)(0,2.4,0.188)(0,2.4,0)(0,2.1,0)
\pstThreeDLine[linecolor=black](0,3,0)(0,3,0.188)(0,3.3,0.188)(0,3.3,0)(0,3,0)
\pstThreeDPut(0,4,0){01/1986}
\pstThreeDLine[linecolor=black](1,0,0)(1,0,1.313)(1,0.3,1.313)(1,0.3,0)(1,0,0)
\pstThreeDLine[linecolor=black](1,0.3,0)(1,0.3,0.563)(1,0.6,0.563)(1,0.6,0)(1,0.3,0)
\pstThreeDLine[linecolor=black](1,0.9,0)(1,0.9,0.375)(1,1.2,0.375)(1,1.2,0)(1,0.9,0)
\pstThreeDLine[linecolor=black](1,1.2,0)(1,1.2,0.375)(1,1.5,0.375)(1,1.5,0)(1,1.2,0)
\pstThreeDLine[linecolor=black](1,2.1,0)(1,2.1,0.188)(1,2.4,0.188)(1,2.4,0)(1,2.1,0)
\pstThreeDLine[linecolor=black](1,3,0)(1,3,0.188)(1,3.3,0.188)(1,3.3,0)(1,3,0)
\pstThreeDPut(1,4,0){02/1986}
\pstThreeDLine[linecolor=black](2,0,0)(2,0,1.435)(2,0.3,1.435)(2,0.3,0)(2,0,0)
\pstThreeDLine[linecolor=black](2,0.3,0)(2,0.3,0.261)(2,0.6,0.261)(2,0.6,0)(2,0.3,0)
\pstThreeDLine[linecolor=black](2,0.6,0)(2,0.6,0.261)(2,0.9,0.261)(2,0.9,0)(2,0.6,0)
\pstThreeDLine[linecolor=black](2,0.9,0)(2,0.9,0.261)(2,1.2,0.261)(2,1.2,0)(2,0.9,0)
\pstThreeDLine[linecolor=black](2,1.2,0)(2,1.2,0.130)(2,1.5,0.130)(2,1.5,0)(2,1.2,0)
\pstThreeDLine[linecolor=black](2,2.1,0)(2,2.1,0.522)(2,2.4,0.522)(2,2.4,0)(2,2.1,0)
\pstThreeDLine[linecolor=black](2,2.7,0)(2,2.7,0.130)(2,3,0.130)(2,3,0)(2,2.7,0)
\pstThreeDPut(2,4,0){01/1987}
\pstThreeDLine[linecolor=black](3,0,0)(3,0,1.174)(3,0.3,1.174)(3,0.3,0)(3,0,0)
\pstThreeDLine[linecolor=black](3,0.3,0)(3,0.3,0.522)(3,0.6,0.522)(3,0.6,0)(3,0.3,0)
\pstThreeDLine[linecolor=black](3,0.6,0)(3,0.6,0.130)(3,0.9,0.130)(3,0.9,0)(3,0.6,0)
\pstThreeDLine[linecolor=black](3,0.9,0)(3,0.9,0.261)(3,1.2,0.261)(3,1.2,0)(3,0.9,0)
\pstThreeDLine[linecolor=black](3,1.2,0)(3,1.2,0.130)(3,1.5,0.130)(3,1.5,0)(3,1.2,0)
\pstThreeDLine[linecolor=black](3,2.1,0)(3,2.1,0.522)(3,2.4,0.522)(3,2.4,0)(3,2.1,0)
\pstThreeDLine[linecolor=black](3,2.4,0)(3,2.4,0.261)(3,2.7,0.261)(3,2.7,0)(3,2.4,0)
\pstThreeDPut(3,4,0){02/1987}
\end{pspicture}

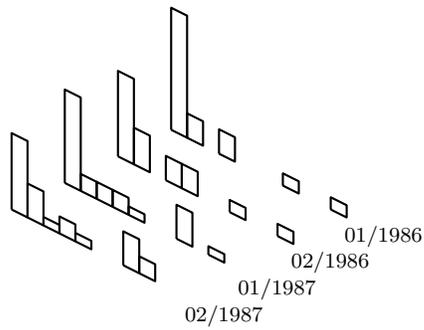
{\noindent \small Figure 48: probability distributions of the normalized betweenness for the first and second semesters of 1986 and 1987.}
\end{minipage}
\hskip 1.3 cm \begin{minipage}{8 cm}
\begin{pspicture}(-3.5,-3)(1,2)
\psset{xunit=1,yunit=1} \scriptsize
\pstThreeDLine[linecolor=black](0,0,0)(0,0,1.789)(0,0.3,1.789)(0,0.3,0)(0,0,0)
\pstThreeDLine[linecolor=black](0,0.3,0)(0,0.3,0.368)(0,0.6,0.368)(0,0.6,0)(0,0.3,0)
\pstThreeDLine[linecolor=black](0,0.6,0)(0,0.6,0.211)(0,0.9,0.211)(0,0.9,0)(0,0.6,0)
\pstThreeDLine[linecolor=black](0,0.9,0)(0,0.9,0.316)(0,1.2,0.316)(0,1.2,0)(0,0.9,0)
\pstThreeDLine[linecolor=black](0,1.2,0)(0,1.2,0.053)(0,1.5,0.053)(0,1.5,0)(0,1.2,0)
\pstThreeDLine[linecolor=black](0,2.1,0)(0,2.1,0.105)(0,2.4,0.105)(0,2.4,0)(0,2.1,0)
\pstThreeDLine[linecolor=black](0,2.4,0)(0,2.4,0.105)(0,2.7,0.105)(0,2.7,0)(0,2.4,0)
\pstThreeDLine[linecolor=black](0,3,0)(0,3,0.053)(0,3.3,0.053)(0,3.3,0)(0,3,0)
\pstThreeDPut(0,5,0){01/1997}
\pstThreeDLine[linecolor=black](1,0,0)(1,0,2)(1,0.3,2)(1,0.3,0)(1,0,0)
\pstThreeDLine[linecolor=black](1,0.3,0)(1,0.3,0.368)(1,0.6,0.368)(1,0.6,0)(1,0.3,0)
\pstThreeDLine[linecolor=black](1,0.6,0)(1,0.6,0.053)(1,0.9,0.053)(1,0.9,0)(1,0.6,0)
\pstThreeDLine[linecolor=black](1,0.9,0)(1,0.9,0.158)(1,1.2,0.158)(1,1.2,0)(1,0.9,0)
\pstThreeDLine[linecolor=black](1,1.2,0)(1,1.2,0.158)(1,1.5,0.158)(1,1.5,0)(1,1.2,0)
\pstThreeDLine[linecolor=black](1,1.5,0)(1,1.5,0.105)(1,1.8,0.105)(1,1.8,0)(1,1.5,0)
\pstThreeDLine[linecolor=black](1,2.4,0)(1,2.4,0.053)(1,2.7,0.053)(1,2.7,0)(1,2.4,0)
\pstThreeDLine[linecolor=black](1,2.7,0)(1,2.7,0.053)(1,3,0.053)(1,3,0)(1,2.7,0)
\pstThreeDLine[linecolor=black](1,3,0)(1,3,0.053)(1,3.3,0.053)(1,3.3,0)(1,3,0)
\pstThreeDPut(1,5,0){02/1997}
\pstThreeDLine[linecolor=black](2,0,0)(2,0,1.966)(2,0.3,1.966)(2,0.3,0)(2,0,0)
\pstThreeDLine[linecolor=black](2,0.3,0)(2,0.3,0.466)(2,0.6,0.466)(2,0.6,0)(2,0.3,0)
\pstThreeDLine[linecolor=black](2,0.6,0)(2,0.6,0.103)(2,0.9,0.103)(2,0.9,0)(2,0.6,0)
\pstThreeDLine[linecolor=black](2,0.9,0)(2,0.9,0.103)(2,1.2,0.103)(2,1.2,0)(2,0.9,0)
\pstThreeDLine[linecolor=black](2,1.2,0)(2,1.2,0.103)(2,1.5,0.103)(2,1.5,0)(2,1.2,0)
\pstThreeDLine[linecolor=black](2,1.5,0)(2,1.5,0.052)(2,1.8,0.052)(2,1.8,0)(2,1.5,0)
\pstThreeDLine[linecolor=black](2,2.1,0)(2,2.1,0.103)(2,2.4,0.103)(2,2.4,0)(2,2.1,0)
\pstThreeDLine[linecolor=black](2,2.4,0)(2,2.4,0.052)(2,2.7,0.052)(2,2.7,0)(2,2.4,0)
\pstThreeDLine[linecolor=black](2,3.3,0)(2,3.3,0.052)(2,3.6,0.052)(2,3.6,0)(2,3.3,0)
\pstThreeDPut(2,5,0){01/1998}
\pstThreeDLine[linecolor=black](3,0,0)(3,0,1.862)(3,0.3,1.862)(3,0.3,0)(3,0,0)
\pstThreeDLine[linecolor=black](3,0.3,0)(3,0.3,0.414)(3,0.6,0.414)(3,0.6,0)(3,0.3,0)
\pstThreeDLine[linecolor=black](3,0.6,0)(3,0.6,0.362)(3,0.9,0.362)(3,0.9,0)(3,0.6,0)
\pstThreeDLine[linecolor=black](3,0.9,0)(3,0.9,0.155)(3,1.2,0.155)(3,1.2,0)(3,0.9,0)
\pstThreeDLine[linecolor=black](3,1.2,0)(3,1.2,0.052)(3,1.5,0.052)(3,1.5,0)(3,1.2,0)
\pstThreeDLine[linecolor=black](3,1.5,0)(3,1.5,0.103)(3,1.8,0.103)(3,1.8,0)(3,1.5,0)
\pstThreeDLine[linecolor=black](3,3.9,0)(3,3.9,0.052)(3,4.2,0.052)(3,4.2,0)(3,3.9,0)
\pstThreeDPut(3,5,0){02/1998}
\end{pspicture}

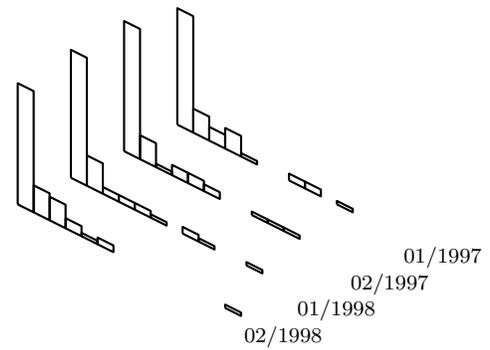
{\noindent \small Figure 49: probability distributions of the normalized betweenness for the first and second semesters of 1997 and 1998.}
\end{minipage}

\vskip 0.6 cm

\begin{minipage}{8 cm}
\begin{pspicture}(-3.5,-3)(1,2)
\psset{xunit=1,yunit=1} \scriptsize
\pstThreeDLine[linecolor=black](0,0,0)(0,0,2.068)(0,0.3,2.068)(0,0.3,0)(0,0,0)
\pstThreeDLine[linecolor=black](0,0.3,0)(0,0.3,0.405)(0,0.6,0.405)(0,0.6,0)(0,0.3,0)
\pstThreeDLine[linecolor=black](0,0.6,0)(0,0.6,0.122)(0,0.9,0.122)(0,0.9,0)(0,0.6,0)
\pstThreeDLine[linecolor=black](0,0.9,0)(0,0.9,0.162)(0,1.2,0.162)(0,1.2,0)(0,0.9,0)
\pstThreeDLine[linecolor=black](0,1.2,0)(0,1.2,0.122)(0,1.5,0.122)(0,1.5,0)(0,1.2,0)
\pstThreeDLine[linecolor=black](0,1.8,0)(0,1.8,0.041)(0,2.1,0.041)(0,2.1,0)(0,1.8,0)
\pstThreeDLine[linecolor=black](0,3,0)(0,3,0.041)(0,3.3,0.041)(0,3.3,0)(0,3,0)
\pstThreeDLine[linecolor=black](0,3.3,0)(0,3.3,0.041)(0,3.6,0.041)(0,3.6,0)(0,3.3,0)
\pstThreeDPut(0,5,0){01/2000}
\pstThreeDLine[linecolor=black](1,0,0)(1,0,2.230)(1,0.3,2.230)(1,0.3,0)(1,0,0)
\pstThreeDLine[linecolor=black](1,0.3,0)(1,0.3,0.284)(1,0.6,0.284)(1,0.6,0)(1,0.3,0)
\pstThreeDLine[linecolor=black](1,0.6,0)(1,0.6,0.203)(1,0.9,0.203)(1,0.9,0)(1,0.6,0)
\pstThreeDLine[linecolor=black](1,0.9,0)(1,0.9,0.081)(1,1.2,0.081)(1,1.2,0)(1,0.9,0)
\pstThreeDLine[linecolor=black](1,1.2,0)(1,1.2,0.041)(1,1.5,0.041)(1,1.5,0)(1,1.2,0)
\pstThreeDLine[linecolor=black](1,2.1,0)(1,2.1,0.041)(1,2.4,0.041)(1,2.4,0)(1,2.1,0)
\pstThreeDLine[linecolor=black](1,2.4,0)(1,2.4,0.041)(1,2.7,0.041)(1,2.7,0)(1,2.4,0)
\pstThreeDLine[linecolor=black](1,2.7,0)(1,2.7,0.041)(1,3,0.041)(1,3,0)(1,2.7,0)
\pstThreeDLine[linecolor=black](1,3.3,0)(1,3.3,0.041)(1,3.6,0.041)(1,3.6,0)(1,3.3,0)
\pstThreeDPut(1,5,0){02/2000}
\pstThreeDLine[linecolor=black](2,0,0)(2,0,2.203)(2,0.3,2.203)(2,0.3,0)(2,0,0)
\pstThreeDLine[linecolor=black](2,0.3,0)(2,0.3,0.342)(2,0.6,0.342)(2,0.6,0)(2,0.3,0)
\pstThreeDLine[linecolor=black](2,0.6,0)(2,0.6,0.076)(2,0.9,0.076)(2,0.9,0)(2,0.6,0)
\pstThreeDLine[linecolor=black](2,0.9,0)(2,0.9,0.114)(2,1.2,0.114)(2,1.2,0)(2,0.9,0)
\pstThreeDLine[linecolor=black](2,1.2,0)(2,1.2,0.114)(2,1.5,0.114)(2,1.5,0)(2,1.2,0)
\pstThreeDLine[linecolor=black](2,1.5,0)(2,1.5,0.038)(2,1.8,0.038)(2,1.8,0)(2,1.5,0)
\pstThreeDLine[linecolor=black](2,1.8,0)(2,1.8,0.076)(2,2.1,0.076)(2,2.1,0)(2,1.8,0)
\pstThreeDLine[linecolor=black](2,3,0)(2,3,0)(2,3.3,0)(2,3.3,0)(2,3,0)
\pstThreeDLine[linecolor=black](2,3.9,0)(2,3.9,0.038)(2,4.2,0.038)(2,4.2,0)(2,3.9,0)
\pstThreeDPut(2,5,0){01/2001}
\pstThreeDLine[linecolor=black](3,0,0)(3,0,2.165)(3,0.3,2.165)(3,0.3,0)(3,0,0)
\pstThreeDLine[linecolor=black](3,0.3,0)(3,0.3,0.456)(3,0.6,0.456)(3,0.6,0)(3,0.3,0)
\pstThreeDLine[linecolor=black](3,0.6,0)(3,0.6,0.038)(3,0.9,0.038)(3,0.9,0)(3,0.6,0)
\pstThreeDLine[linecolor=black](3,0.9,0)(3,0.9,0.076)(3,1.2,0.076)(3,1.2,0)(3,0.9,0)
\pstThreeDLine[linecolor=black](3,1.8,0)(3,1.8,0.038)(3,2.1,0.038)(3,2.1,0)(3,1.8,0)
\pstThreeDLine[linecolor=black](3,2.1,0)(3,2.1,0.038)(3,2.4,0.038)(3,2.4,0)(3,2.1,0)
\pstThreeDLine[linecolor=black](3,2.4,0)(3,2.4,0.114)(3,2.7,0.114)(3,2.7,0)(3,2.4,0)
\pstThreeDLine[linecolor=black](3,2.7,0)(3,2.7,0.038)(3,3,0.038)(3,3,0)(3,2.7,0)
\pstThreeDLine[linecolor=black](3,3,0)(3,3,0.038)(3,3.3,0.038)(3,3.3,0)(3,3,0)
\pstThreeDPut(3,5,0){02/2001}
\end{pspicture}

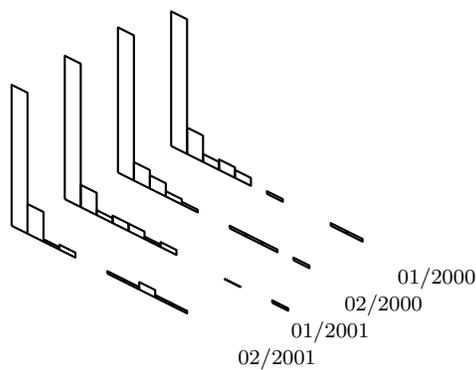
{\noindent \small Figure 50: probability distributions of the normalized betweenness for the first and second semesters of 2000 and 2001.}
\end{minipage}
\hskip 1.3 cm \begin{minipage}{8 cm}
\begin{pspicture}(-3.5,-3)(1,2)
\psset{xunit=1,yunit=1} \scriptsize
\pstThreeDLine[linecolor=black](0,0,0)(0,0,2.120)(0,0.3,2.120)(0,0.3,0)(0,0,0)
\pstThreeDLine[linecolor=black](0,0.3,0)(0,0.3,0.261)(0,0.6,0.261)(0,0.6,0)(0,0.3,0)
\pstThreeDLine[linecolor=black](0,0.6,0)(0,0.6,0.293)(0,0.9,0.293)(0,0.9,0)(0,0.6,0)
\pstThreeDLine[linecolor=black](0,0.9,0)(0,0.9,0.065)(0,1.2,0.065)(0,1.2,0)(0,0.9,0)
\pstThreeDLine[linecolor=black](0,1.2,0)(0,1.2,0.033)(0,1.5,0.033)(0,1.5,0)(0,1.2,0)
\pstThreeDLine[linecolor=black](0,1.5,0)(0,1.5,0.033)(0,1.8,0.033)(0,1.8,0)(0,1.5,0)
\pstThreeDLine[linecolor=black](0,2.1,0)(0,2.1,0.098)(0,2.4,0.098)(0,2.4,0)(0,2.1,0)
\pstThreeDLine[linecolor=black](0,2.4,0)(0,2.4,0.033)(0,2.7,0.033)(0,2.7,0)(0,2.4,0)
\pstThreeDLine[linecolor=black](0,2.7,0)(0,2.7,0.033)(0,3,0.033)(0,3,0)(0,2.7,0)
\pstThreeDLine[linecolor=black](0,3,0)(0,3,0.033)(0,3.3,0.033)(0,3.3,0)(0,3,0)
\pstThreeDPut(0,4.3,0){01/2007}
\pstThreeDLine[linecolor=black](1,0,0)(1,0,2.250)(1,0.3,2.250)(1,0.3,0)(1,0,0)
\pstThreeDLine[linecolor=black](1,0.3,0)(1,0.3,0.326)(1,0.6,0.326)(1,0.6,0)(1,0.3,0)
\pstThreeDLine[linecolor=black](1,0.6,0)(1,0.6,0.163)(1,0.9,0.163)(1,0.9,0)(1,0.6,0)
\pstThreeDLine[linecolor=black](1,0.9,0)(1,0.9,0.065)(1,1.2,0.065)(1,1.2,0)(1,0.9,0)
\pstThreeDLine[linecolor=black](1,1.5,0)(1,1.5,0.033)(1,1.8,0.033)(1,1.8,0)(1,1.5,0)
\pstThreeDLine[linecolor=black](1,2.4,0)(1,2.4,0.098)(1,2.7,0.098)(1,2.7,0)(1,2.4,0)
\pstThreeDLine[linecolor=black](1,2.7,0)(1,2.7,0.033)(1,3,0.033)(1,3,0)(1,2.7,0)
\pstThreeDLine[linecolor=black](1,3.3,0)(1,3.3,0.033)(1,3.6,0.033)(1,3.6,0)(1,3.3,0)
\pstThreeDPut(1,4.3,0){02/2007}
\pstThreeDLine[linecolor=black](2,0,0)(2,0,2.120)(2,0.3,2.120)(2,0.3,0)(2,0,0)
\pstThreeDLine[linecolor=black](2,0.3,0)(2,0.3,0.489)(2,0.6,0.489)(2,0.6,0)(2,0.3,0)
\pstThreeDLine[linecolor=black](2,0.6,0)(2,0.6,0.163)(2,0.9,0.163)(2,0.9,0)(2,0.6,0)
\pstThreeDLine[linecolor=black](2,0.9,0)(2,0.9,0.033)(2,1.2,0.033)(2,1.2,0)(2,0.9,0)
\pstThreeDLine[linecolor=black](2,1.2,0)(2,1.2,0.033)(2,1.5,0.033)(2,1.5,0)(2,1.2,0)
\pstThreeDLine[linecolor=black](2,2.4,0)(2,2.4,0.033)(2,2.7,0.033)(2,2.7,0)(2,2.4,0)
\pstThreeDLine[linecolor=black](2,2.7,0)(2,2.7,0.130)(2,3,0.130)(2,3,0)(2,2.7,0)
\pstThreeDPut(2,4.3,0){01/2008}
\pstThreeDLine[linecolor=black](3,0,0)(3,0,2.315)(3,0.3,2.315)(3,0.3,0)(3,0,0)
\pstThreeDLine[linecolor=black](3,0.3,0)(3,0.3,0.293)(3,0.6,0.293)(3,0.6,0)(3,0.3,0)
\pstThreeDLine[linecolor=black](3,0.6,0)(3,0.6,0.098)(3,0.9,0.098)(3,0.9,0)(3,0.6,0)
\pstThreeDLine[linecolor=black](3,0.9,0)(3,0.9,0.065)(3,1.2,0.065)(3,1.2,0)(3,0.9,0)
\pstThreeDLine[linecolor=black](3,1.5,0)(3,1.5,0.033)(3,1.8,0.033)(3,1.8,0)(3,1.5,0)
\pstThreeDLine[linecolor=black](3,2.4,0)(3,2.4,0.163)(3,2.7,0.163)(3,2.7,0)(3,2.4,0)
\pstThreeDLine[linecolor=black](3,2.7,0)(3,2.7,0.033)(3,3,0.033)(3,3,0)(3,2.7,0)
\pstThreeDPut(3,4.3,0){02/2008}
\end{pspicture}

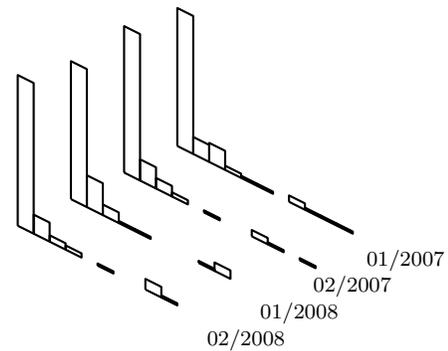
{\noindent \small Figure 51: probability distributions of the normalized betweenness for the first and second semesters of 2007 and 2008.}
\end{minipage}

\vskip 0.5 cm

Figure 52 shows the evolution in time of the three centrality measures whose frequency distributions have been just shown plus the Eigenvector centrality, which takes into account the degree of neighbouring nodes when calculating the importance of a node. In the figure, geographical regions are specified, and one may notice that Europe concentrates most of the centralities. The peaks are smeared out in the node strength representations, and are more evident in the betweenness centrality evolution. The countries of Oceania (Australia and New Zealand) show remarkable centralities, as do some key Pacific Asian countries. The USA is, surprisingly, not very central in these representations.



\begin{center}
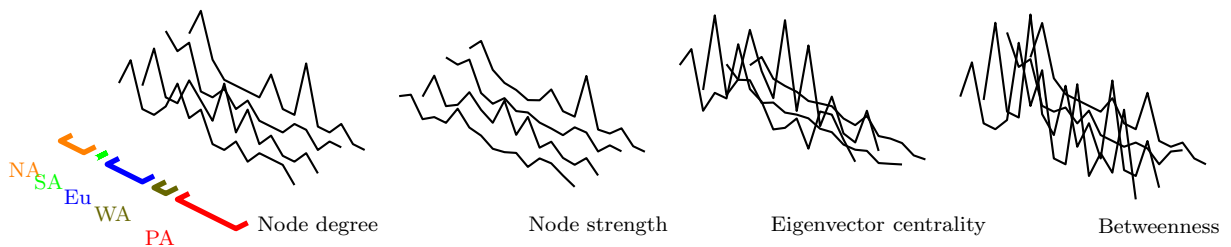
\noindent Figure 52a: first and second semesters of 1986 and 1987.
\end{center}

%

\vskip 0.4 cm

\begin{center}
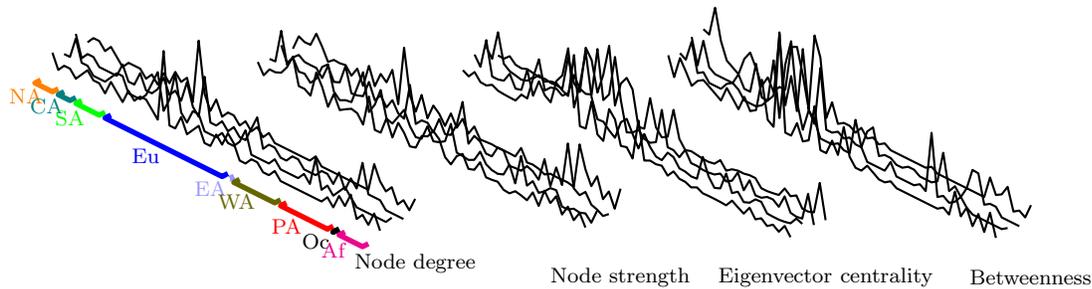
\noindent Figure 52b: first and second semesters of 1997 and 1998.
\end{center}

%

\vskip 1 cm

\begin{center}
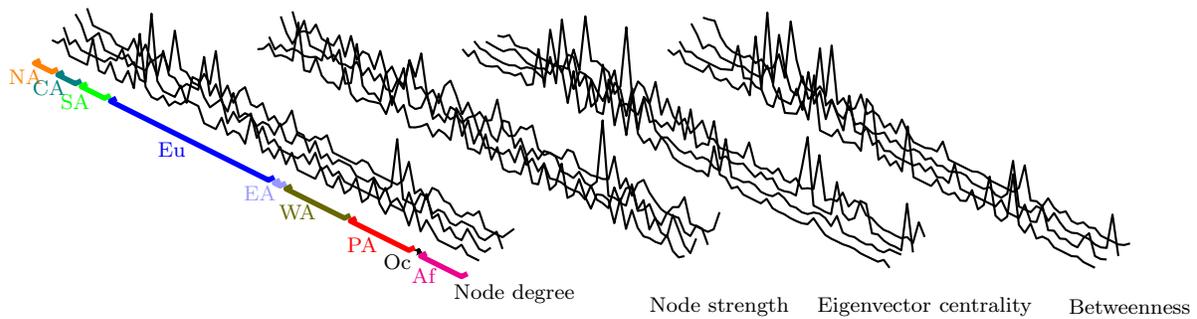
\noindent Figure 52c: first and second semesters of 2000 and 2001.
\end{center}

%

\vskip 1.6 cm

\begin{center}
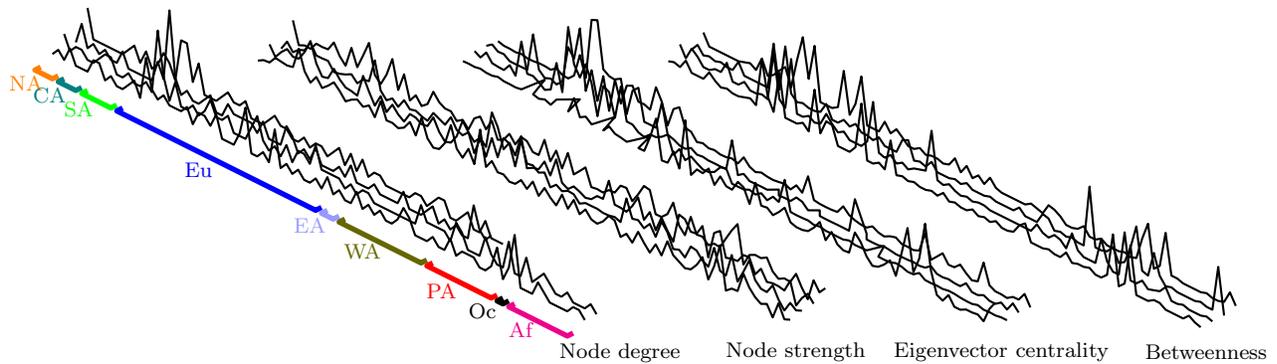
Figure 52d: first and second semesters of 2007 and 2008.
\end{center}

\vskip 0.2 cm

\noindent Figure 52: evolution of some centrality measures of stock exchange indices in time. From right to left, each graphs shows the evolution for the first and second semesters of, respectively, a) 1986 to 1987, b) 1997 to 1998, c) 2000 to 2001, and d) 2007 to 2008. The graphs also discriminate indices by geographic region: NA (orange) stands for North America, CA (darkgreen) for Central America and the Caribbean, SA (green) for South America, Eu (blue) for Europe, EA (light blue) for Eurasia, WA (brown) for Western Asia, PA (red) for Pacific Asia, Oc (black) for Oceania, and Af (magenta) for Africa.

\section{Conclusion}

Most of the previous works dealing with the minimum spanning tree representation of networks in finance only dealt with the most interacting nodes, what was done by choosing the most liquid stocks in a stock exchange, or the ones used in order to build an official index of the same stock exchange. So, poorly connected nodes never arised as an important issue, although some authors have taken this into account \cite{pruned1}, using another sort of pruning. In the approach of this article, I followed the path of limiting the connections to those that were strong, meaning they were related with distances below a threshold that was obtained by simulations with randomized data, and to those that were not strong, but that were very enduring in time. This made it possible to obtain diagrams that show sensible relations between the many indices that were studied. The graphs showed that the connections between stock market indices are strongly influenced by the geographic positions of the stock exchanges, as there are well defined clusters related with continents, reinforcing results previously obtained by regression methods \cite{h33}. They also show that cultural relations or international treaties have a strong effect in stock market indices correlations, as one can see by the strong connections between Israel and South Africa with Europe, and between Greece and Cyprus, as examples. The centrality of France has been detected by other authors \cite{h19}, and the closeness between Russia and Turkey was detected in \cite{dependance2}.

An unexpected result is the relative isolation of the USA stock markets, strongly linked with the other North and South American stock markets, but weakly linked with Europe. This is probably related with the fact that stock markets operate at different hours, so that the trading hours of most European markets intersect the opening hours of American ones only slightly, and Asian markets have null intersections with their American counterparts. How deep this affects the correlations between markets and how this discrepancy should be dealt with is a matter for further study. Preliminary studies made by the author show that there is no considerable change in the minimum spanning trees of world indices if one considers the stock market indices of countries that have null intersection with the NYSE lagged by one day, which means one compares the S\&P 500 and Nasdaq indices with the next day indices of those markets. One remarkable exception is the index from New Zealand, which becomes much more connected with the USA indices when lagged by one day. The separation between Central Europe and North America persists even when one laggs the whole of Europe by one day. The exception is the FTSE 100 from the UK, which in this case gets more connected with America then with Europe. A model developed in \cite{Rudi1}-\cite{Rudi2} for the analysis of intraday data, where comparing data taken at different times (or ticks) is an important issue, may also be applied to the lagging problem of international stock market indices.

The other factor to be observed is the nature of the centrality measures in a network. They are built in such a way that, if many nodes move together, then one or more of them are better representatives of the collective movement, like France with respect to the European market nodes. Centrality does not mean, necessarily, that the node is more influential. The S\&P 500 (or the Dow Jones), for example, is watched by all stock market dealers in the world, although it is not very central in the minimum spanning trees. This is not due to the particularities of the minimum spanning tree representation, as other work [ref], which deals with asset graphs based on thresholds, show. What is needed is a measure of the impact a market has on others, a one-directional measure that takes into account the volume a stock market trades, and also the solidity and inertia of the market. Some measures which handle causality are partial correlation \cite{Dror1}-\cite{Dror4}, Granger causality \cite{caus1}-\cite{dependance1}, and transfer of entropy \cite{entropy1}-\cite{entropy3}, but none of them takes into account the size of a market.

Models of how a crisis may propagate on a network of stock markets have been built using epidemics models or econometric models, with some success \cite{shocks1}-\cite{shocks8}. Contagion models consider centrality or k-core number as key features in the propagation of crises.

This work contributes to a better understanding of the minimum spanning tree representation when dealing with weakly interacting indices, and offers a way to prun some of them away from the network. The process shows that weakly interacting nodes rarely survive for long, so that, in practice, it makes little difference if one pruns them away based only on a distance threshold obtained by randomizing the original data. It also shows how the world stock markets evolved in the past decades, mainly in the proximity of crises, and how stock markets became more integrated with time.

Further work, as cited above, shall deal with lagg effects between markets, with the causality chain between them, and with the building of a better model that explains a little of the complex interactions between markets, how contagion between them happens, and maybe how to devise policies for cutting such contagions short.

\vskip 0.6 cm

\noindent{\Large \bf Acknowledgements}

\vskip 0.4 cm

The author acknowledges the support of this work by a grant from Insper, Instituto de Ensino e Pesquisa. I am also grateful to Siew Ann Cheong, Rudi Schäffer, Dror Kenett, and Sergio Giovanetti Lazzarini, for useful discussions, and to the attenders and organizers of the Econophysics Colloquium 2010. This article was written using \LaTeX, all figures were made using PSTricks, and the calculations were made using Matlab, Ucinet and Excel. All data are freely available upon request on leonidassj@insper.edu.br.

\appendix

\section{Stock Market Indices}

The next table (table 1) shows the stock market indices we used, their original countries, the symbols we used for them in the main text, and their codes in Bloomberg. In the tables, we use ``SX'' as short for ``Stock Exchange''. Some of the indices changed names and/or method of calculation and are designated by the two names, prior to and after the changing date.

\[ 
\begin{array}{|l|l|c|c|} \hline \text{Index} & \text{Country} & \text{Symbol} & \text{Code in Bloomberg} \\ \hline \text{\bf North America} \\ \hline \text{S\&P 500} & \text{United States of America} & \text{S\&P} & \text{SPX} \\ \text{Nasdaq Composite} & \text{United States of America} & \text{Nasd} & \text{CCMP} \\ \text{S\&P/TSX Composite} & \text{Canada} & \text{Cana} & \text{SPTSX} \\ \text{IPC} & \text{Mexico} & \text{Mexi} & \text{MEXBOL} \\ \hline \text{\bf Central America} \\ \hline \text{Bolsa de Panama General} & \text{Panama} & \text{Pana} & \text{BVPSBVPS} \\
\hline \end{array} \]

\[ \begin{array}{|l|l|c|c|} \hline \text{Index} & \text{Country} & \text{Symbol} & \text{Code in Bloomberg} \\ \hline \text{\bf Eastern Europe} \\ \hline \text{PX or PX50} & \text{Czech Republic} & \text{CzRe} & \text{PX} \\ \text{SAX} & \text{Slovakia} & \text{Slok} & \text{SKSM} \\ \text{Budapest SX Index} & \text{Hungary} & \text{Hung} & \text{BUX} \\ \text{BELEX 15} & \text{Serbia} & \text{Serb} & \text{BELEX15} \\ \text{CROBEX} & \text{Croatia} & \text{Croa} & \text{CRO} \\ \text{SBI TOP} & \text{Slovenia} & \text{Slov} & \text{SBITOP} \\
\hline \end{array} \]

\[ \begin{array}{|l|l|c|c|} \hline \text{Index} & \text{Country} & \text{Symbol} & \text{Code in Bloomberg} \\ \hline \text{\bf Asia-Pacific} \\ \hline \text{Nikkei 25} & \text{Japan} & \text{Japa} & \text{NKY} \\ \text{Hang Seng} & \text{Hong Kong} & \text{HoKo} & \text{HSI} \\ \text{Shangai SE Composite} & \text{China} & \text{Chin} & \text{SHCOMP} \\ \text{MSE TOP 20} & \text{Mongolia} & \text{Mong} & \text{MSETOP} \\ \text{TAIEX} & \text{Taiwan} & \text{Taiw} & \text{TWSE} \\ \text{KOSPI} & \text{South Korea} & \text{SoKo} & \text{KOSPI} \\ \text{SET} & \text{Thailand} & \text{Thai} & \text{SET} \\ \text{VN-Index} & \text{Vietnam} & \text{Viet} & \text{VNINDEX} \\ \text{KLCI} & \text{Malaysia} & \text{Mala} & \text{FBMKLCI} \\ \text{Straits Times} & \text{Singapore} & \text{Sing} & \text{FSSTI} \\ \text{Jakarta Composite Index} & \text{Indonesia} & \text{Indo} & \text{JCI} \\ \text{PSEi} & \text{Philipines} & \text{Phil} & \text{PCOMP} \\ \hline \text{\bf Central and Southern Africa} \\ \hline \text{Ghana All Share Index} & \text{Ghana} & \text{Ghan} & \text{GGSEGSE} \\ \text{Nigeria SX All Share Index} & \text{Nigeria} & \text{Nige} & \text{NGSEINDX} \\ \text{NSE 20} & \text{Kenya} & \text{Keny} & \text{KNSMIDX} \\ \text{DSEI} & \text{Tanzania} & \text{Tanz} & \text{DARSDSEI} \\ \text{FTSE/Namibia Overall} & \text{Namibia} & \text{Nami} & \text{FTN098} \\ \text{Gaborone} & \text{Botswana} & \text{Bots} & \text{BGSMDC} \\ \text{FTSE/JSE Africa All Share} & \text{South Africa} & \text{SoAf} & \text{JALSH} \\ \text{SEMDEX} & \text{Mauritius} & \text{Maur} & \text{SEMDEX} \\ \hline \end{array} \]
\hskip 2.4 cm Table 1: names, codes, and abreviations of the stock market indices used in this article.

\section{Minimum spanning trees in three dimensions}

In what follows, I show the three dimensional pictures of the minimum spanning tree, both original and pruned, in three dimensions, using an algorithm that represents as best as possible the real distances between indices. Note that some indices that seem to be far apart in the two dimensional view of the minimum spanning trees are in fact close together, like some african indices displayed in the next figures.

%

\vskip 1.3 cm

\noindent Figure 53: three dimensional view of the minimum spanning tree for the first semester of 1986. The second figure is the pruned spanning tree.

\vskip 0.2 cm

%

\vskip 0.9 cm

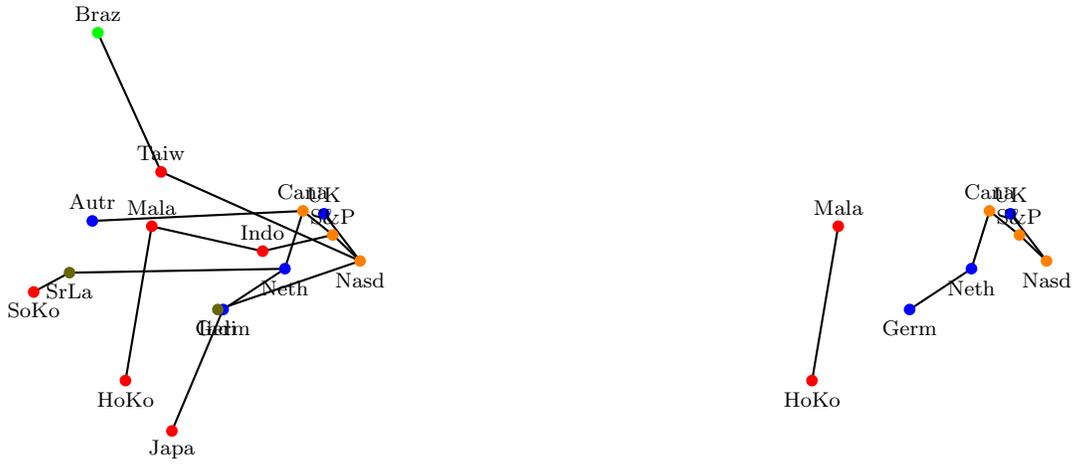
\noindent Figure 54: three dimensional view of the minimum spanning tree for the second semester of 1986. The second figure is the pruned spanning tree.

%

\noindent Figure 55: three dimensional view of the minimum spanning tree for the first semester of 1987. The second figure is the pruned spanning tree.

%

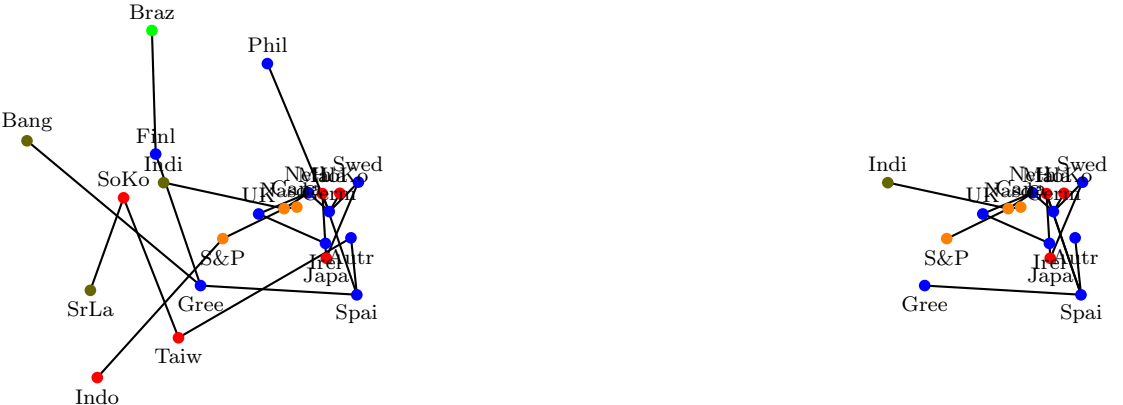
\noindent Figure 56: three dimensional view of the minimum spanning tree for the second semester of 1987. The second figure is the pruned spanning tree.

\newpage 

%

\vskip 1.9 cm

\noindent Figure 57: three dimensional view of the minimum spanning tree for the first semester of 1997. The second figure is the pruned spanning tree.

%

\vskip 1.4 cm

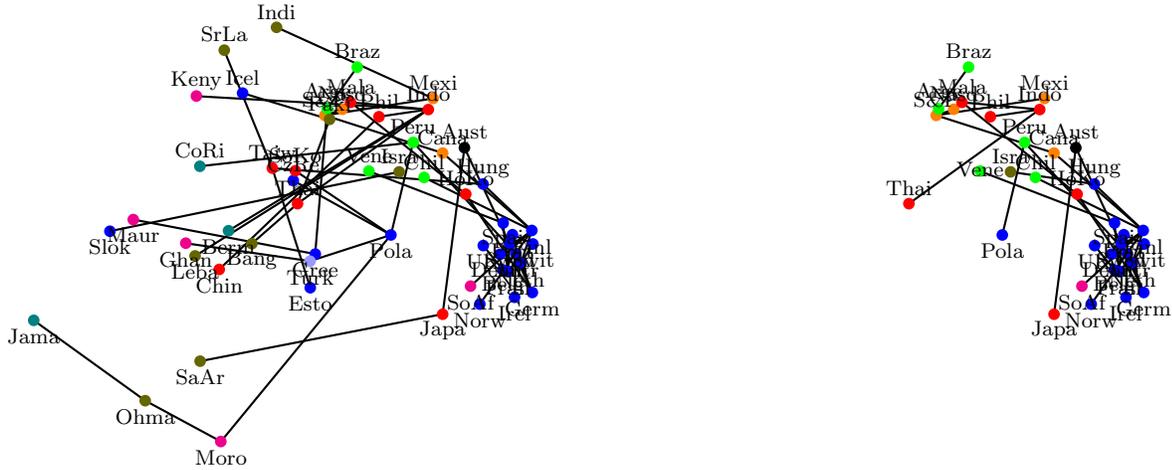
\noindent Figure 58: three dimensional view of the minimum spanning tree for the second semester of 1997. The second figure is the pruned spanning tree.

%

\vskip 1.5 cm

\noindent Figure 59: three dimensional view of the minimum spanning tree for the first semester of 1998. The second figure is the pruned spanning tree.

\newpage 

%

\vskip 1.5 cm

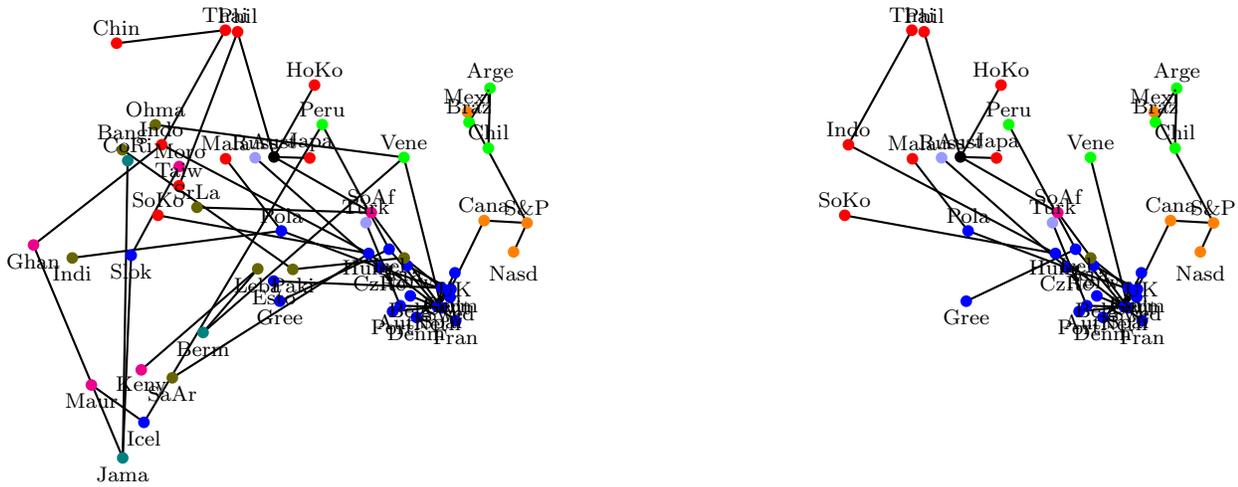
\noindent Figure 60: three dimensional view of the minimum spanning tree for the second semester of 1998. The second figure is the pruned spanning tree.

%

\vskip 0.8 cm

\noindent Figure 61: three dimensional view of the minimum spanning tree for the first semester of 2000. The second figure is the pruned spanning tree.

%

\vskip 0.8 cm

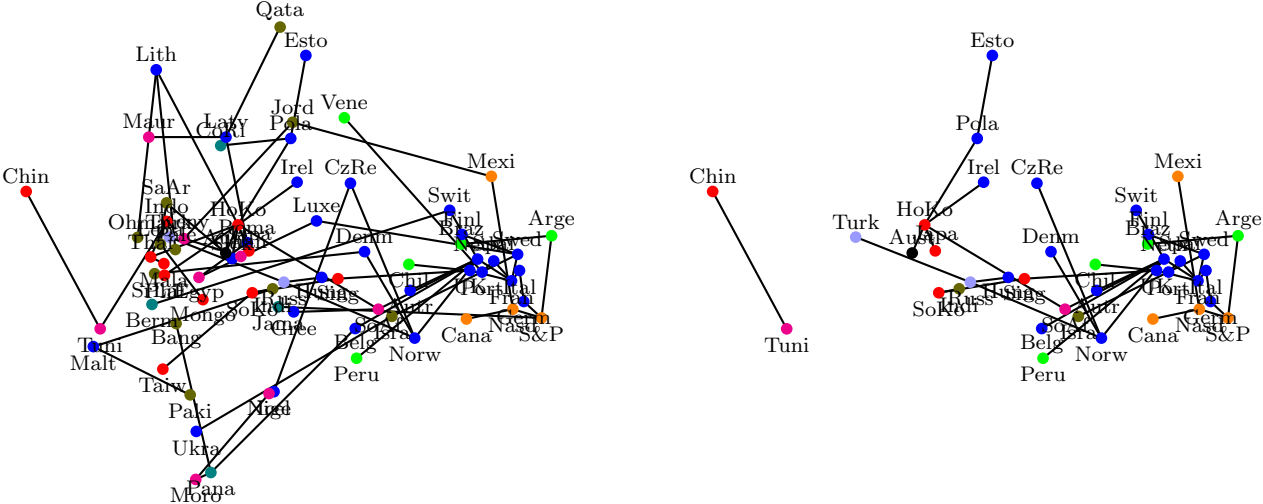
\noindent Figure 62: three dimensional view of the minimum spanning tree for the second semester of 2000. The second figure is the pruned spanning tree.

\newpage

%

\vskip 0.8 cm

\noindent Figure 63: three dimensional view of the minimum spanning tree for the first semester of 2001. The second figure is the pruned spanning tree.

%

\vskip 1.7 cm

\noindent Figure 64: three dimensional view of the minimum spanning tree for the second semester of 2001. The second figure is the pruned spanning tree.

\newpage

%

\vskip 1.5 cm

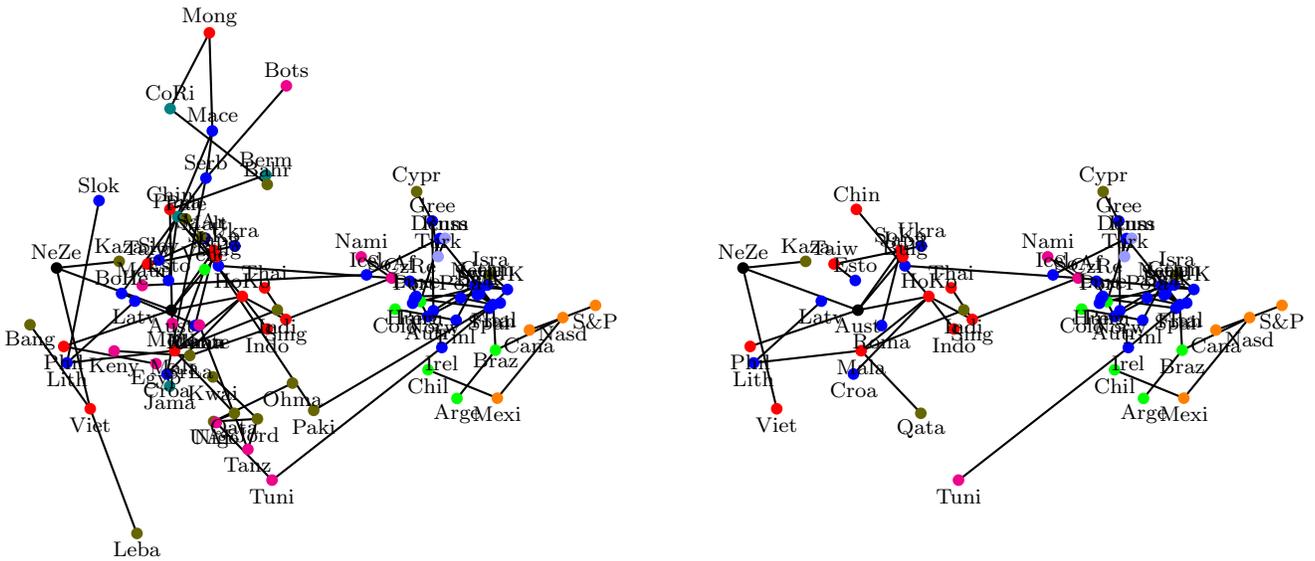
\noindent Figure 66: three dimensional view of the minimum spanning tree for the second semester of 2007. The second figure is the pruned spanning tree.

\newpage

%

\vskip 1.5 cm

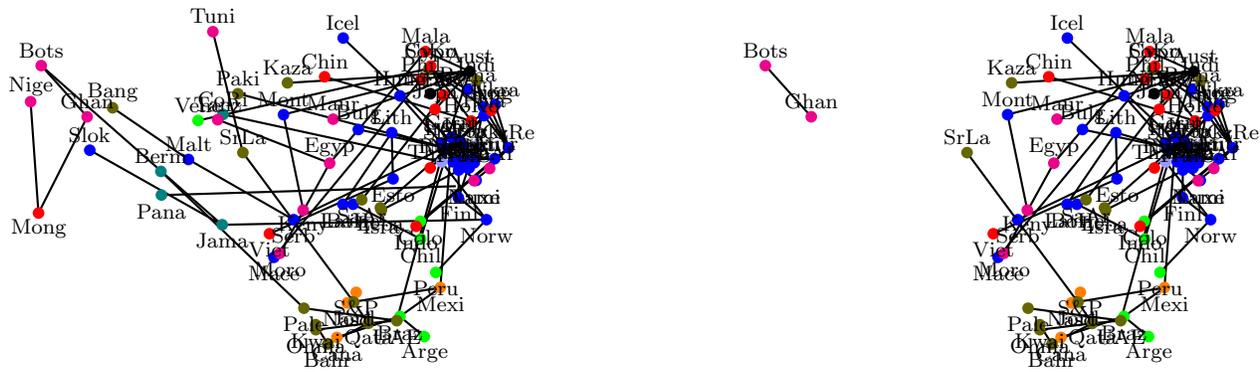
\noindent Figure 68: three dimensional view of the minimum spanning tree for the second semester of 2008. The second figure is the pruned spanning tree.










\end{document}